\documentclass[aps, prx, twocolumn,reprint, superscriptaddress]{revtex4-2}

\usepackage{soul}
\usepackage{amsmath}
\usepackage{amssymb}
\usepackage{bm}
\usepackage{physics}
\usepackage{verbatim}
\usepackage[scr]{rsfso}
\usepackage{mwe}
\usepackage{graphicx}
\usepackage{overpic}
\usepackage[dvipsnames]{xcolor}
\definecolor{urlblue}{RGB}{6,69,173}

\usepackage[
	colorlinks=true,
	linkcolor=blue,
	anchorcolor=blue,
	citecolor=blue,
	urlcolor=blue,
	pdftitle={graviton_fci!}
]{hyperref}
\usepackage{tikz}
\usetikzlibrary{decorations.markings}

\tikzset{
  ->-/.style={
    decoration={markings, mark=at position #1 with {\arrow{latex}}},
    postaction={decorate}
  }
}

\newcommand{\connectarrow}[3][0.8]{
  \fill #2 circle (1pt);
  \fill #3 circle (1pt);
  \draw[thick, ->-=#1] #2 -- #3;
}

\newcommand{\rr}{\mathbf{r}}
\newcommand{\ex}{\mathbf{e}_{x}}
\newcommand{\ey}{\mathbf{e}_{y}}

\newcommand{\mbf}[1]{\mathbf{#1}}

\renewcommand{\paragraph}[1]{\textit{#1}.---}

\makeatletter
\newcommand{\thetitle}{\@title}
\makeatother

\begin{document}
\setstcolor{blue}
\title{Chiral Graviton Modes in Fermionic Fractional Chern Insulators}

\author{Min Long}
\thanks{these authors contribute equally}
\affiliation{Department of Physics and HK Institute of Quantum Science \& Technology, The University of Hong Kong, Pokfulam Road,  Hong Kong SAR, China}
\affiliation{State Key Laboratory of Optical Quantum Materials, The University of Hong Kong, Pokfulam Road,  Hong Kong SAR, China}

\author{Zeno Bacciconi}
\thanks{these authors contribute equally}
\affiliation{SISSA --- International School for Advances Studies, via Bonomea 265, 34136 Trieste, Italy}
\affiliation{ICTP --- The Abdus Salam International Centre for Theoretical Physics, Strada Costiera 11, 34151 Trieste, Italy}

\author{Hongyu Lu}
\affiliation{New Cornerstone Science Lab, Department of Physics, The University of Hong Kong, Pokfulam Road, Hong Kong SAR, China}
\affiliation{HK Institute of Quantum Science \& Technology, The University of Hong Kong, Pokfulam Road, Hong Kong SAR, China}
\affiliation{State Key Laboratory of Optical Quantum Materials, The University of Hong Kong, Pokfulam Road,  Hong Kong SAR, China}

\author{Hernan B. Xavier}
\affiliation{ICTP --- The Abdus Salam International Centre for Theoretical Physics, Strada Costiera 11, 34151 Trieste, Italy}
\affiliation{SISSA --- International School for Advances Studies, via Bonomea 265, 34136 Trieste, Italy}

 \author{Zi Yang Meng}
 \email{zymeng@hku.hk}
\affiliation{Department of Physics and HK Institute of Quantum Science \& Technology, The University of Hong Kong, Pokfulam Road,  Hong Kong SAR, China}
\affiliation{State Key Laboratory of Optical Quantum Materials, The University of Hong Kong, Pokfulam Road,  Hong Kong SAR, China}

\author{Marcello Dalmonte}
 \email{mdalmont@ictp.it}
\affiliation{ICTP --- The Abdus Salam International Centre for Theoretical Physics, Strada Costiera 11, 34151 Trieste, Italy}
\affiliation{Dipartimento di Fisica e Astronomia, Università di Bologna, via Irnerio 46, I-40126 Bologna, Italy}

\date{\today}
\begin{abstract} 
Chiral graviton modes are hallmark collective excitations of Fractional Quantum Hall (FQH) liquids. However, their existence on the lattice, where continuum symmetries that protect them from decay are lost, is still an open and urgent question, especially considering the recent advances in the realization of Fractional Chern Insulators (FCI) in transition metal dichalcogenides and rhombohedral pentalayer graphene. Here we present a comprehensive theoretical and numerical study of graviton-modes in fermionic FCI, and thoroughly demonstrate their existence. We first derive a lattice stress tensor operator in the context of the fermionic Harper-Hofstadter(HH) model which captures the graviton in the flat band limit. Importantly, we discover that such lattice stress-tensor operators are deeply connected to correlation hole dynamics, readily generalizable to generic Chern bands. We then explicitly show the adiabatic connection between FQH and FCI chiral graviton modes by interpolating from a low flux HH model to a Checkerboard lattice model that hosts a topological flat band. In particular, using state-of-the-art matrix product state and exact diagonalization simulations, we provide strong evidence that chiral graviton modes are long-lived excitations in FCIs despite the lack of continuous symmetries and the scattering with a two-magnetoroton continuum. By means of a careful finite-size analysis, we show that the lattice generates an intrinsic decay rate for the graviton mode which is small compared to its energy. We discuss the relevance of our results for the exploration of graviton modes in FCI phases realized in solid state settings, experimental utility on diagnosis of fractional topological phases, as well as cold atom experiments.

\end{abstract}
\maketitle
\section{Introduction}

The discovery of Fractional Chern Insulators (FCIs) has opened a new frontier in the exploration of topological order on the lattice~\cite{Tang2011_flat_chern_band, Sun2011_flat_chern_band, Neupert2011_flat_chern_band, Sheng2011_FQAH_checkerboard_fermion, Regnault2011_FCI,Xiao2011_quantum_hall,wuAdiabatic2012,KaiDasSarma_prl2011_flatbandsCB}. Recent experimental breakthroughs - including the discovery of fractional quantum anomalous Hall (FQAH) states~\footnote{We note in this paper the FQAH and FCI are synonyms.} in two-dimensional moir\'e
materials~\cite{Cai2023_signature_fqah_mote2,Park2023_observation_fqah_mote2,Zeng2023_thermo_evidence_fqah_mote2,Xu2023_Observation_FQAH_tMote2,Lu2024_FQAH_multilayer_graphene} - have further motivated studies in understanding both material-related formation mechanisms, as well as fundamental questions about the relation between FCI and their fractional quantum Hall (FQH) cousins~\cite{Regnault2011_FCI,Wu2012Zoology,Hongyu_prl2024_thermodinamicFCI,luContinuous2025,Lu2024Interaction}, various work has been done on the similarity between the Landau Levels and Chern band~\cite{Roy2014_band_geometry_fci,Wang2021_geometry_flatband,Ledwith2023_vortexability,Parameswaran2012Fractional,Parameswaran2013flatband,Wu2013Bloch,Sterdyniak2013Series,Wu2012Wannier}, and the adiabatic relation between FCI and FQH ground state(in terms of translational symmetry, entanglement, ground state dynamics and generalized Pauli principle)~\cite{Qi2011Generic, wuAdiabatic2012,Liu2013Adiabatic,Bernevig2012Emergent,Scaffidi2012Adiabatic,Barkeshli2012nematic,Luchli2013Hierarchy,Jian2013Crystal}. While FCIs share many phenomenological features with FQH liquids---such as fractionalized quasiparticles and topological ground-state degeneracies---they differ crucially in the fact that they lack a continuum magnetic field and the associated continuous translational and rotational symmetries. As a result, one of the central challenges in the field is to identify whether geometrical properties and collective excitations of FQH states survive, in a robust and universal way, when the perfectly flat Landau Level (LL) is replaced by a topological band with nonuniform quantum geometry and Berry curvature~\cite{shenMagnetorotons2024,Hongyu_prl2024_thermodinamicFCI,repellin2014,Lu2024Interaction,Dong2018Charge,longSpectra2025}.

A fundamental property of FQH liquids is the existence of emergent metrics \cite{haldane2009hall,haldane2011geometrical, YangHaldane_prb2012_metric,ippoliti2018geometry,golkar2016spectral} whose long-wavelength fluctuations give rise to gapped chiral modes with a charge-neutral and spin-2 character, often referred to as gravitons \cite{gromov2017bimetric,liu2018quench,liou2019chiral,nguyen2022multiple,kumar2022neutral,yang2016acoustic,liuResolving2024,Balram_2024,Yang_prl2012_modelwavefunctions_graviton,Nguyen2021Dirac,Wang2022Analytic,wang2023Geometric,Haldane2021Graviton,pu2024microscopic}. In another aspect, the graviton mode is understood in terms of the magnetoroton, which is the lowest lying neutral excitation in the FQH system, and reach it's minimum at finite momentum. The long wavelength limit of magnetoroton, albeit embedded in continuum, is a well-defined quasi-particle called graviton.
The FQH graviton-mode is an almost unique example of well defined excitation often embedded in a continuum spectrum: in continuous space, they are thought to be protected by emergent conservation laws~\cite{golkar2016spectral}, and are captured by the spectral response of chiral components of the stress tensor \cite{nguyen2014lowest,nguyen2021probing,nguyen2022multiple,kumar2024}. Very recent experiments using polarized inelastic light scattering have now reported direct evidence of these chiral graviton modes, establishing their existence as a genuine, measurable property of FQH liquids~\cite{liang2024evidence}.Apart from the theoretical importance, the graviton mode provide possible new bulk probe for the characterization of fractional topological phases~\cite{Haldane2021Graviton,Liu_prl2025_nonabelianFCI}.

\begin{figure*}
    
        \begin{overpic}
    [width=\linewidth]{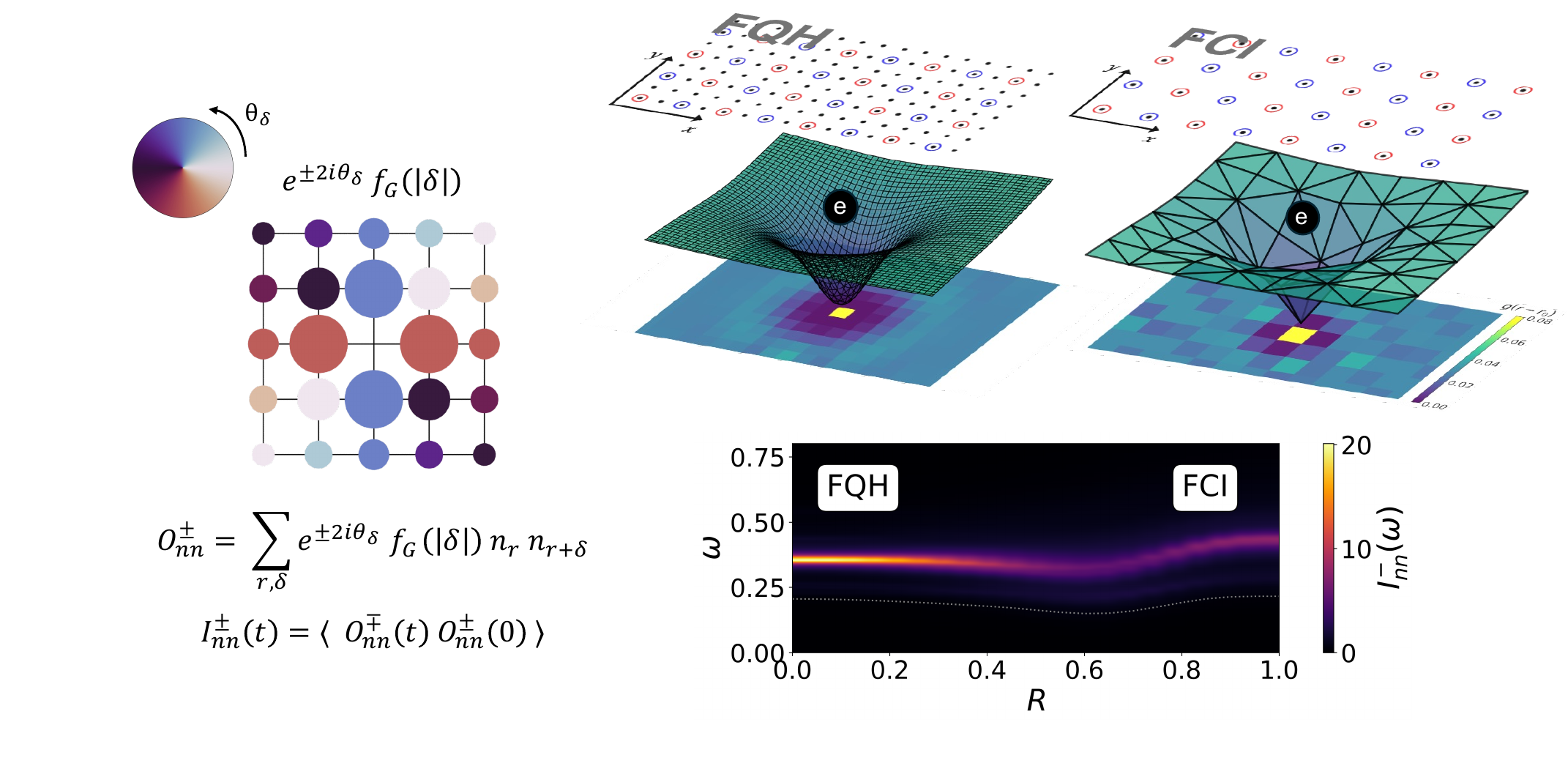}
    \put(5,48){(a)}
    \put(35,48){(b)}
    \put(67,48){(c)}
    \put(45,23){(d)}
    \end{overpic}

    \vspace*{-8mm} 
    \caption{\textbf{Chiral graviton operator on lattice and adiabatic connection between FQH and FCI.} (a) Representation of the chiral lattice operator structure which captures the graviton dynamics in generic FCI phases. The definition of the operator is given in Eq.~\eqref{eq:definition_Onn}. In the figure, we plot $O^\pm_{nn}$ on a single site. The color represents the chiral phase factor $e^{\pm 2 i \theta_\delta}$ attached to the density quadrupolar operator and the size of the circle labels the weigth $f_G(|\delta|)$. (b-c) Top layer depicts the lattice geometries used for  FQH-like states on low flux HH models ($n_\phi=1/8$) and FCI states on a CB lattice. The bottom layer shows the MPS numerically evaluated density-density static correlations $\langle n_r n_{r'} \rangle / \langle  n_{r'} \rangle \langle n_r \rangle $ featuring a ``correlation hole" which is then artistically represented in the middle layer. (d) Computed spectral function of the chiral graviton operator defined in panel (a) across the adiabatic interpolation from FQH to FCI limits (same data are also shown in Fig.~\ref{fig:adiabatic_graviton} (a)). The two-magnetoroton energy scale is shown as a dotted gray line.
    }
    \label{fig:sketch_fig1}
\end{figure*}
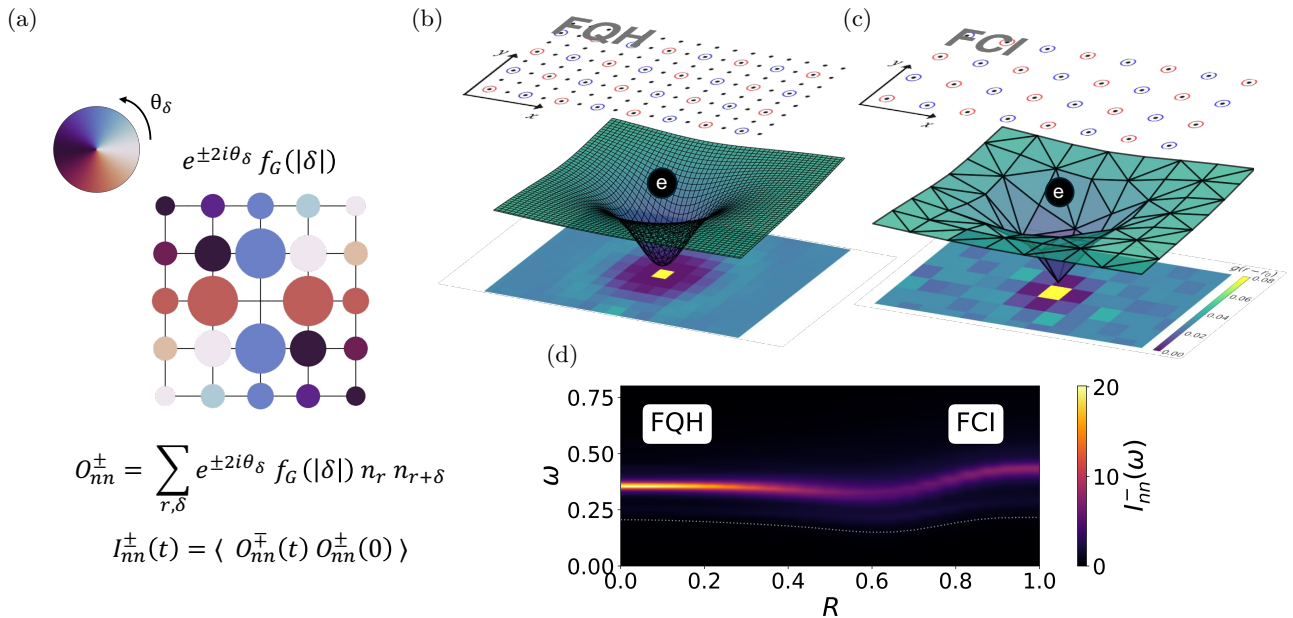

A natural question arises: do analogous chiral graviton-like excitations exist in lattice realizations of FCIs? In other words, can an FCI support long-lived, chiral, spin-2 collective modes despite the discrete lattice symmetries and spatially varying Berry curvature that characterize its underlying single-particle states? Resolving this question is of fundamental importance at least for  the following three reasons: 1) it clarifies whether the geometric response encoded in the FQH liquid remains a defining feature of fractional topological matter beyond the continuum limit;
2) it provides insights on the role of Chern bands quantum geometry in the stabilization of FCI phases; 3) it establishes the potential of new dynamical probes \cite{liang2024evidence} in the context of identifying FCI phases beyond the current standards in Moir\'e materials and synthetic quantum matter \cite{leonard2023realization,lunt_prl2024_fqhfermions,Wang_science2024_realization_photonfqh}.

In this work, we address the aforementioned questions by a combined numerical and analytical study of chiral graviton modes from a lattice $\nu=1/3$ FCI state to continuum-like $\nu=1/3$ FQH liquid. To do so we construct explicit lattice graviton operators that continuously connect the stress-tensor description of continuum FQH liquids to that of lattice FCIs and demonstrate with state-of-the-art numerical simulations their efficacy in detecting chiral graviton modes (see schematics in Fig. \ref{fig:sketch_fig1}). Starting from a fermionic Harper–Hofstadter (HH) model at fractional magnetic filling, we derive a lattice expression of the chiral stress tensor operator~\cite{xavier2025chiralgravitonslattice} and find that it is equivalent to a quadrupolar operator built solely from local density correlators~\cite{longSpectra2025}. The latter can then be used  to faithfully probe the graviton on generic Chern bands with non-uniform quantum geometry. We utilize large-scale exact diagonalization (ED) and matrix-product-state (MPS) simulations, to analyze the graviton spectral functions across a controlled interpolation from the Landau-level regime to a flat-band Chern insulator~\cite{wuAdiabatic2012}. These results provide the first evidence - to the best of our knowledge - of a clear adiabatic continuity between geometric excitations in FQH, and spectral properties of FCI, providing a new viewpoint on FCIs spectral properties. 

Our analytical and numerical results, pictorially represented in Fig.~\ref{fig:sketch_fig1},  demonstrate that 

\begin{itemize}

    \item A well-defined Chiral graviton mode exist in FCI phases and is adiabatically connected to FQH graviton-modes, establishing the graviton as a unifying geometric excitation of FQH and its FCI cousins.
    \item The lifetime of FCI graviton-modes is large despite being embedded in a two magnetoroton continuum: its spectral peak can be clearly distinguished. Band-mixing effects do not introduce qualitative differences. 
    \item The chiral spectral response of the graviton mode can be used to witness the presence of the FCI phase, giving a new probe to be used in experimental realizations with cold atoms systems as well as Moir\'e materials.
\end{itemize}


The rest of the paper is organized as follows: In Sec.\ref{sec:II}, we derive chiral stress tensor operators for a fermionic HH model which naturally flow to their LL counterparts. Furthermore we analytically show that quadrupolar lattice density correlators also recover the same LL limit. In Sec.\ref{sec:III} we numerically show this equivalence by considering the HH model at different fluxes. Importantly we introduce an analysis on the numerically obtained graviton spectra that highlights how the decay rate of the graviton mode into magnetoroton continua decreases as the LL limit is taken. In Sec.\ref{sec:IV} we numerically show the adiabatic connection between graviton modes of FQH phase on such low flux HH models and that of an FCI phase on a checkerboard(CB) lattice. By large scale numerical simulations, we argue that the graviton decay rate remains small also in the thermodynamic limit. We also compare the chiral graviton spectra inside the FCI phase with that of nearby phases. Finally, Sec.\ref{sec:V} makes connections of our results and understanding to the ongoing experimental probes of the graviton modes and other characteristic features of FCIs in solid state settings as well as cold atoms.

\section{Graviton lattice operators}
\label{sec:II}
The magnetoroton mode, first described by Girvin–MacDonald–Platzman (GMP)~\cite{gmp1986magnetoroton}, is a typical neutral and gapped collective mode in fractional quantum Hall (FQH) liquids. and has been recently generalized to periodic potential case ~\cite{Kousa2025moire,Wolf2025Intraband}. It corresponds to a collective density excitation, hence can be built with projected density operators, and generally displays a minimum at finite wavevectors. In the long-wavelength limit, however, the projected density operator gradually becomes ineffective, due to charge conservation ($\bar{\rho}(\mathbf{q}=0)$ is just the total particle number) and dipole moment conservation in the LLL (freezing the cyclotron motion and leaving only the noncommuting guiding-center degrees of freedom). 
 Therefore, after an expansion in powers of the wavevector, the linear contribution is forbidden, and the leading nontrivial term appears at order $q^2$, corresponding to a quadrupolar (spin-2) operator associated with fluctuations of the intrinsic metric~\cite{Yang2025Quantum}.
In the FQH liquids with broken time-reversal symmetry (under magnetic field), the two spin-2 chiralities of the quadrupolar mode are typically non-degenerate and do not carry the same weight. Since such spin-2 excitations can be viewed as deformations of the FQH emergent metric~\cite{haldane2011geometrical,gromov2017bimetric,golkar2016spectral}, they are referred to as (chiral) gravitons.

In the case of FQH states defined on the lowest Landau level (LLL), graviton-mode excitations have been argued to be captured by measuring spectral functions of chiral stress tensor operators \cite{nguyen2014lowest,liou2019chiral,nguyen2022multiple}, i.e., capturing the dynamical response of the emergent FQH metric to external perturbations. A direct coupling to the chiral kinetic stress tensor of the FQH liquid has then been shown to appear in inelastic Raman scattering \cite{nguyen2021probing}, resulting in the recent direct observation of chiral graviton modes \cite{liang2024evidence}. 

The exploration of graviton-modes away from LL systems, such as in lattice FCI systems, has been more recently intensely investigated~\cite{xavier2025chiralgravitonslattice,longSpectra2025,wangDynamics2025,shenMagnetorotons2024}, but a comprehensive theory of emergent metrics and graviton-mode excitations in the absence of continuum translational and rotational symmetry is still missing. As such, different approaches and choices of operators to describe the graviton spectra have been put forward, and no consensus has been reached. For example, some works favor the existence of graviton modes in FCI~\cite{shenMagnetorotons2024,longSpectra2025}, whereas some argue that they will decay into different scattering channels due to the discrete lattice symmetry and therefore not long-lived~\cite{wangDynamics2025}.

In this section, we generalize (and in fact, combine) to the fermionic case the approaches of Refs.~\cite{xavier2025chiralgravitonslattice,longSpectra2025}, leveraging an explicit continuum limit of a lattice model (Sec. \ref{sec:II_continuumlimit}) to build an explicit lattice stress tensor operator (Sec. \ref{sec:II_stresstensor}), and propose a graviton operator based on short range density correlations (Sec. \ref{sec:II_densitygrav}). The latter will be then used to probe generic topological bands away from the Landau level. Finally, we end the section by introducing the numerical methods used in this paper to probe the aforementioned operators (Sec. \ref{sec:II_numerics}).

\subsection{Continuum limit of the Fermionic Harper-Hofstadter model}\label{sec:II_continuumlimit}

We consider the fermionic Harper-Hofstadter model~\cite{hofstadter1976energy} on a square lattice with $n_\phi=1/q$ flux per plaquette, as this is the simplest lattice model that features LL physics in the small flux limit, and $\nu = 1/3$ lowest band filling:
\begin{align}
    \hat{H}_{hh}&=\hat{H}_0+\hat{H}_V\;,
\end{align}
with:
\begin{align}
    \hat{H}_0&=-t \sum_{\langle i ,j\rangle} \left(e^{i\phi_{ij}} \, c^\dagger_i c_j +\mathrm{h.c.}\right)\\ 
    \hat{H}_V&= \frac{V}{2}\sum_{ij}f(|r_{i}-r_j|) n_in_j
\end{align}
where $\langle ij\rangle$ denotes nearest neighbor sites $i$ and $j$, $\phi_{ij}= -\phi_{ji}$ encodes the vector potential (will be specified later) and $f(r_{ij})$ the two-body interaction terms. This model is closely related with the cold atom systems pierced by synthetic gauge fields~\cite{jaksch2003creation,aidelsburger2013realization,mancini2015observation,hafezi2007fractional}. It hosts the FQH state that has been widely studied from numerical sides~\cite{Bauer2016QuantumGeometry, Motruk2016Density,gerster2017fractional}, and realized experimentally~\cite{leonard2023realization,wang2024realizing}

In the \textit{continuum} limit, i.e., small flux $n_\phi\to0$, and for finite range interactions $f(r>r_c)=0$, the system at low energies can be expanded in terms of a continuum fermionic field:
\begin{align}
    \psi(r) \sim \frac{c_i}{a_0}+\cdots  \qquad \mathrm{with}\qquad r=(x_i,y_i) \; ;
\end{align}
where $a_0$ is the lattice spacing. The resulting continuum Hamiltonian is then written as:
\begin{align}\label{eq:hamiltonian_continuum}
    \mathcal{H}= &\int d^2r \;\frac{1}{2m}\psi^\dagger D_aD^a \psi  + \;V_1\;\rho\nabla^2 \rho \;;
\end{align}
where $\rho=\psi^\dagger \psi$. Here, the kinetic energy is expressed in terms of the effective mass $m=\frac{1}{2ta_0^2}$ and the covariant derivative $D^a =\partial_a -A_a$ with $\boldsymbol{B}=\nabla \times \boldsymbol{A}$. The interaction strength of the first Haldane pseudopotential is parametrized by an energy scale $V_1$ which depends on the specificity of the lattice interactions $V f(r_{ij})$ and its range $r_c$. In the continuum, two Ward identities are present, which concern the conservation of charge density $\rho=|\psi|^2$ and momentum:
\begin{align}
\partial_t\rho= \boldsymbol{\nabla}\cdot \boldsymbol{j}\qquad
    \partial_t (m \boldsymbol{j}) = \boldsymbol{\nabla}\cdot \boldsymbol{\overleftrightarrow{\mathcal{T}}} + \boldsymbol{f}_{ext}
    \label{eq:eom}
\end{align}
where $m\boldsymbol{j}$ is the particle momentum, whose time derivative is bound to the divergence of stress tensor $\boldsymbol{\overleftrightarrow{\mathcal{T}}} $ and external force $\boldsymbol{f}_{ext}$. 
\subsection{Lattice stress tensor}\label{sec:II_stresstensor}
In order to get a lattice discretization of the stress tensor operator we follow the procedure introduced in Ref. \cite{xavier2025chiralgravitonslattice}. In particular, to guide our ansatz we seek to match Eq.~\eqref{eq:eom} with the Heisenberg equation for the current:
\begin{align}\label{eq:heisenberg_eom}
    \partial_tj^a = i[H,j^a].
\end{align}
We are interested in a low-energy equivalence, but the above lattice equation holds many contributions, including inter-band kinetic terms corresponding to inter-Landau Level terms in the continuum. Note that in order to cancel the external force $f_{ext}$ we can repeat the commutator using only the free part of the Hamiltonian $H_0$. In the continuum this procedure would cancel the Lorentz force ($f_{ext}^a\propto\epsilon_{ab} j^b$), while on the lattice it takes into account also other forces ($f_{ext}^a\propto j^a $) generated by the lattice at the free particle level which are expected to be controlled by $n_\phi$. Furthermore, in order to find a suitable lattice expression which recovers the form of Eq. \eqref{eq:eom} from Eq. \eqref{eq:heisenberg_eom} we must symmetrize the obtained commutator using translation and inversion symmteries.

For the sake of concreteness, we here focus on the case where only nearest neighbor interactions are considered, i.e., $f(a_0)=1$ and 0 otherwise. Furthermore, we stay in the limit $V\gg t$, where we can project the Hamiltonian $H_{hh}$ to a constrained Hilbert space for the fermions, where nearest neighbor pairs are forbidden. As such, the effect of the interaction is treated by imposing new geometric constraints on the fermionic operators $\bar{c}$ and  $\bar{c}^\dagger$, which obey a fermion exclusion algebra given by the condition $n_in_j=0$ for all nearest-neighbor pairs $i$ and $j$.

The interaction terms are hence naturally incorporated, rendering a constrained hopping Hamiltonian,
\begin{align}
    \bar{H}_{hh}=-t \sum_{\langle i ,j\rangle} \left(e^{i\phi_{ij}} \, \bar{c}^\dagger_i \bar{c}_j +\mathrm{h.c.}\right) .
    \label{eq:proj_ham}
\end{align}
Under the projection, current operators become $\bar{j}_i^a = -(\bar{c}^\dagger_i \bar{c}_{i+e_a}e^{i\phi_{ij}}-h.c.), a\in \{x,y\}$. The stress tensor can then be evaluated by the procedure outlined above and as detailed in App.\ref{sec:appA}. 

Here we report as an example how the diagonal terms $T_{aa}$ with $a=x,y$ are calculated. %
To simplify the analysis we split the Hamiltonian into terms that
generate motion along $x$ and $y$ directions: $\bar{H} = \bar{H}_x + \bar{H}_y$. In this
way the diagonal components $T_{aa}$ can be readily identified. We are left with 
\begin{equation}
    \begin{aligned}
    T^{xx}_{\rr+\ex} - T^{xx}_\rr = \partial_x T^{xx}_\rr &= i[\bar{H}^x,\bar{j}_\rr^x]-i[H^x_0,j_\rr^x], \\ 
    T^{yy}_{\rr+\ey} - T^{yy}_{\rr} = \partial_y T^{yy}_\rr &= i[\bar{H}^y,\bar{j}_\rr^y]-i[H^y_0,j_\rr^y], \\ 
    \end{aligned}
    \label{eq:commutators}
\end{equation}
where $j_i^a = -i(c^\dagger_i c_{i+e_a}e^{i\phi_{ij}}-h.c.)$, $a\in \{x,y\}$, is the discretized current operator on lattice. As detailed in App.\ref{sec:appA}, the evaluation of the above commutator give rise to correlated hopping terms which we must symmetrize under inversion in order to recover a translational invariant form for the stress tensor $T_\rr^{aa}$. A similar procedure is then followed for the off-diagonal term $T_\rr^{xy}$ (App.\ref{sec:appA}).  
The final result for all components of $T^{ab}_\rr$ consists of a specific combination of correlated hoppings which respect the nearest neighbour constraint introduced above. In order to write its full expression, we introduce the shorthand notation
\begin{equation}
    \begin{aligned}
    &B_{\rr}^{\ex}=e^{i\phi_{\rr}^x}c^\dagger_{\rr+\ex} c_\rr,\\
    &A_\rr^{\ex,\ey} = e^{i\phi_\rr^{-x} +i\phi_\rr^y}\;c^\dagger_{\rr+\ey}n_\rr c_{\rr-\ex}.
\end{aligned}
\end{equation}
Here $\phi_\rr^a$ is abbreviation for $\phi_{\rr+\mathbf{e}_a,\rr}$, $a\in\{x,y\}$. Then,
the diagonal components of the stress tensor, e.g., $T_{\rr}^{xx}$, are written as:
\begin{align}
    T^{xx}_{\rr}= A_\rr^{x,x}
    +\frac{B_{\rr-\ex}^x}{2}\left(\sum_{d\rr=\pm \ey,\ex} B_{\rr+d\rr}^{\ex} \right)+\mathrm{h.c}.
\end{align}
The other one $T^{yy}_{\rr}$ follows by replacing $x$ with $y$. The full off-diagonal components are instead:
\begin{align}
    T^{xy}_{\rr} &=\frac{1}{2} \sum_{d\rr=\pm \ex,\pm \ey} s_{d\rr} s_{d\rr'} A_\rr^{d\rr,d\rr'}+\\ \nonumber
    & +\frac{1}{2}\sum_{d\rr=\pm \ex,\pm \ey} s_{d\rr}s_{d\rr'}  \left(B_{\rr+d\rr'}^{d\rr'} +  B_{\rr+d\rr}^{d\rr'}\right)B_{\rr-d\rr}^{d\rr}  +\mathrm{h.c}
\end{align}
where we introduce $d\rr'$ as a shorthand for a counterclockwise rotation by $90^\circ$ of $d\rr$ (i.e. $\ex' =\ey$ and $\ey'=-\ex$) and $s_{d\rr}$ the sign of direction $d\rr$ (i.e. $s_{\ex}=+1$ and $s_{-\ex}=-1$).

Finally, the chiral components of the stress-energy tensor at zero momentum then read:
\begin{align}
    O^\pm_s =\sum_r T_r^{xx} -T^{yy}_r \pm 2i T^{xy}_r
    \label{eq:op_s}
\end{align}
We remark that such expressions are derived for strong ($V\gg t$) nearest neighbor interactions and close to the continuum limit ($n_\phi\to0$) of the HH model. In this limit the operator $O_s^\pm$ will correspond to the continuum stress tensor\cite{nguyen2014lowest} plus band mixing contributions. As we will inspect numerically in Sec. \ref{sec:III_spectra}, the operator $O_s^\pm$ projected to the lowest band flows to the continuum stress tensor of Eq. \ref{eq:hamiltonian_continuum}.   

\subsection{Quadrupolar density graviton operator}\label{sec:II_densitygrav}
A different approach to capture the dynamics of graviton-modes on a discrete lattice was introduced in the context of bosonic FCI on the honeycomb lattice~\cite{longSpectra2025}. As such operator structure has been empirically observed to capture the correct physics, we here generalize its form and analytically connect it to the stress tensor in the continuum limit. 

We start with the important physical insight that, in the continuum limit, the emergent FQH metric controls the shape of short-range correlations in the liquid \cite{haldane2011geometrical}. Therefore, as we move from the FQH limit to the FCI limit, we need an operator that properly captures the structure of short-range correlation holes~\cite{Qiu2012Model,liou2019chiral}, present in both continuum (FQH) and lattice (FCI) (see Fig. \ref{fig:sketch_fig1}). A minimal example that is short-range, quadrupolar and chiral (see Fig.~\ref{fig:sketch_fig1}(a)) reads
\begin{equation}
    \begin{aligned}
    O^{\pm}_{nn} & =\sum_{r_i,\delta} e^{\pm 2i \arg{[\delta]}}f_G(\delta) n_{r_i} n_{r_i+\delta}. 
    \end{aligned}
\label{eq:definition_Onn}
\end{equation}
This operator explicitly captures the dynamics of short-range density-density correlations and the finite chiral spin structure expected from continuum graviton mode excitations. In particular, we explicitly show here that such density-density operator in the continuum limit flows to the stress tensor operators known in the continuum. 
In the continuum limit, assuming $f_G$ smooth over $l_B$, we have
\begin{align}
    n_in_{i+\delta} \to \rho(r_i)\rho(r_i+\delta).
\end{align}
Going to momentum space $\rho(q)=\int d^2r\, e^{iqr} \rho(r)$, and express the summation on $\mbf{\delta}$ into intergral in polar coordinate $\boldsymbol{\delta} = (\delta,\theta) $

we have
\begin{align}
    O^{\pm}_{nn} & \to\sum_{\boldsymbol{q}}\rho(\boldsymbol{q})\rho(-\boldsymbol{q})\int d^2\delta \;\;e^{i\boldsymbol{q}\cdot\boldsymbol{\delta}} e^{\pm 2i \arg{[\delta]}}f_G(|\delta|). 
    \label{eq: density_op LL proj}
\end{align} 
For a proper comparison, we also use the projection to the LLL:
\begin{equation}
\begin{aligned}
\mathcal{P}_{LLL}\rho(\boldsymbol{q}) \mathcal{P}_{LLL} &= \mathcal{P}_{LLL}\sum_i e^{-i\boldsymbol{q}\cdot \boldsymbol{r_i}} \mathcal{P}_{LLL} \\ &
    = \mathcal{P}_{LLL}\sum_i e^{-\frac{iq}{2}z_j^* - \frac{iq^*}{2}z} \mathcal{P}_{LLL}   \\ & = e^{-iq\frac{d}{d z} - \frac{iq^*}{2}z}  =  e^{-\frac{1}{4}|ql_B|^2}\bar{\rho}(\boldsymbol{q}). \\
\end{aligned}
\end{equation}
The projected density operator is $\bar{\rho}(\boldsymbol{q}) = \sum_i e^{-\frac{iq}{2}z_j^*}e^{- \frac{iq^*}{2}z}  $~\cite{gmp1986magnetoroton,haldane2011geometrical} and the form factor $e^{-\frac{1}{4}|ql_B|^2}$ could be evaluated using Baker-Campbell-Hausdorff formula. 

The integration over the azimuth could be explicitly evaluated by Taylor-expanding $e^{i\boldsymbol{q}\cdot\boldsymbol{\delta}}$ to order $\boldsymbol{\delta}$.

\begin{equation}
     \sum_\delta f_G(\boldsymbol{\delta}) e^{2i\theta}e^{i\boldsymbol{q}\cdot\boldsymbol{\delta}} = \sum_{\boldsymbol{\delta},n} \frac{f_G(\delta)}{n!} e^{\pm 2i \theta} (i\boldsymbol{\delta} \cdot \boldsymbol{q})^n.
\end{equation}

By exchanging the order of summation over $n$ and integration over $\delta$, the integral over $\delta$ can be evaluated directly. Labeling the contribution by order as $V_G^n(\boldsymbol{q})$ and substituting into Eq.~\eqref{eq: density_op LL proj}, we derive
\begin{equation}
     O_{nn}^-\to \sum_n \sum_\mathbf{q} V_G^n(\boldsymbol{q}) e^{-\frac{1}{2}(ql_B)^2} \bar{\rho}_q\bar{\rho}_{-q}.
\end{equation}
And 
\begin{equation}
    V_G^n(q) = \left \{ \begin{aligned}
         &0, ~~  n = 0,1 \\
         &C_n(q_x \pm iq_y)^2 |\boldsymbol{q}|^{n-2}, ~~ n>1
    \end{aligned} \right.
\end{equation}
Where $C_n$ is a constant without $\mathbf{q}$ dependence but is a functional of $f_G$. For bosons, the leading contribution would be the $n=2$ corresponding to the contact interaction (which onsite hard core condition what has no momentum dependence). For fermions, this contribution automatically vanishes and the leading one becomes the $n=3$ corresponding to the shortest possible range interaction, i.e. the $V_1$ pseudopotential.
Therefore, we arrive at the correspondence of the density graviton operator and the continuum stress tensor operator.

The real-space operator $O_{nn}^\pm$ not only offers direct generalizability to any lattice scenario, but also a physically transparent interpretation of the graviton-mode as a quadrupolar distortion of short-range density correlations. We note that the correspondence between $O_{nn}^{\pm}$ and the continuum metric tensor can also be derived using the holomorphic property of LLL, which is detailed in App.\ref{App: derivation_complex}.

\subsection{Numerical spectrum evaluation}
\label{sec:II_numerics}
In order to probe the spectral content of the operators described so far and detect the graviton-modes on microscopic models, we will calculate spectral functions for the operators defined as
\begin{align}\label{eq:IO_definition}
    I_O^\sigma (\omega)= \frac{1}{\mathcal{N}_-}\sum_n|\bra{n}O^\sigma\ket{0}|^2 \delta_\eta(\omega-E_n+E_0),
\end{align}
where $O$ is either the lattice stress tensor (subscript $s$) defined in Eq.~\eqref{eq:op_s} or the density-density quadrupolar (subscript $nn$) operator given by Eq.~\eqref{eq:definition_Onn}, $\ket{n}$ are many-body eigenstates, $E_n$ their energies, and $2\eta$ the full-width-half-maximum (FWHM) of the Lorentzian broadening used for the delta function. The normalization $\mathcal{N}_-$ is with respect to the negative chirality, and is defined as:
\begin{align}\label{eq:normalization_def}
    \mathcal{N}_-= \int_{-\infty}^{\Lambda_{lb}} d\omega\;\sum_n|\bra{n}O^-\ket{0}|^2 \delta_\eta(\omega-E_n+E_0),
\end{align}
with $\Lambda_{lb}$ a cutoff that selects weight in the lowest band. Note that as $O$ is an extensive operator, $\mathcal{N}_-$ will be extensive. Instead of directly calculating each eigenstate $\ket{n}$, which can be challenging if $n$ is large, we will use different methods. For all ED spectra, we will use the Lanczos continued-fraction method \cite{ED_spectralfunctions_koch,Weisse_rmp2008_kpm} to directly access the spectral function in Eq. (\ref{eq:IO_definition}) as the imaginary part of a Green's function, where $\eta$ naturally enters as an imaginary shift of the frequency $\omega$. 

Further, as detailed in App. \ref{sec:app_edtricks}, we crucially average over different ground states and twisted boundary conditions on the torus in order to reduce finite size effects and obtain smooth spectra. In MPS simulations, the spectrum is obtained via the Fourier transform of such time-ordered Green's function:
\begin{equation}
\begin{aligned}
         I_O^\pm (\omega) &= \frac{1}{\mathcal{N}_-}  \int_0^T \ e^{i(\omega+i\eta) t } \big( \langle O^\mp(t)O^\pm\rangle \\ &
        -\langle O^\pm \rangle \langle O^\mp\rangle\big)dt.
\end{aligned}
\end{equation}
The bracket $\langle \bullet \rangle$ denotes the expectation value on the ground state. To compute the time-ordered Green's function, we first obtain the ground state via DMRG~\cite{White1992,White1993}, then time evolve $O^\pm |G\rangle$ using TDVP~\cite{Haegeman2011,Haegeman2016}. Note here that the resolution of the spectra is not only controlled by $\eta$ but also by the total evolution time $T$. We will always use $T\gtrsim2\eta^{-1}$ such that, at fixed $\eta$, the result does not depend on $T$. Further details of the MPS implementation and convergence test are discussed in App.\ref{sec:appD}.

We also examine the operator dependence of the spectrum in App. \ref{sec:app_fG}. The graviton peak in the negative chirality sector is robust. The chirality of the signal is also clear, albeit demonstrating modest operator-dependent behavior; the peak energy is stable.

\section{Fermionic Harper Hofstadter graviton spectra}
\label{sec:III}
In this section we present numerical results for the spectra of graviton operators discussed in the previous section in the context of the fermionic HH model. We start with a direct comparison of the spectra of two operators as a function of the flux per plaquette $n_\phi$ (Sec. \ref{sec:III_spectra}) and then introduce an analysis on such finite size spectral functions which allow us to extract an intrinsic lifetime of the graviton-mode (Sec. \ref{sec:HHnum_lifetime}). We will present results for  finite range interaction $f(r)=1/r$ for $r\leq r_c$ and 0 otherwise (the HH square lattice spacing is set to $a_0 = 1$). In all cases we will keep the operator $O_{nn}^\pm$ the same, i.e. $f_G(r)=1/r$ up to $r_G=2\sqrt{2}$ and 0 otherwise. We study square lattice systems of size $L_x \times L_y$ at $\nu=1/3$ magnetic filling, i.e., $N=\nu n_\phi L_xL_y$ particles.

\subsection{Numerical testing of $O^\pm_{nn}$ and $O^\pm_s$}\label{sec:III_spectra}
The derivation in Sec. \ref{sec:II} suggests that, in the LL limit $n_\phi\to 0$, the two lattice graviton operators $O^\sigma_n$ and $O^\sigma_s$ flow to the same continuum operator. When $n_\phi$ is finite, however, they are not in general the same.

\begin{figure}[t]
    \centering
    \begin{overpic}
    [width=\linewidth]{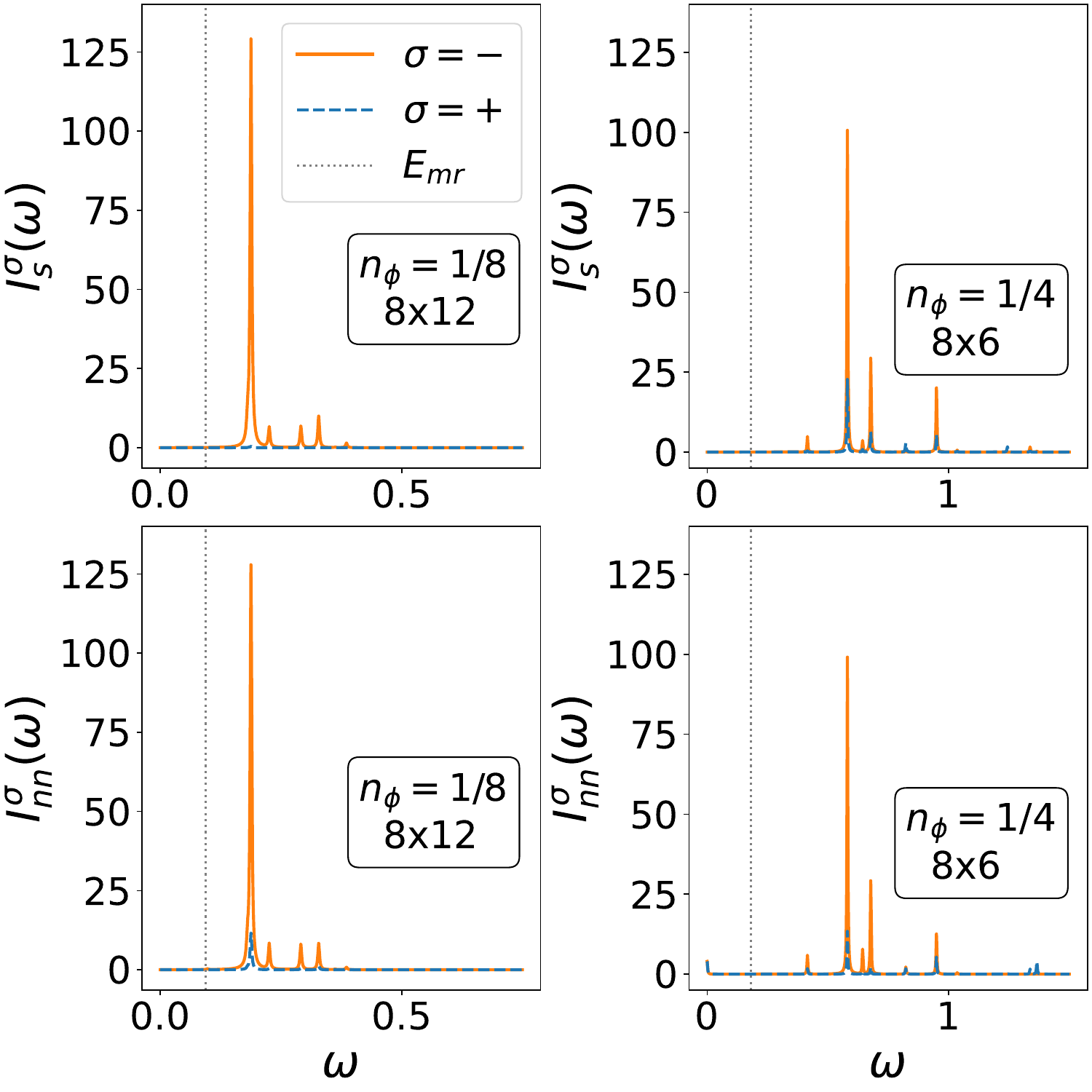}
    \put(15,94){(a)}
    \put(15,48){(c)}
    \put(92,94){(b)}
    \put(92,48){(d)}
    \end{overpic}
    \caption{\textbf{Full ED spectra of $O^\pm_{nn}$ and $O^\pm_{s}$ at different fluxes}. Chiral graviton full ED spectra on the fermionic $\nu=1/3$ Harper Hofstadter model with (a,b) lattice stress tensor operator $O^\pm_s$ and (c,d) density-density operator $O^\pm_{nn}$. Two fluxes are considered, in (a,c) on the left $n_\phi=1/8$ while in (b,d) on the right $n_\phi=1/4$. $N=4$ for all panels and broadening $\eta=0.002$. Vertical dotted line is twice the energy gap given by the magnetoroton energy $E_{mr}$. Interaction parameters are $V=10$ and $r_c=1$.}
    \label{fig:hh_currentvsdensity}
\end{figure}
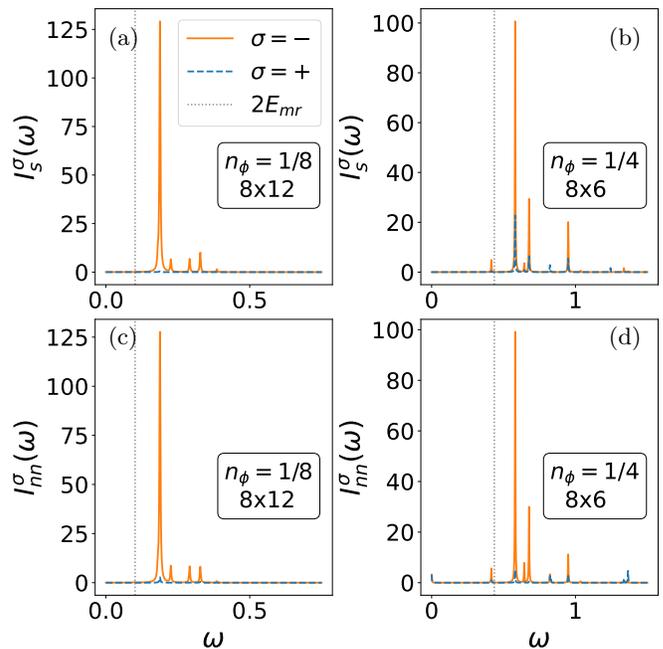

We start with a precise evaluation of the spectra with full ED (no band truncation) of small systems with $N=4$ particles for large nearest neighbor interactions ($V=10$ $r_c=1$) where the stress tensor operator is $O_s^\pm$ is derived. In Fig. \ref{fig:hh_currentvsdensity} we show results for the two lattice graviton operators, (a,b) top panels lattice stress tensor $O_s^\pm$ and (c,d) bottom panel density-density operator $O_{nn}^\pm$, for two fluxes (a,c) $n_\phi=1/8$ on the left and $n_\phi=1/4$ (b,d) on the right. Note that all spectra are normalized according to Eq. \eqref{eq:IO_definition} with $\eta=0.002$. The spectral weight is peaked and chiral in all cases, even though at $n_\phi=1/4$ the weight is in general more distributed. Closer to the LL limit $n_\phi=1/8$ (left panels), the lattice stress tensor $O^\pm_s$ (panel (a)) slightly better captures the chirality of the mode with respect to the density operator (panel (c)). At $n_\phi=1/4$ (right panels), the situation switches and the chirality is instead better captured by the density-density operator (see panel (b) vs panel (d) ). It is also important to point out that the lattice stress-tensor operator $O_s^\pm$ used for the top panels is strictly derived in the continuum limit for only nearest-neighbor interactions. While the justification for the density-density operator is also clear only in the continuum limit, these results show that the latter is less sensitive to band dispersion and Berry curvature variations, which are indeed sizable at $n_\phi=1/4$.

While not directly shown in the figure, also the total normalization of the spectral functions (not just up a lowest band cutoff $\Lambda_{lb}$) supports the above analysis. For the operator $O_s^-$ ($O_{nn}^-$) the weight in the lowest band relative to the total one up to $\omega=+\infty$ is $22\% \,(30\%)$ at $n_\phi=1/8$ and $3\% \,(15\%)$$n_\phi=1/4$. 

We also remark that the system sizes of Fig. \ref{fig:hh_currentvsdensity} have only $N=4$ particles, leading to total Hilbert space sizes for the two fluxes of $\mathcal{N}\sim 2\cdot10^4$ and $\mathcal{N}\sim 3\cdot10^5$. However, as the lowest band only has $\mathcal{N}_{l.b.}\sim 45$ states in a particular momentum sector, these results cannot give a quantitative indication on the fate of the mode in the thermodynamic limit.

\begin{figure}
    \centering
    \begin{overpic}
    [width=\linewidth]{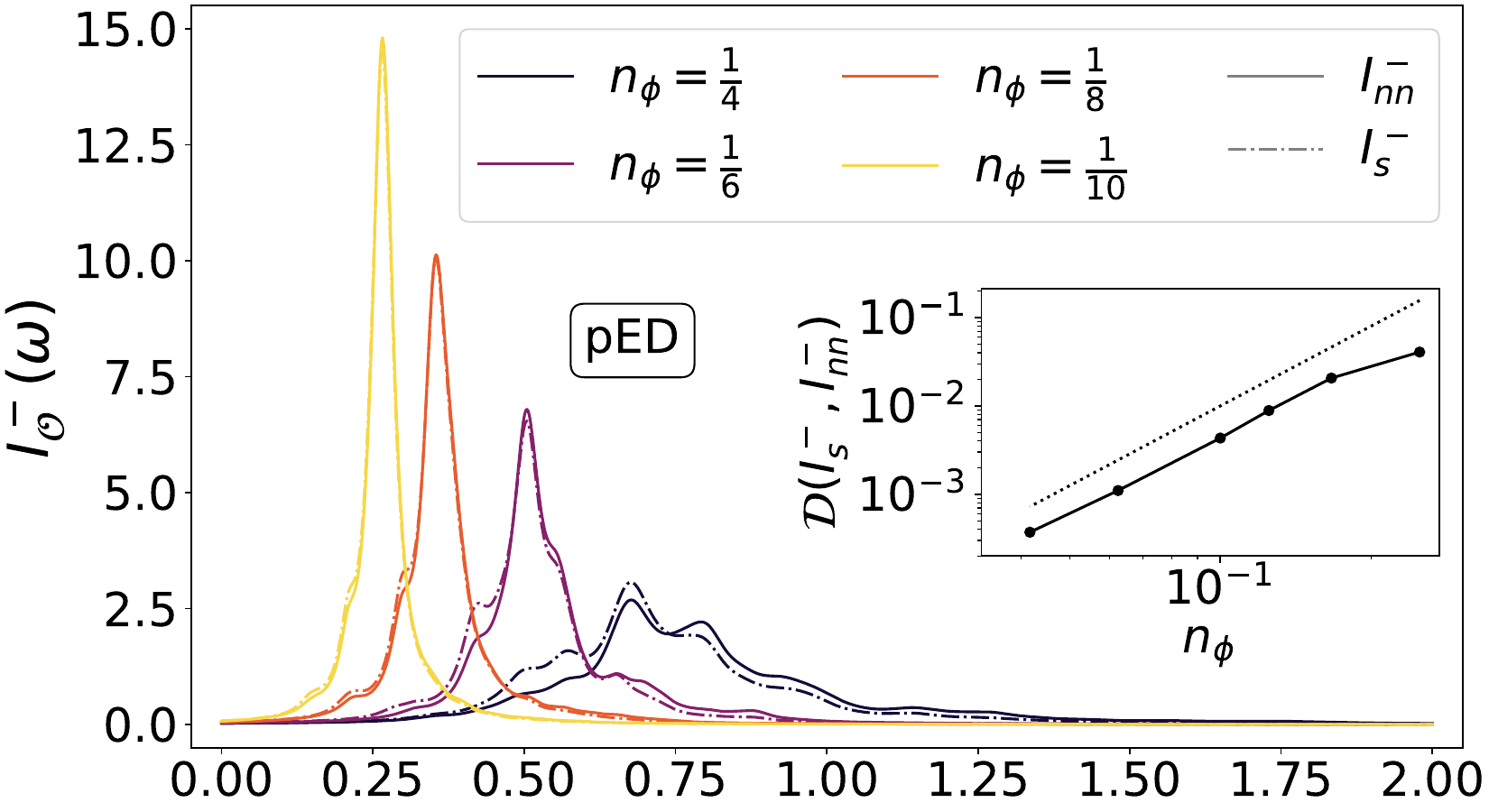}
    \put(15,44){(a)}
    \end{overpic}
    \begin{overpic}
     [width=\linewidth]{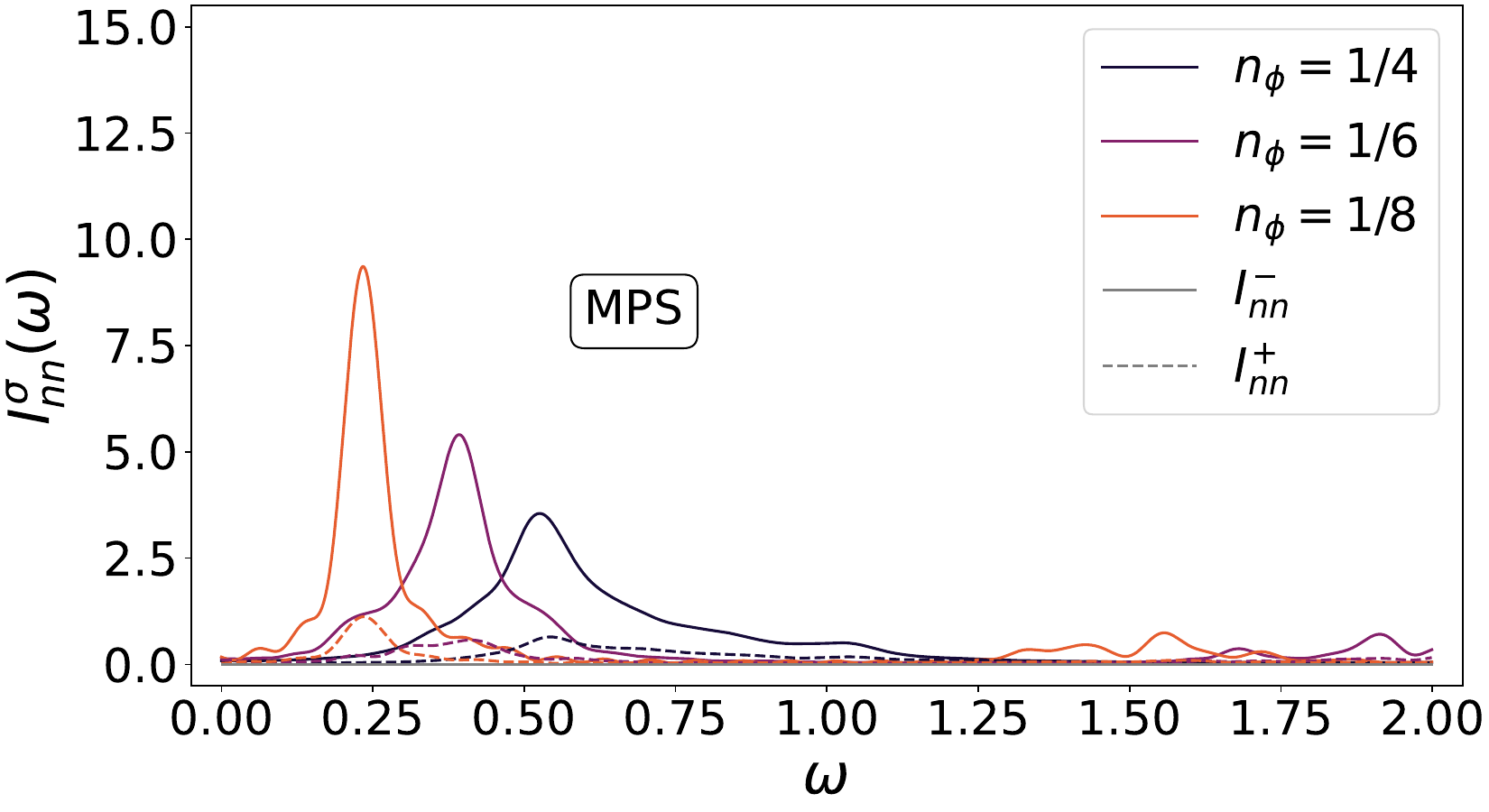}
    \put(15,44){(b)}
    \end{overpic}
    \caption{\textbf{Continuum limit of $O^\pm_{nn}$ and $O^\pm_s$}. (a) pED graviton spectra of the Harper-Hofstadter model at $N=8$ particles for the lattice stress tensor (dashed lines) and the density-density operator (full lines). Different fluxes $n_\phi$ are shown. Inset: Overlap distance between the two graviton mode operators as the continuum limit $n_\phi\to 0$ is approached (dotted line is $\propto n_\phi^3$). (b) MPS result of graviton spectrum for $n_\phi = 1/4,1/6,1/8$ fluxes obtained from cylinders of the size $L_x\times L_y = 24\times 6,~48\times 9,~36\times 8$ cylinders using density-density operator with full and dashed line labels the chiralities of the graviton. We choose $\eta=0.16n_\phi$ in both panels so that the ratio broadening over graviton energy remains roughly constant. }
    \label{fig:hh_continuumlimit}
\end{figure}

In order to access larger system sizes, we now present projected ED (pED) results on the lowest band and real-space MPS results without band truncation. While the pED neglects band mixing effects which can produce shifts in the graviton energy (see App. \ref{app:band_truncation}) and broaden the peak when interaction strength is large, MPS does not and provides a sanity check to the pED results. In order to allow for a meaningful study with pED, we here use as interaction range $r_c=2$ and smaller overall strength $V=2$. In Fig. \ref{fig:hh_continuumlimit} we show pED spectra (chirality $\sigma=-1$) for a set of fluxes $n_\phi=1/q$ with $q=4,6,8,10$ and systems of $N=8$ particles (respectively $8\times12$, $12\times 12$, $16\times 12$ and $20\times 12$). The spectra are regularized with a flux-dependent $\eta=0.08n_\phi$ such that the precision relative to the graviton energy $\omega_G\propto n_\phi$ remains roughly constant. 

Importantly, the two operators, i.e., quadrupolar density-density correlators $O^\pm_{nn}$  (full lines) and lattice stress tensor $O^\pm_s$ (dashed lines), show very similar features even at fairly large fluxes $n_\phi\simeq1/6-1/4$ when the magnetic length $l_B$ approaches the length scale of lattice spacing. To provide a more quantitative statement, we show in the inset of Fig. \ref{fig:hh_continuumlimit} an overlap distance between the two graviton spectra:
\begin{align}
    \mathcal{D}(I_s^-,I_{nn}^-)=
    1-|\braket{\psi^-_s}{\psi^-_{nn}}|
\end{align}
where:
\begin{align}
    \ket{\psi^-_{O}}=\frac{1}{\sqrt{\bra{0}O^+O^-\ket{0}}}O^-\ket{0}
\end{align}
is the normalized graviton wavefunction for a particular graviton operator $O$. The distance is already quite small at $n_\phi=1/4$ and tends to 0 as $n_\phi^3$ in the continuum limit. These results confirm that both operators flow to the same operator in the continuum. Since the operator $O_{nn}^-$ is defined with an effective range which goes to zero in the continuum limit $r_c/l_B\to 0$, we conclude both operators flow to the continuum stress tensor of ultra-short-range interactions (Eq. \eqref{eq:hamiltonian_continuum}) when projected to the lowest band. Note that away from the continuum limit $O_{nn}$ still has very large overlap with the lattice stress stensor $O_s$.

In Fig. \ref{fig:hh_continuumlimit}(b), we show MPS results for only the density-density graviton operator $O^\pm_{nn}$ at different fluxes $n_\phi$. The volumes realized by MPS are bigger $N= 12-24$, with a strongly anisotropic aspect ratios $L_x\times L_y=24\times6,48\times9,36\times8$ respectively for $n_\phi=1/4, 1/6,1/8$. The clearer effect of band mixing is a reduction of the graviton energy $\omega_G$ (see also App. \ref{app:band_truncation}). Here we are in a regime where $\omega_G\sim \omega_B /5$ with $\omega_B$ the bandwidth. Secondly, the peak is a bit broadened despite the spectral broadening parameter $\eta$ being the same in the two panels (a) and (b). We attribute this to the anisotropic and edge effect of the cylinder boundary. 
\begin{figure*}
    \centering
         \begin{overpic}[width=\linewidth]{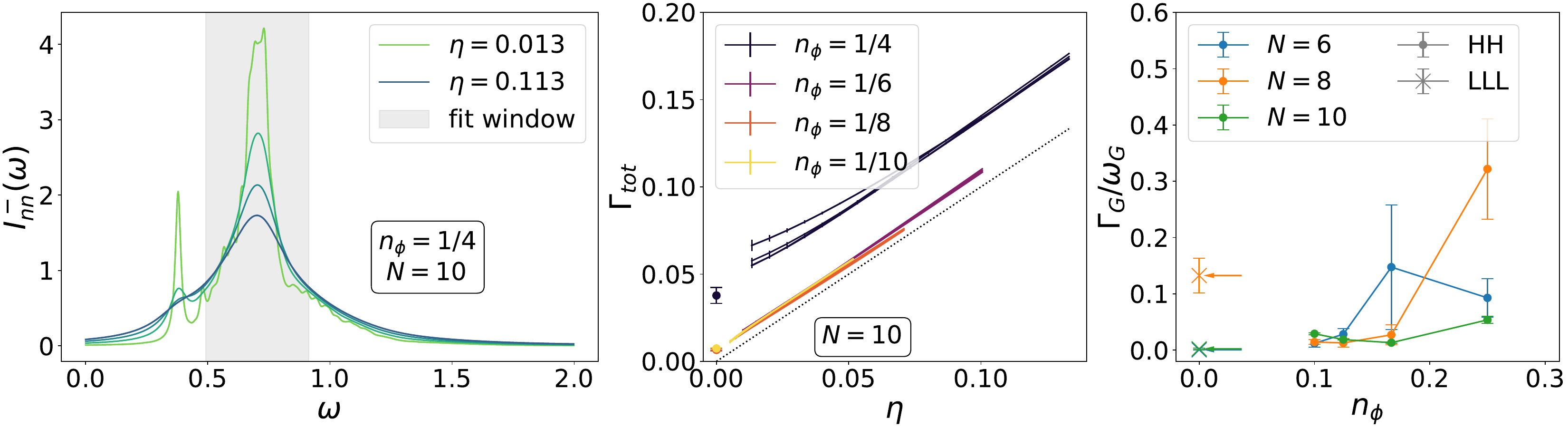}
    \put(6,24){(a)}
    \put(65,24){(b)}
    \put(76,15){(c)}
    \end{overpic}
    \caption{\textbf{Life time analysis on the Harper Hofstadter model}. Analysis on the lifetime of the graviton mode from finite size ED spectra on the $\nu=1/3$ Harper Hofstadter model at different fluxes $n_\phi$. (a) Example of how the spectra changes for different regularization parameters $\eta$. (b) Dependence of $\Gamma_{tot}$ on $\eta$ for different fluxes $n_\phi$ for $N=10$ particles. The dashed black line represents the infinite lifetime scenario $\Gamma_{tot}=\eta$, while the $\eta=0$ points are the estimates for $\Gamma_G$. The lines with the same color come from different fitting energy windows
    $\omega \in[(1-\delta)\omega^* ,(1+\delta)\omega^*]$
where $\omega^*\simeq\omega_G$ corresponds to the maximum of the spectral function and  $\delta\in [0.15,0.2,0.25,0.3]$  (c) Graviton estimated intrinsic decay rate $\Gamma_G$  relative to its energy $\omega_G$ as a function of $n_\phi$ for $N=6,8,10$. At $n_\phi=0$ we also show results for continuum LLL with $V_1$ pseudopotential interaction (cross marker and arrow).}
    \label{fig:lifetime_hh}
\end{figure*}
\subsection{Lifetime of the graviton-mode on Harper-Hofstadter}\label{sec:HHnum_lifetime}

A central question regarding the graviton-mode on lattice FQH systems concerns its lifetime. The main mechanism of decay is attributed to scattering into the two magnetoroton continuum. In FQH systems, this decay is thought to be inhibited by the guiding center rotational symmetry \cite{haldane2011geometrical} present in the ground state wavefunctions and believed to extend to low frequencies. On the lattice, this is not the case \cite{wangDynamics2025} as the broken continuum translation and rotational symmetries give rise to interactions which do not conserve the relative angular momentum between guiding centers. It has then been argued \cite{wangDynamics2025} that the allowed decay into two magnetorotons eventually leads to zero lifetime for the graviton-mode in the thermodynamic limit when the density of states around $\omega_G$ increases. 

We present here a comprehensive analysis of the graviton-mode spectra (Fig. \ref{fig:lifetime_hh}), intending to extract not only an energy $\omega_G$ but also a decay rate $\Gamma_G$ (inverse lifetime) for the graviton mode. Since the graviton in the presented cases is hidden in a two-magneto-roton continuum, its weight in finite sizes will be distributed among an extensive number of many-body eigenstates for any finite value of the graviton/two-magnetoroton scattering matrix element. In order to correctly recover this weight in a single finite-lifetime peak, we must introduce a finite broadening $\eta$ larger than the spacing between neighboring many-body eigenstates. 

Now, if the operator $O^-$ actually acts as the creation operator of the graviton, we expect it's green function to show a pole at the graviton energy as:
\begin{align}\label{eq:greensfunction}
    G_{O^-}(\omega+i\eta)\simeq Z\frac{1}{\omega+i\eta-\omega_G  +\Sigma(\omega) }+...
\end{align}
where $G_{O^-}(t)=i\theta(t)\langle [O^+(t),O^-(0)]\rangle$ is the retarded Green's function, $\omega_G$ the graviton energy, $Z$ an operator dependent quasi-particle weight and $\Sigma(\omega)$ a possible self-energy corrections coming from interactions with the two magnetoroton continuum. Assuming the latter is smooth around the graviton energy, we can approximate it with its constant part:
\begin{align}\label{eq:selfenergy}
    \Sigma(\omega)\simeq i \Gamma_G
\end{align}
where we neglect any real part which can be reabsorbed into $\omega_G$. Note that Eq. \ref{eq:selfenergy} defines an intrinsic decay rate for the graviton $\Gamma_G$. Following this idea, the spectral function $I^-_{nn}(\omega)=\frac{1}{\pi}\Im G_{O^-_{nn}}(\omega)$ can be fitted around the graviton energy as:
\begin{align}
    I_0(\omega,\eta)= \frac{Z(\eta)}{\pi}\frac{\Gamma_{tot}(\eta)}{(\omega-\omega_G)^2+\Gamma_{tot}(\eta)} + a_0(\eta)+a_1(\eta)\omega 
\end{align}
where $\{\Gamma_{tot},Z,\omega_G\}$ are free $\eta$-dependent parameters which capture the graviton contribution and $\{a_0,a_1\}$ can account for a smooth background contribution that takes into account the small asymmetry around the peak. If the graviton is a good quasi-particle captured by $O^-_{nn}$, then Eq. \eqref{eq:greensfunction}\eqref{eq:selfenergy} hold and we also expect:
\begin{align}\label{eq:decay_extrapolation}
    \Gamma_{tot}(\eta)\simeq c\,\eta+\Gamma_G
\end{align}
with $c\simeq1$. 

The fit above is used in Fig. \ref{fig:lifetime_hh} to extract the intrinsic graviton decay rate $\Gamma_G$. As detailed in App. \ref{sec:app_lifetime_extraction}, we do the above procedure for different fitting windows $\omega\in[\omega_{min},\omega_{max}]$ around the main peak (an example is shown in Fig. \ref{fig:lifetime_hh}(a) as a shaded area). This provides different estimations on $\Gamma_{tot}(\eta)$, shown in Fig. \ref{fig:lifetime_hh}(b) with which one can linearly extrapolate to $\eta=0$ to extract the decay rate $\Gamma_G$. Indeed all lines $\Gamma(\eta)$ are close to being linear ($c\simeq 1$) as in Eq. \eqref{eq:decay_extrapolation}. The zero-decay line $\Gamma_{tot}=\eta$ is shown as a guide as a dotted black line. Errorbars account for variations in both frequency window $[\omega_{min},\omega_{max}]$ used for the fit and in extrapolation range $[\eta_{min},\eta_{max}]$ (see App. \ref{sec:app_lifetime_extraction} for more details).

In panel (c) of Fig.~\ref{fig:lifetime_hh} we finally show the extracted graviton decay rate in the HH model (circle markers) relative to its energy $\Gamma_G/\omega_G$ as a function of $n_\phi$ for different particle numbers $N=6,8,10$. Here we also include at $n_\phi=0$ results from continuum LLL (cross markers) on an $L\times L$ torus with $V_1$ pseudopotential interaction \cite{liou2019chiral} where the lifetime is thought to be very large \footnote{To the best of our knowledge, apart from the experimentally reported $\nu=1/3$ graviton peak width in continuum LLL of about $2\Gamma_G\sim 30\,\mu eV$ with $\omega_G\sim 0.65 \,\mathrm{m}eV$ \cite{liang2024evidence}, no other estimate is currently available in the literature}. Despite the large error bars and oscillations with system size, we can see a clear trend of a decreasing decay rate $\Gamma_G$ (increased lifetime) as the continuum is approached $n_\phi\to0$. Importantly by almost doubling the system size ($N$ from 6 to 10) we do not see any marked increase in the intrinsic relative decay rate $\Gamma_G/\omega_G$. On a minor note, we also remark that the different aspect ratios realized by different $n_\phi$ at fixed $N$ are also likely to play a role in the apparent trend of $\Gamma_G$. For example at $N=10$ and $n_\phi=1/4,1/6,1/8,1/10$ the system sizes are respectively $12\times 10$, $12\times15$, $16\times15$ and $20 \times15$. In general the chosen geometry is the one with aspect ratio closest to 1 that fits the magnetic unit cell $n_\phi^{-1} \times 1$ of the Landau gauge Harper Hofstadter model.

\section{Adiabatic connection of graviton from FQH to FCI}
\label{sec:IV}
Having understood basic lattice effects on the FQH side based on the HH model, we now turn to FCI phases. In the following, we will study the \textit{adiabatic} connection between graviton modes of a widely studied checkerboard lattice model for FCI \cite{Hongyu_prl2024_thermodinamicFCI,Sheng2011_FQAH_checkerboard_fermion,KaiDasSarma_prl2011_flatbandsCB,wuAdiabatic2012} and a low flux $n_\phi=1/8$ Harper-Hofstadter type model (which we label as HH$^*_{\frac{1}{8}}$ on the following), which we have shown will approximate continuum LL physics. Then, we perform lifetime analysis on the FCI graviton. Finally, we discuss the interaction dependence of the graviton, especially its behavior when the FCI-FL(Fermi Liquid) phase transition happens.

\subsection{Interpolation from LL to a generic Chern band}
\begin{figure}
    \centering
    \begin{overpic}[width=1\linewidth]{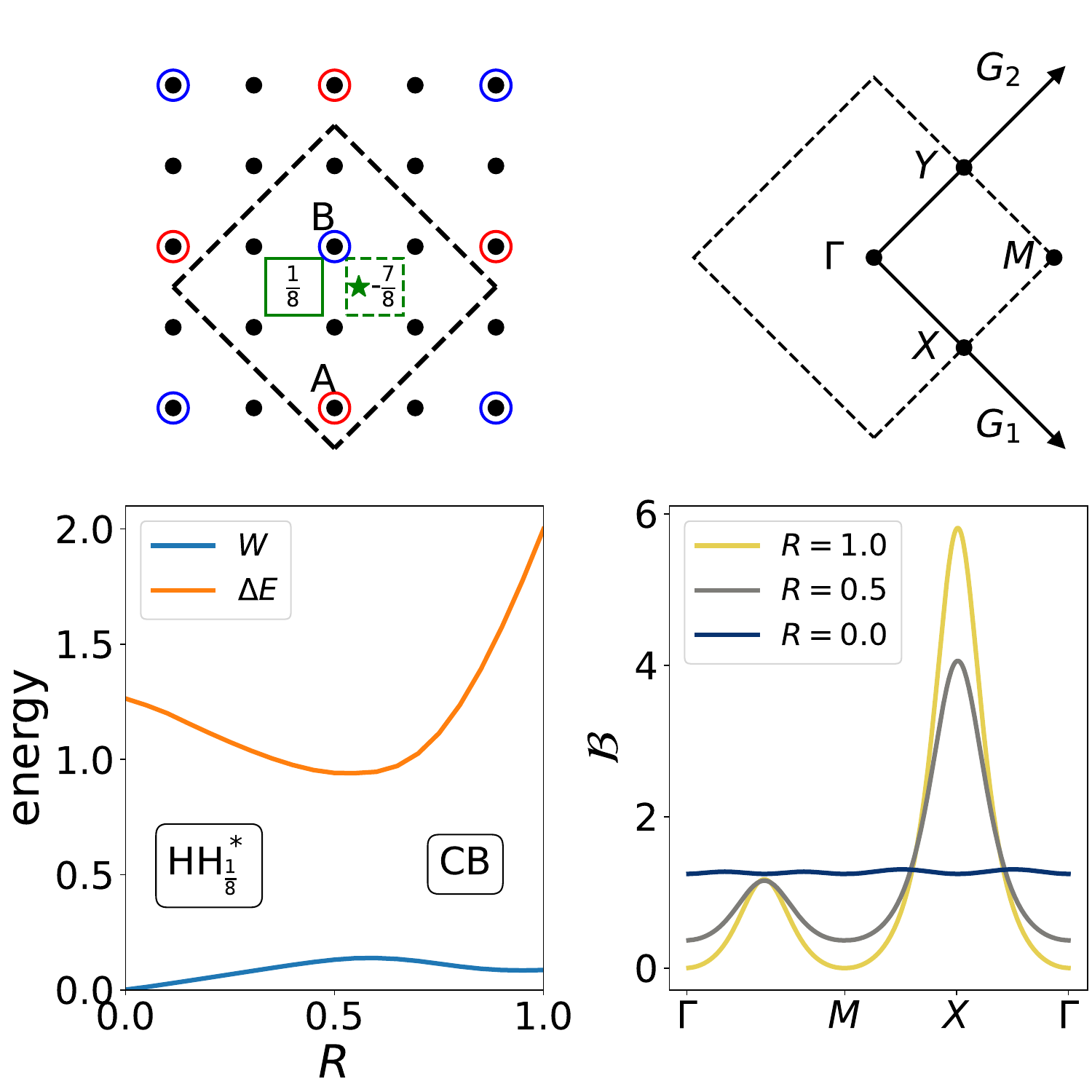}
    \put(5,90){(a)}
    \put(65,90){(b)}
    \put(40,49){(c)}
    \put(92,49){(d)}

\end{overpic}
    \caption{\textbf{Interpolation from HH$^*_{\frac{1}{8}}$ model to a CB lattice}. (a) Square lattice Hofstadter model with $1/8$ flux per plaquette. In the position of the green star, we insert $-2\pi$ flux to make the total unit cell (dashed black line) have zero net flux. The A and B sublattices of the checkerboard sites are encircled in red and blue, respectively. (b) First Brillouin zone of the lattice with high symmetry points labeled. (c) The band gap $\Delta E$ and band width $W$ of the lowest band as we tune from the the HH$^*_{\frac{1}{8}}$ limit to the CB limit. The energy is in units of $t=1$. (d) Berry curvature along the high symmetry point at different interpolation ratios $R$.}
    \label{fig:8site_inter_scheme}
\end{figure}

The first important ingredient to design an FQH to FCI path is the realization of adiabatic interpolation at the energy band level, where an isolated Chern band shall be intact throughout the path. We follow a procedure of \textit{lattice embedding}, designed in Ref. \cite{wuAdiabatic2012}, which is represented schematically in Fig. \ref{fig:8site_inter_scheme}(a). 

\begin{figure*}
    \centering
    \begin{overpic}[width=0.48\linewidth]{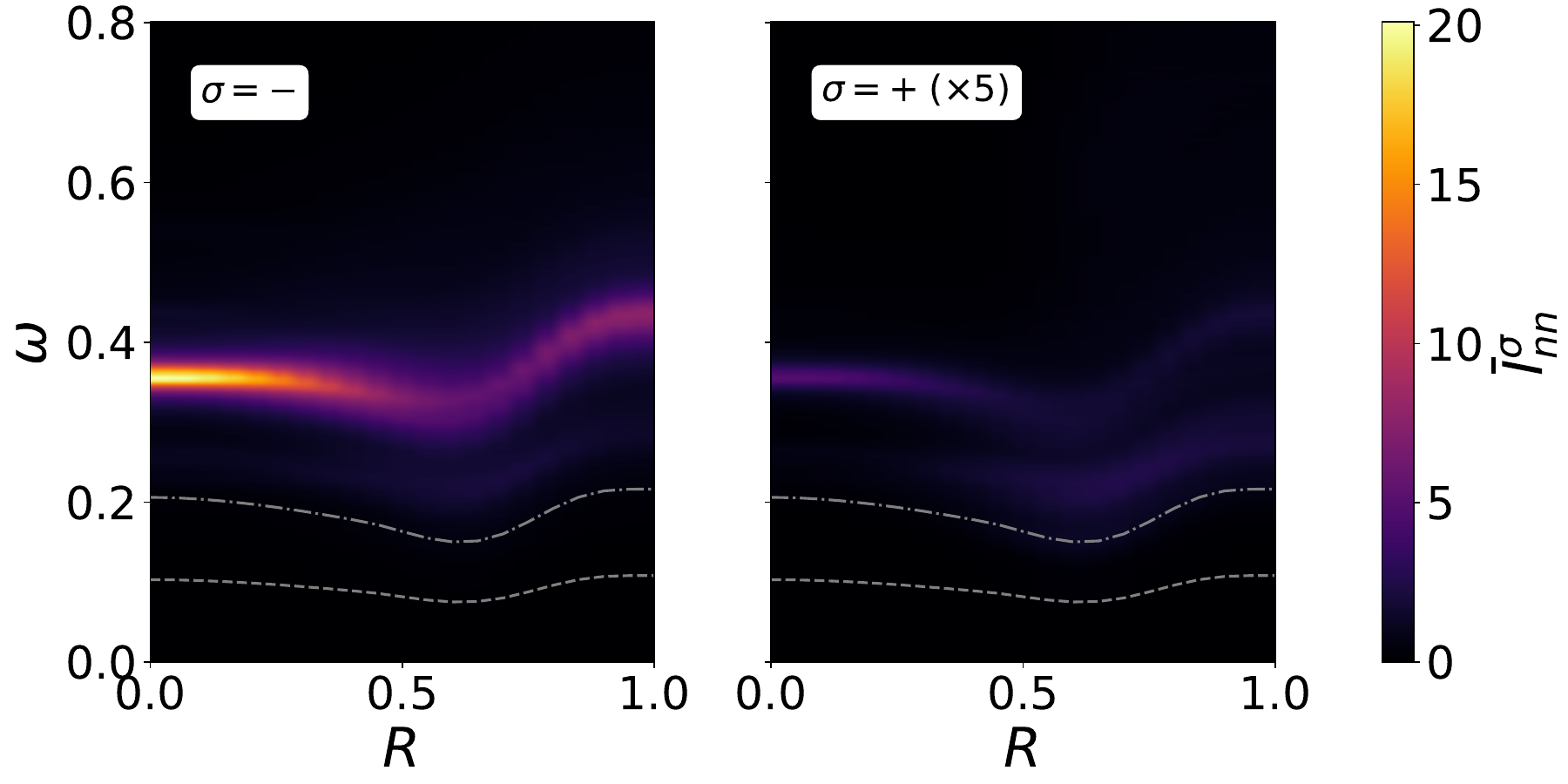}
    \put(35,43){\textcolor{white}{(a)}}
    \put(75,43){\textcolor{white}{(b)}}
    \end{overpic}
    \begin{overpic}[width=0.48\linewidth]{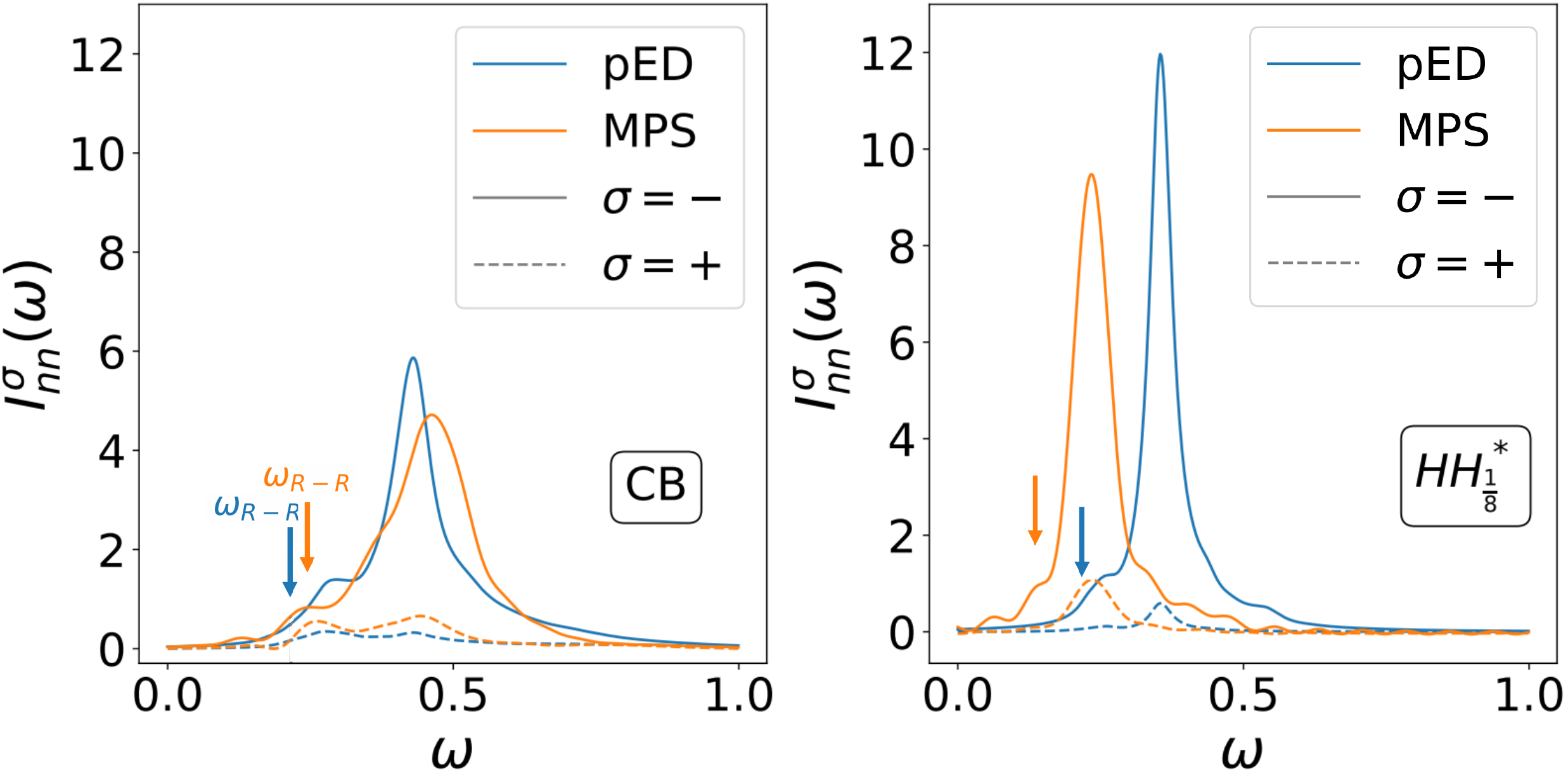}
    \put(10,45){(c)}
    \put(60,45){(d)}
    \end{overpic}
    \caption{\textbf{Graviton spectra from the FQH limit ($R=0$) to the FCI limit ($R=1$)}. (a,b) Adiabatic connection of the two graviton modes between $R=0$ and $R=1$ obtained with pED. The $\sigma=-$ chirality (a) always dominates the $\sigma=+$ chirality (b), whose signal is enhanced by a factor $\times5$ for visibility. We also show the first excited state in the ground state momentum sector (dashed line) and twice its value (dashed-dotted line) as an estimate for the magnetoroton energy and the start of the 2 magnetoroton continuum. (c,d) Comparison between band pED (blue) and MPS (orange) results with chirality resolution ($\sigma=-$ full, $\sigma=+$ dashed) for two limits $R=1$ CB (c) and $R=0$ HH$^{*}_{\frac{1}{8}}$ (d). The estimated 2-magnetoroton continuum energies $\omega_{R-R}$ from pED and DMRG are annotated with arrows. All the spectra $\bar{I}^\sigma_{nn}$ are normalized as discussed in the text. On the CB side, $N=10$ for pED and $N = 24$ for DMRG, while on the HH$^{*}_{\frac{1}{8}}$ side, $N=10$ for pED and $N = 12$ for DMRG. The system sizes are discussed in the text.}
    \label{fig:adiabatic_graviton}
\end{figure*}

The basic idea of lattice embedding is to define a checkerboard super-lattice on top of an underlying square lattice where the HH$^*$ Hamiltonian is defined (see Fig. \ref{fig:8site_inter_scheme}(a)). Specifically, the gauge choice of HH$^*$ respects the translational symmetry of the CB lattice, which is different from the traditional \textit{Landau gauge} choice of the HH model and has a total zero flux per unit cell \cite{wuAdiabatic2012}. Thereby, we label it as HH$^*$. The single-particle particle Hamiltonian $H_0(R)$ across the path $R\in[0,1]$ is defined as:
\begin{align}
    H_0(R)= R\, H_{0,cb}+(1-R) \,H_{0,hh}
\end{align}
where:
\begin{align}
 &H_{0,hh}=-t \sum_{\langle i,j\rangle} e^{i\phi_{ij}} c^\dagger_i c_j +h.c.\\
&H_{0,cb}= -t'\sum_{\langle I,J\rangle} e^{i\phi_{IJ}} c^\dagger_J c_J -t''\sum_{\langle\langle I,J\rangle\rangle} e^{i\phi_{IJ}} c^\dagger_J c_J +\nonumber\\
&\qquad \qquad-t'''\sum_{\langle\langle\langle I,J\rangle \rangle\rangle} e^{i\phi_{IJ}} c^\dagger_J c_J +h.c.
\end{align}
Therefore, $R = 0$ and $R= 1$ corresponds to the HH$^*_{\frac{1}{8}}$ and the CB limit separately. Here we use a convention where small indices $i,j$ indicate the underlying square lattice (black dots in Fig. \ref{fig:8site_inter_scheme}(a)), while capital indices $I,J$ specify sites only in the checkerboard superlattice (red and blue circled black dots in Fig. \ref{fig:8site_inter_scheme}(a)). Then $\langle \dots\rangle$ ,$\langle\langle \dots\rangle\rangle$, and $\langle \langle\langle \dots\rangle\rangle\rangle$ indicate nearest neighbors, next nearest neighbors, and next to next nearest neighbors. Note that $\langle I,J\rangle$ should be understood as nearest neighbour on the checkerboard superlattice. We fix $t=t'=1$, $t''=1/(2+\sqrt{2})$ and $t'''=-1/(2+2\sqrt{2})$ \cite{Hongyu_prl2024_thermodinamicFCI}. These choices of tight-binding parameters realize the adiabatic path where the lowest band evolves from a LL-like band to a generic Chern band, as shown in Fig.~\ref{fig:8site_inter_scheme} (c) and (d). 

Importantly, the phases $\phi_{i,j}$ and $\phi_{I,J}$ give a non-trivial topological nature to the band structure. In order to interpolate to CB lattice with 0 net flux, $-2\pi$ flux is threaded in the Harper-Hofstadter square lattice  to make it 0 flux as well, which we signal with a green star in  Fig.~\ref{fig:8site_inter_scheme}(a). The insertion of $-2\pi$ flux in our model is simply a gauge choice that will leave the energy band unchanged while inducing a gauge transformation to the states\cite{wuAdiabatic2012}. The phases $\phi_{i,j}$ then keep track of the fact that in each small plaquette there is a flux $n_\phi=1/8$, except for the one that encloses the $-2\pi$ flux, which has $n_\phi=-7/8$. Moreover, the gauge choice respects the translation symmetry of the CB lattice. Once a gauge that realizes this is fixed, the longer range hoppings $\phi_{I,J}$ are also fixed according to the Stokes theorem of the vector potential. For detailed construction of the periodic gauge and Hamiltonian, see App. \ref{app:gauge_ham}.

In Figure \ref{fig:8site_inter_scheme}(c) we then show how the bandwidth of the lowest band $W$ and the band-gap to higher bands $\Delta E$ evolve as a function of $R$. The lowest band is always well separated from the higher ones, hence it maintains its total Chern number throughout the $R$-path. However, as explicitly shown in Fig.  \ref{fig:8site_inter_scheme}(d), the $k$-space distribution of the Berry curvature strongly differs at the $R=0$ and $R=1$ points. Indeed, at the $R=0$ point the lowest band mimics a LLL, with uniform Berry curvature and almost no dispersion ($W\sim10^{-3}$), while $R=1$ captures the essential deformations of a Chern band, i.e., a non-trivial quantum geometry, while maintaining a flat band character ($W\simeq0.08 \ll \Delta E$).

The last ingredient needed for an actual FQH to FCI path is the interaction. We use the same interaction form throughout the interpolation:
\begin{align}
   H_{V}= \frac{V}{2} \sum_{i\neq j} f(|r_{ij}|) n_in_j
\end{align}
where $f(r)=1/r$ for $r<r_c$ and 0 otherwise, with a finite range $r_c=2$ corresponding to nearest neighbors on the checkerboard super-lattice. Unless stated otherwise, we use $V=2$ so that the nearest neighbor interaction for the checkerboard super-lattice is $V_{AB} = 1$. The presence of an FCI ground state across the whole path is shown in App. \ref{app:adiabatic_gs}.

\subsection{Graviton-mode adiabatic connection}
We now provide numerical evidence for the adiabatic connection between the graviton mode in the FQH limit $R=0$ (HH$^*_{\frac{1}{8}}$) and the FCI limit $R=1$ (CB). We use the density-density graviton operator definition given in Eq.~\eqref{eq:definition_Onn}, as this readily generalizes to any lattice scenario. In particular, we use $f_G(r)=1/r$ for $r<r_G=2\sqrt{2}$ and zero otherwise, such that the operator includes NNN density-density correlations on the super-lattice (8 neighboring sites in total and corresponds to 24 neighboring sites for the underlying lattice), hence able to resolve the chirality even at the CB limit. Choosing different $f_G(r)$ amounts in small qualitative differences on the spectra (see App. \ref{sec:app_fG}). We remark that this stability with the choice of the operator is important and confirms that the operator $O^-_{nn}$ has large overlap with the unknown stress tensor across the whole path.

In Figure \ref{fig:adiabatic_graviton} we show a combination of pED and MPS results for the normalized graviton spectral functions $I^\sigma_{nn}(\omega)$. We remind the reader that the normalization for both MPS and pED results is with respect to the total intra-band contribution ($\omega <\Lambda_{lb}$) of the $\sigma =-$ chirality. 

In Fig.~\ref{fig:adiabatic_graviton}(a,b), we show the two chiralities of the graviton mode along the path, obtained via pED for a system of $6\times5$ unit cells (240 sites) with $N=10$ particles (broadening $\eta=0.01$). The spectral function of the negative chirality $\sigma=-$ (a) shows a clear peak whose frequency evolves continuously with $R$, providing evidence for an adiabatic connection between the two limits (same data are shown in Fig.~\ref{fig:sketch_fig1}(d) ). The chirality of the mode also remains well defined across the full path. In these panels, we also show the first excited state energy (dashed gray line) in the ground state momentum sector and twice of its value (dashed-dotted gray line) as a function of $R$. The latter signals the energy scale where the two-magnetoroton continuum starts. The graviton peak lies well within this continuum, making its decay into two magneto-rotons possible, as discussed in Sec. \ref{sec:HHnum_lifetime}. While the graviton peak is sharper when close to the FQH limit $R=0$, its lifetime evolves continuously along the path and remains finite close to the FCI limit $R=1$.

We focus on these two limits  $R=1$ CB (Fig.~\ref{fig:adiabatic_graviton} (c)) and $R=0$  HH$^{*}_{\frac{1}{8}}$ (Fig.~\ref{fig:adiabatic_graviton} (d)) and give a comparison of graviton spectra obtained with pED (blue lines) and MPS (orange lines). The former are obtained for $N=10$ particles on a torus, while the latter are on cylinder with $N=24$ on CB side and $N=12$ on HH$^*_{\frac{1}{8}}$ side. The dotted-dashed gray line represents the onset of the two magnetoroton continuum as twice the pED excited state energy. We first note that the large-scale MPS results and pED are in good agreement. We notice small shifts to the graviton energy in the HH$^*_{\frac{1}{8}}$ case and a slightly increased peak width(decay rate), which we respectively attribute to mild band mixing effects and scattering from the open cylinder edges (see similar discussion in Sec. \ref{sec:III_spectra}). Apart from a strong graviton-mode peak, we can here also appreciate a finite weight on the two-magnetoroton continuum, whose energy for MPS case is estimated using the relative graviton energy ratio to pED, relatively stronger in the FCI case (CB) than in the FQH case (HH$^*_{\frac{1}{8}}$) compared with the graviton peaks, but always below the graviton-mode peak. 
\begin{figure}
    \centering
        \begin{overpic}[width=1\linewidth]{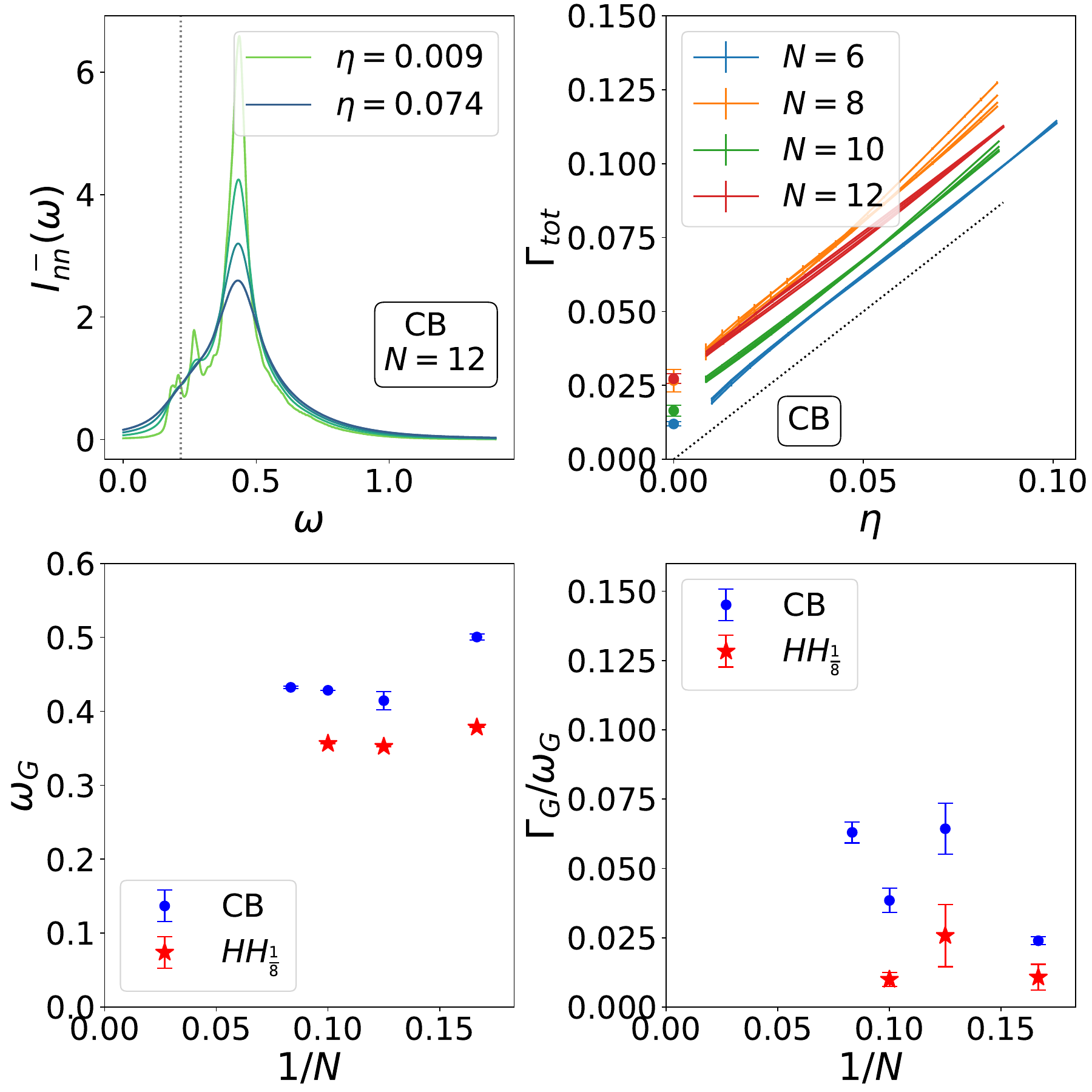}
    \put(10,95){(a)}
    \put(90,95){(b)}
    \put(10,45){(c)}
    \put(90,45){(d)}

\end{overpic}
    \caption{\textbf{Lifetime analysis of graviton on FCI limit}. (a) pED graviton spectra for different regularization of the spectral function $\eta$ for $R=1$ at $N=10$. The dashed gray line represents an estimate for the two magneto-roton energy. (b) Dependence of the total decay rate $\Gamma_{tot}$ on $\eta$ for different particle numbers for $R=1$ in pED. The dashed black line represents the perfect scenario $\Gamma_{tot}=\eta$, while the $\eta=0$ points the estimates for $\Gamma_G$.  The lines with the same color come from different fitting energy windows
    $\omega \in[(1-\delta)\omega^* ,(1+\delta)\omega^*]$
where $\omega^*\simeq\omega_G$ corresponds to the maximum of the spectral function and  $\delta\in [0.15,0.2,0.25,0.3]$. (c) Graviton peak position and (d) estimated graviton lifetime as a function of system size for the FCI $R=1$ case (blue) and the HH case $R=0$ (red).  the size of the k-meshgrid in BZ that is used for finite size scaling ((c) and (d)) from small to large is $3\times 6$,$4\times 6$,$5\times 6$,$6\times 6$}
    \label{fig:lifetime_analysis}
\end{figure}
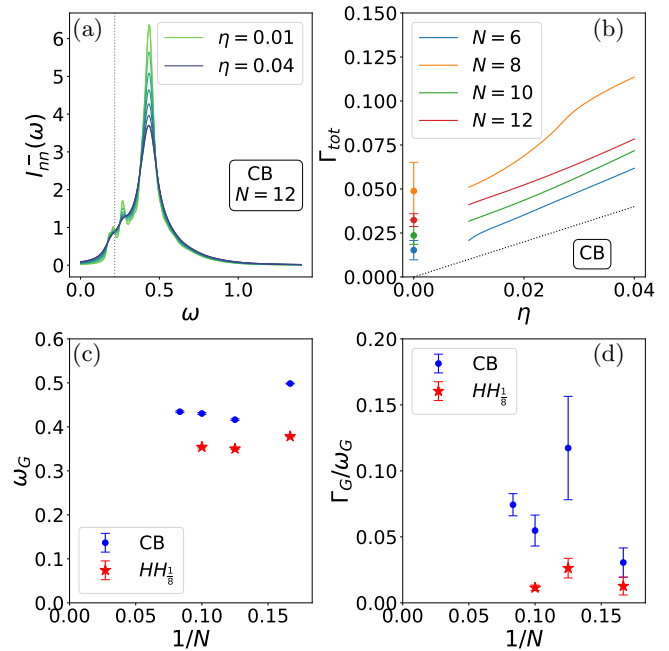

\subsection{Lifetime of the graviton-mode on FCI}

We now repeat the quantitative analysis on the graviton intrinsic decay rate $\Gamma_G$ (Fig.~\ref{fig:lifetime_analysis}) as in Sec. \ref{sec:HHnum_lifetime}. We use pED results, as this excludes edge effects and strongly anisotropic aspect ratios of MPS simulations. The qualitative agreement of pED and MPS can be directly seen from the comparison of the graviton spectra (see Fig.~\ref{fig:adiabatic_graviton})(c)). Focusing on the two limits $R=0$ and $R=1$, first, we examine the graviton spectra at $R=1$ (CB) for different broadening parameters $\eta$ (Fig.~\ref{fig:lifetime_analysis} (a)) and the dependence of the total decay rate $\Gamma_{tot}$ on $\eta$ for different system sizes. The linear dependence $\Gamma_{tot}\simeq \Gamma_G+\eta$ is clear. For the extraction of $\Gamma_G$ and its error bar $\delta \Gamma_G$ we still follow what is outlined in Sec. \ref{sec:HHnum_lifetime} and detailed in App.\ref{sec:app_lifetime_extraction}. The result is shown as points at $\eta=0$ in Fig.~\ref{fig:lifetime_analysis} (b). Then in Fig.~\ref{fig:lifetime_analysis} (c), we show the $1/N$ dependence of the graviton energy $\omega_G$ and in Fig.~\ref{fig:lifetime_analysis} (d) the relative intrinsic decay rate $\Gamma_G/\omega_G$. While oscillations in the decay rate are present, these become less pronounced for larger $N$, suggesting a finite value in the FCI case (CB) of order $\Gamma_G/\omega_G\sim 0.07$ in the $N\to\infty$ limit. A finite but smaller intrinsic decay rate $\Gamma_G/\omega_G \sim0.02$ is present also for the HH$^*_{\frac{1}{8}}$ case, compatible with the pure HH results of Sec. \ref{sec:HHnum_lifetime}.

As highlighted by the dashed gray line in Fig. \ref{fig:lifetime_analysis}(a) the graviton peak lies well above the two magnetoroton energy. Indeed, as explicitly shown in App. \ref{sec:app_dos}, the gravitons keep lying inside the excitation continuum in finite size calculation, the density of states at the graviton energy here is large (for $N=12$ roughly $10^4$ states per unit frequency)  and very likely gets most of its contributions from 3-4 magnetoroton states. Therefore, its long lifetime should not be attributed to a low density of states below the two magnetoroton continuum as in the case explored in Ref. \cite{wangDynamics2025} but on a suppression of scattering matrix elements which render the graviton-mode a true quasi-particle hidden in a continuum. Overall our data provide strong evidence for a finite and large lifetime in the CB point, we find it unlikely that a divergence in the decay rate develops at larger system sizes. 
A detailed analysis of the scattering process of the graviton and its guiding center symmetry would help illuminate the thermodynamic behavior of the graviton mode, which we leave to future study.

We also interpolate HH$^*_{\frac{1}{4}}$ Hamiltonian to a CB lattice (see App.\ref{sec:appC}), the band gap remains intact and much larger than the band width. In HH$^*_{\frac{1}{4}}$ interpolation, the CB side has a relatively flatter band while the HH side has smoother quantum geometry. The graviton chirality is resolved in both limits. Interestingly, compared with $1/4$ HH$^*$, the graviton in the CB limit has a longer lifetime and decays more slowly as the interaction increases. This scenario is different from $1/8$ HH$^*$ interpolation as there is competition between the effect of band flatness and quantum geometry on the lifetime of the graviton. This comparison highlights the complication of the interplay between band flatness and quantum geometry on the properties of graviton-modes in FCIs.


\subsection{Interaction dependence}

We now study the dependence of the graviton spectra on the strength of interactions $V$. A priori, this does not correspond to a rigid energy scale shift as it can compete with the bandwidth of the lowest band $W$, especially in the FCI limit. In particular, having fixed hopping amplitudes and hence the bandwidth $W$ (see Fig. \ref{fig:8site_inter_scheme}), we tune the interaction scale $V$ keeping the same finite range interactions $f(r)=1/r$ for $r<2$ and zero otherwise.

\begin{figure}
    \begin{overpic}
 [width=\linewidth]{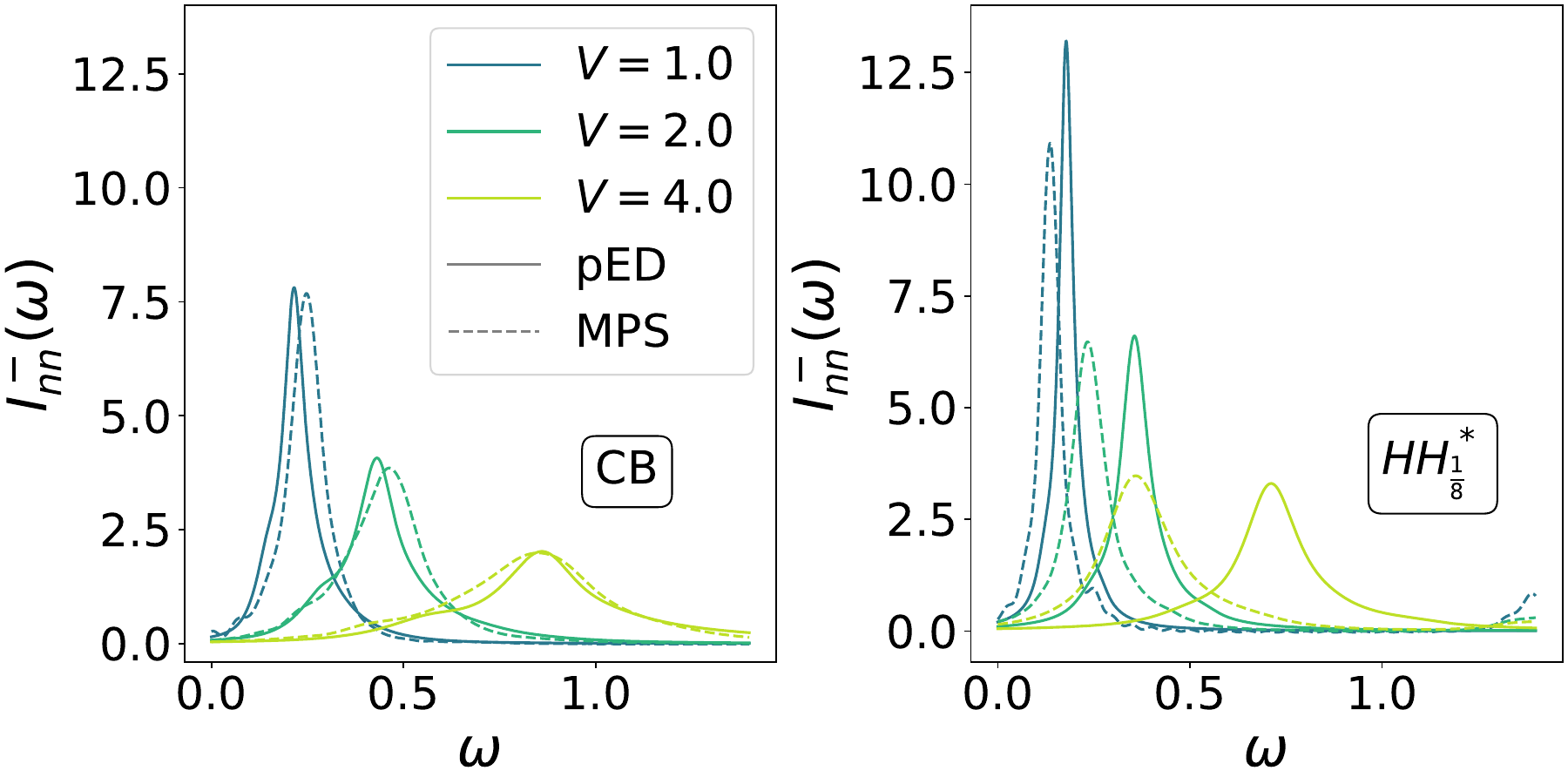}
    \put(13,45){(a)}
    \put(90,45){(b)}
    \end{overpic}
    \caption{\textbf{Interaction dependence of the graviton-mode}. (a) Comparison between pED ($N=10$) and MPS ($N=24$) results at $R=1$ (FCI limit at CB side) for different interaction strengths ($\eta=0.01V$). (b) same comparison as (a) but for $R = 0$ in HH$^*_{\frac{1}{8}}$ side between pED ($N=10$) and MPS ($N=12$). $\eta=0.01V$ in all panels. }
    \label{fig:Vdependence}
\end{figure}

As large interaction strengths can increase band-mixing effects, in Fig. \ref{fig:Vdependence} we compare spectra obtained with both pED (full lines, $N=10$ torus) and MPS (dashed lines, $N=24$ cylinder) at different values of $V$ for the CB (a) and HH$^*_{\frac{1}{8}}$ (b) cases. The results are again qualitatively in agreement, with both graviton energy and decay rate increasing with $V$. This confirms that below the band-gap $\omega_G\lesssim \Delta E$ band mixing is not qualitatively important. We also highlight that \textit{band} mixing does not in principle correspond to the same physics of Landau Level mixing. At the CB point, the first excited band is highly dispersive and of opposite Chern number, while the HH$^*_{\frac{1}{8}}$ case mimics LLs up to the third band. We further note that at the CB point (panel (a)) the band mixing seems to have less effect with respect to the HH$^*_{\frac{1}{8}}$ point (panel (b)). Indeed, while the total band-gap is comparable (see Fig. \ref{fig:8site_inter_scheme} (c)), the actual gap at the $\Gamma$ point where the graviton lives is much larger in the CB case $\Delta E_\Gamma \sim 6$ than in the HH$^*_{\frac{1}{8}}$ case $\Delta E_\Gamma \sim 1.3$, which largely eliminate the band mixing effect on CB side. 

\begin{figure}
\centering
        \begin{overpic}
 [width=\linewidth]{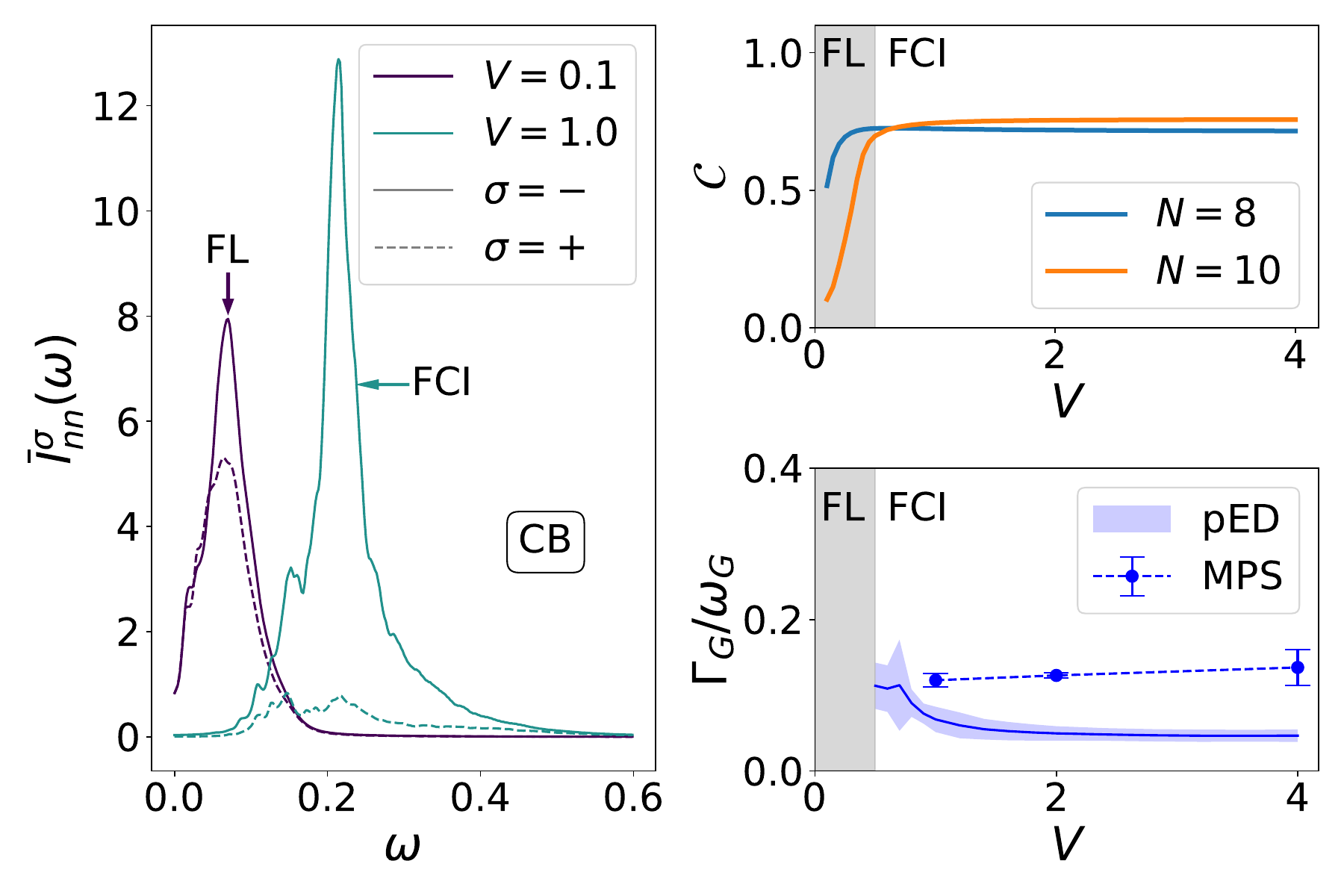}
 \put(13,60){(a)}
  \put(70,57){(b)}
  \put(70,23){(c)}
    \end{overpic}
    \caption{\textbf{Spectroscopic signature of the FCI and FL(Fermi Liquid) phase in the CB point}. (a) Comparison between spectral functions inside the FCI phase ($V=1$) and in the Fermi Liquid phase ($V=0.1$) \cite{Sheng2011_FQAH_checkerboard_fermion}. Results obtained with pED ($N=10$) and $\eta=0.005$. (b) Relative chirality of the integrated spectral function $\mathcal{C}=|N_--N_+|/(N_++N_-)$, where $N_\sigma$ is the integrated spectrum weight at $\sigma\in\{+,- \}$ sector. (c) Relative decay rate $\Gamma_G/\omega_G$ of the chiral graviton-mode as a function of $V$ in the FCI phase, obtained with pED (full line with shaded area) and MPS (dashed line). The pED line is an average between the $N=8$ and $N=10$ sizes, with the shaded area representing the error of this average. MPS results have a fixed $N = 24$.}
    \label{fig:Vdependence_transition}
\end{figure}

We now focus more closely on the CB point in Fig.~\ref{fig:Vdependence_transition}. Here we extend our analysis in a wider regime of interactions $V$, including $V\lesssim V_c=0.5$ where the system is in a Fermi Liquid (FL) phase \cite{Sheng_natcom2011_fciCB}. In panel (a), we compare two spectra obtained in the two phases FL and FCI, using a relative normalization to the spectra at $V=1$. The FCI shows the feature of so far analyzed graviton response (peaked, chiral, and gapped) while the FL shows a response which is mostly not chiral, gapless, and not strongly peaked. In panel (b), we show how the spectrum chirality $\mathcal{C}=|N_--N_+|/(N_++N_-)$ (see Eq. \eqref{eq:normalization_def} for definition), witnesses the phase transition, and thereby can be used to distinguish the two phases. 

In panel (c), we further quantify the relative decay rate $\Gamma_G/\omega_G$ dependence on $V$ inside the FCI phase and approaching the transition point. Here, the pED data are represented as a shaded region which averages the results at $N=8$ and $N=10$ obtained with the procedure introduced in Sec. \ref{sec:HHnum_lifetime} and App. \ref{sec:app_lifetime_extraction}. For MPS spectra (at fixed $N=24$) we extract the decay rate and its error again in the same way. At large values of $V$, the relative decay becomes independent of $V$, which implies the linear dependence of $\Gamma_G$ on $V$ as the graviton energy $\omega_G$ is proportional to $V$. The $V$ linear dependence of $\Gamma_G$ indicates that the interactions dominate the graviton-magnetoroton pair scattering matrix element. This is to be expected as the strength of the non-rotational invariant projected interactions, which generate graviton-magnetoroton pairs scattering, must also be proportional to $V$, in agreement with the discussion introduced in Ref. \cite{wangDynamics2025}. Closer to the FL phase the relative decay rate of the graviton seems to increase. In this regime, we also have that bandwidth ($W\sim 0.09$) becomes comparable with the interaction scale $V$, possibly giving rise to a different scattering matrix element. Whether this is a genuine new scattering channel, an increase in graviton-magnetoroton pair scattering, or just a finite-size effect coming from the proximity to another phase is an investigation we leave for future investigation.

\section{Conclusions }
\label{sec:V}

We have presented a unified analytical and numerical study of chiral graviton modes in fermionic fractional Chern insulators, addressing fundamental questions concerning the fate of geometric collective excitations beyond the continuum Landau level (LL) scenario. Our workhorse is a construction of explicit lattice stress tensor operator and lattice quadrupolar density correlators~\cite{xavier2025chiralgravitonslattice,longSpectra2025} at the level of the HH$^*$ model. We have validated the phenomenological lattice quadrupolar density correlators in the LL limit and track their spectral response across a controlled interpolation between fractional quantum Hall (FQH) and fractional Chern insulator (FCI) regimes. We have demonstrated that chiral graviton modes persist as long-lived excitations in lattice fractionalized phases.

The key result of our analysis is the explicit establishment of an adiabatic connection between FQH and FCI gravitons, described by the quadrupolar dynamics of short-range density correlations of these phases. Starting from a low-flux Harper-Hofstadter model that faithfully reproduces continuum LL physics and interpolating to a CB topological flat band with nonuniform quantum geometry and imperfect flatness, we showed that the graviton mode evolves continuously in energy, chirality, and lifetime. This firmly identifies the graviton as a robust geometric excitation shared by both continuum and lattice fractional topological phases, despite the absence of continuous translational and rotational symmetries on the lattice, that prevent an explicit identification of putative conserved charges. 

Importantly, we have provided strong evidence that FCI gravitons displays a relatively long lifetime. While lattice effects may ultimately allow decay into the two-magnetoroton continuum, our finite-size analysis reveals an intrinsic decay rate that remains small compared to the graviton energy. This establishes that the loss of guiding-center rotational symmetry does not immediately destroy the graviton as a meaningful excitation, but instead induces a controlled broadening that reflects lattice-induced non-uniform quantum geometry.

Another key conceptual insight of this work is the identification of short-range quadrupolar density-density correlators as faithful probes of graviton dynamics. We showed that a chiral quadrupolar density operator not only reproduces the continuum stress-tensor response in the Landau level limit, but also remains effective deep in the FCI regime. This real-space perspective clarifies that graviton excitations govern the dynamics of quadrupolar distortions in the short-range correlation holes on the lattice. Moreover, it justifies the use of density-based probes amenable to experimental platforms such as cold-atom quantum simulators \cite{xavier2025chiralgravitonslattice,leonard2023realization,impertro2024realization}.

Our results also open several promising directions for future work, that encompass solid state settings, cavity QED materials, and cold atom experiments. The demonstrated sharp and chiral graviton spectral response can be used as a dynamical witness of FCI phases. Circularly polarized Raman scattering experiments \cite{liang2024evidence} have been demonstrated to be a viable tool for GaAs heterostructures, however, similar spectroscopy information can be recovered from the formation of graviton-polariton excitations \cite{bacciconi_prx2025_gravitonpolaritons} in the context of cavity QED with subwavelength set-ups \cite{andolina2025quantumelectrodynamicsgraphenelandau}. In fact, our findings here directly motivate experiments with FCI embedded in cavities at frequencies comparable to the magnetoroton band~\cite{appugliese2022breakdown}. 

Overall, we stress that, while for solid-state systems our discoveries of long-lived graviton mode would complement and enrich more standard experiments for transition metal dichalcogenides and rhombohedral multilayer graphene \cite{Cai2023_signature_fqah_mote2,Park2023_observation_fqah_mote2,Zeng2023_thermo_evidence_fqah_mote2,Xu2023_Observation_FQAH_tMote2,Lu2024_FQAH_multilayer_graphene}, for cold-atom systems \cite{leonard2023realization}, where transport properties are harder to measure, detecting signatures of a chiral graviton-mode would be a key spectroscopic tool. Such probes are not only alternative routes to demonstrate genuine properties of FQH and FCI systems, but, importantly, work extremely well already at mesoscopic volumes (less than ten fermions), some of which could even be accessible to remarkable single tweezer experiments~\cite{lunt_prl2024_fqhfermions}.

On the theoretical side, the adiabatic connection with continuum physics indicates the existence of an associated metric in FCIs, which we have shown to linked to the shape of short-range correlations; it would also be interesting to directly build an effective theory for an emergent metric in the FCI point\footnote{We thank Bo Yang for pointing out this perspective to us.}. Then, similarly to what was done in the continuum, 
the study of graviton-modes in non-abelian phases~\cite{liou2019chiral,Haldane2021Graviton} or more complex abelian phases \cite{nguyen2022multiple,BalramPapic_PRX2022} realized on the lattice is also a natural extension, relevant for both bosonic cold-atom systems \cite{palm2021pfaffian,BoeslKnap_prb2022_characterizing_nonabelian} and fermionic quantum materials \cite{Chen_natcom2025_nonabelian,Aidan_PRL2024_nonabelian,Cheong-Eung_prb2024_nonabelian}. Going back to the continuum scenarios, it would also be interesting to apply the graviton lifetime analysis introduced here to different FQH states \cite{liou2019chiral,nguyen2022multiple,BalramPapic_PRX2022,herviou2024numerical} and provide a numerically controlled estimate for it. This could be used for example to understand to what extent the observed linewidth of the graviton resonance in Ref.~\cite{liang2024evidence} is a disorder-induced effect. On the other side, the exploration of the lattice version of FQH-FQHN(fractional quantum hall nematic) transition driven by the closing of the graviton gap is also interesting~\cite{pu2024microscopic}.

Taken together, our work establishes chiral gravitons as a unifying and experimentally relevant feature of fractionalized quantum phases in lattice settings, and provides a concrete framework for their detection, characterization, and future exploration.

\begin{acknowledgements}
We thank Kai Sun for introducing us the HH$^*$-CB interpolations. Z.B., M.D. and H. X. B. thank Titas Chanda and Dam Thanh Son for previous collaboration on the topic. We thank Yuzhu Wang, Dung Xuan Nguyen, Ajit C. Balram, Bo Yang, and Kun Yang for their constructive communications and suggestions on the manuscript.

ML and ZYM acknowledge the support from the Research Grants Council (RGC) of Hong Kong (Project Nos. AoE/P-701/20, 17309822, HKU C7037-22GF, 17302223, 17301924), the ANR/RGC Joint Research Scheme sponsored by RGC of Hong Kong and French National Research Agency (Project No. A\_HKU703/22), and the HKU Seed Funding for Strategic Interdisciplinary Research “Many-body paradigm in
quantum moiré material research”. ML thanks ICTP for its kind hospitality
through the Project "Wave-function Networks: Probe and understand quantum many-body systems
via network and complexity theory - WaveNets", funded by the European Union (Grant Agreement n. 101087692). We thank HPC2021 system under the Information Technology Services at the University of Hong Kong~\cite{hpc2021}, as well as the Beijing Paratera Tech Corp., Ltd~\cite{paratera} for providing HPC resources that have contributed to the research results reported within this paper. 
M.~D. was partly supported by the QUANTERA DYNAMITE PCI2022-132919, by the EU-Flagship programme Pasquans2, by the PNRR MUR project PE0000023-NQSTI, the PRIN programme (project CoQuS), and by the ERC Consolidator grant WaveNets (Grant agreement ID: 101087692). 

\end{acknowledgements}

\appendix

\section{Derivation of lattice stress tensor}
\label{sec:appA}
In this section, we provide a derivation of the lattice stress tensor operator. 

As explained in the main text, the stress tensor is obtained via matching the derivatives of the stress tensor with the equation of motion of the current operator.

\begin{equation}
    \begin{aligned}
    T^{xx}_{\rr+\ex} - T^{xx}_\rr = \partial_x T^{xx}_\rr &= i[\bar{H}^x,\bar{j}_\rr^x]-i[H^x_0,j_\rr^x], \\ 
    T^{yy}_{\rr+\ey} - T^{yy}_{\rr} = \partial_y T^{yy}_\rr &= i[\bar{H}^y,\bar{j}_\rr^y]-i[H^y_0,j_\rr^y], \\ 
    T^{yx}_{\rr+\ey} - T^{yx}_{\rr} = \partial_y T^{yx}_i &= i[\bar{H}^y,\bar{j}_\rr^x]-  i[H^y_0,j_\rr^x]
    \end{aligned}
    \label{eq:commutators}
\end{equation}

\begin{figure}[h!]
    \centering
    \includegraphics[width=\linewidth]{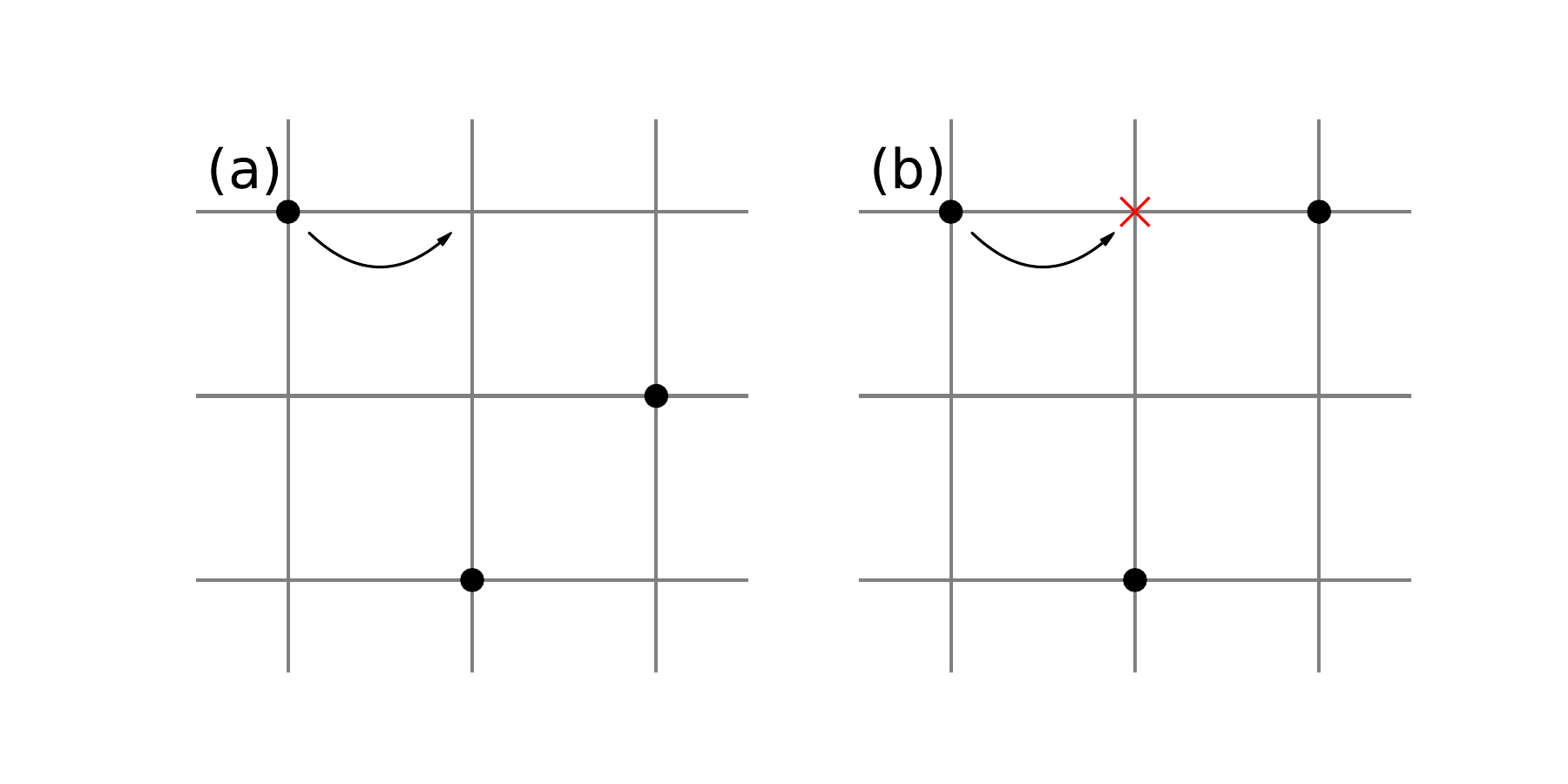}
    \caption{Geometric constraint for $\bar{c}^\dagger_i,\bar{c}_i$ in $V\rightarrow \infty$ limit with $V$ the nearest neighboring interaction. Hoppings of type (a) are allowed. Hoppings of type (b) will result in neighboring occupation and energy penalty $V$, and thus are forbidden. }
    \label{fig:SI_currentI}
\end{figure}

Where $H$ and $\mathbf{j}$ are written in terms of the projected operators $\bar{c}^\dagger_i,\bar{c}_i$ in $V\rightarrow \infty$ limit, where the occupation of neighboring sites is forbidden. As shown in Fig.~\ref{fig:SI_currentI}, hoppings that will result in occupation of neighboring sites will gain an energy penalty $V$, which is forbidden in $V\rightarrow \infty$. Therefore, we add the geometric constraint on the projected fermionic operators that will bring a $V$ penalty. 
\begin{equation}
   \forall \{ \langle k , j \rangle \} : \langle \bar{c}_i^\dagger \bar{c}_j\bar{c}_k^\dagger\bar{c}_l\rangle  = 0
    \label{eq:SI_geo_constraint}
\end{equation}
This expression is not normal-ordered; however, in evaluating the commutators like $[\bar{H}^x,\bar{j}_{\boldsymbol{r}}^x]$, we always encounter 2-point correlators of this type. Therefore, it is sufficient for us to use these types of geometric constraints.

\begin{figure}[th!]
    \centering
    \includegraphics[width=\linewidth]{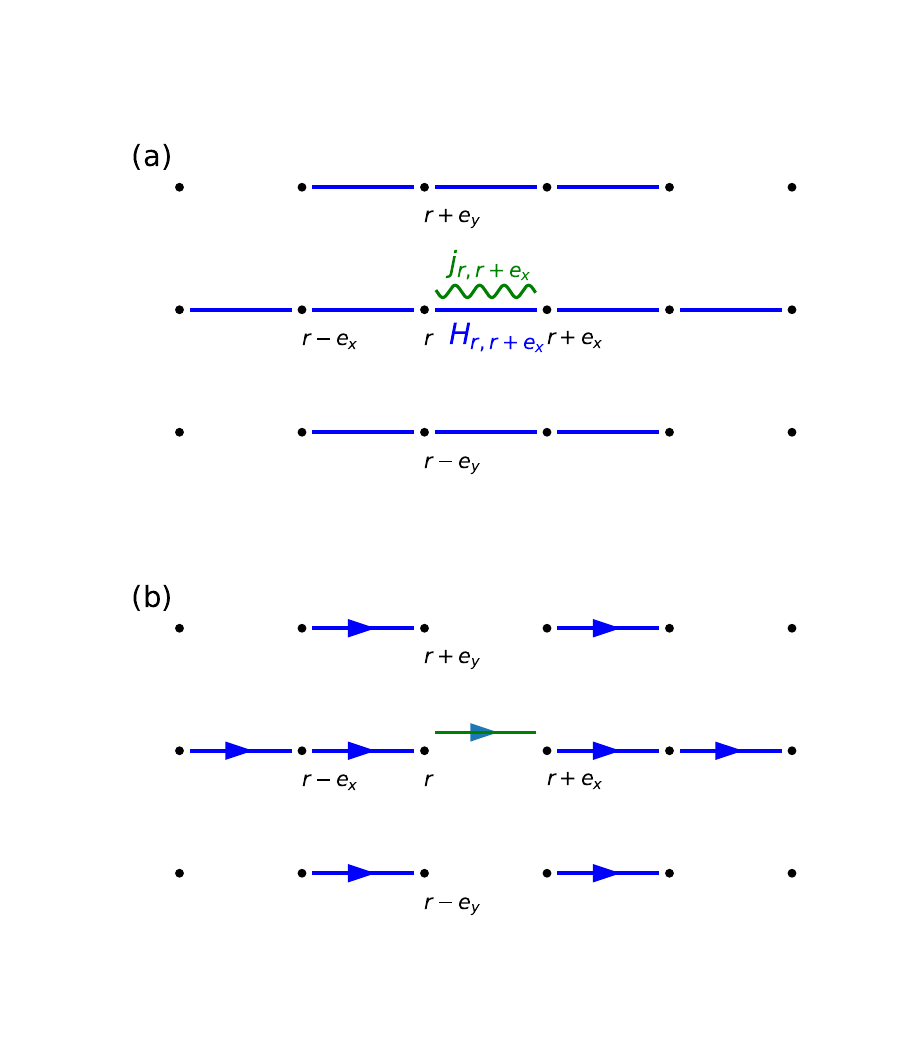}
    \caption{Relevant commutators that will contribute to $[\bar{H}^x,\bar{j}_\rr^x]$. The blue lines represent $\bar{H}^x_r$, and the green wavy line labels $\bar{j}_r^x$}
    \label{fig:sketch_current_II}
\end{figure}

 To derive $T^{xx}$, we first sort out the terms in $\bar{H}^x$ that could have non-zero commutators with $\bar{j}^x_\mathbf{r}$. Due to the geometrical constraint of Eq~\ref{eq:SI_geo_constraint}. Next nearest terms are included, as depicted in Fig.~\ref{fig:sketch_current_II}(a), where we label the relevent elements of $H^x$ in computing $[\bar{H}^x,\bar{j}^x_r]$ with $\bar{j}_r^a = -i(\bar{c}^\dagger_r \bar{c}_{r+e_a}e^{i\phi_{r}^x}-\bar{c}^\dagger_{r+e_a} \bar{c}_{r}e^{-i\phi_{r}^x})$.We use a diagrammatic way to lighten the notation.

\begin{equation}
\begin{array}{cc}
     & \tikz[baseline=-.4ex]{\connectarrow{(0,0)}{(0.5,0)}; 
        \node at (0,-0.2) {$i$};
        \node at (0.5,-0.2) {$j$};
        }   = \bar{c}_j^\dagger \bar{c}_i e^{\theta_{ji}} = B_{ij} \\
     &  \tikz[baseline=-.4ex]{\connectarrow{(0,0)}{(0.5,0)}; 
        \node at (0,-0.2) {$i$};
        \node at (0.5,-0.2) {$j$};
        \connectarrow{(1,0)}{(1,-0.5)};
        \node at (1.2,0) {$k$};
        \node at (1.2,-0.5) {$l$};
        }   = \bar{c}_j^\dagger \bar{c}_i \bar{c}_l^\dagger \bar{c}_k e^{\theta_{ji}}e^{\theta_{lk}} = B_{ij} B_{kl} \\ &
        (\tikz[baseline=-.4ex]{\connectarrow{(0,0)}{(0.5,0)}; 
        \node at (0,-0.2) {$i$};
        \node at (0.5,-0.2) {$j$};
        })^\dagger = \tikz[baseline=-.4ex]{\connectarrow{(0.5,0)}{(0,0)}; 
        \node at (0,-0.2) {$i$};
        \node at (0.5,-0.2) {$j$};
        }~,~ B_{ij}^\dagger = B_{ji}
\end{array}
\label{eq:diagram_notation}
\end{equation}

It is straight forward to see $i[\bar{H}^x,\bar{j}^x_r] = [\bar{H}^x,  \tikz[baseline=-.4ex]{\connectarrow{(0.,0.)}{(0.5,0)}; 
        \node at (0,-0.2) {$i$};
        \node at (0.5,-0.2) {$j$};
        } - \tikz[baseline=-.4ex]{\connectarrow{(0.5,0)}{(0,0)}; 
        \node at (0,-0.2) {$i$};
        \node at (0.5,-0.2) {$j$};
        }]$ =  $[H^x, \tikz[baseline=-.4ex]{\connectarrow{(0.,0)}{(0.5,0)}; 
        \node at (0,-0.2) {$i$};
        \node at (0.5,-0.2) {$j$};
        }] + h.c.$ 

For $[H^x, \tikz[baseline=-.4ex]{\connectarrow{(0.,0)}{(0.5,0)}; 
        \node at (0,-0.2) {$i$};
        \node at (0.5,-0.2) {$j$};
        }]$. The commutators could be computed considering the geometric constraints. Here we compute $[B_{r+e_y-e_x,r+e_y}+B_{r+e_y,r+e_y-e_x},B_{r,r+e_x}]$ as an example.
\begin{equation}
    \begin{aligned}
       &[B_{r+e_y-e_x,r+e_y}+B_{r+e_y,r+e_y-e_x},B_{r,r+e_x}] \\ & = [B_{r+e_y-e_x,r+e_y},B_{r,r+e_x}] + [B_{r+e_y,r+e_y-e_x},B_{r,r+e_x}] \\ 
       &=B_{r+e_y-e_x,r+e_y}B_{r,r+e_x} - B_{r,r+e_x}B_{r+e_y-e_x,r+e_y} + \\ &B_{r+e_y,r+e_y-e_x}B_{r,r+e_x} - B_{r,r+e_x}B_{r+e_y,r+e_y-e_x} \\ &
        = B_{r+e_y-e_x,r+e_y}B_{r,r+e_x}
    \end{aligned}
\end{equation}
The other three terms contribute zero because of the geometrical constraint. In diagram language,
\begin{equation}
\begin{aligned}
            [ \tikz[baseline=-.4ex]{\connectarrow{(-0.5,0.5)}{(0.,0.5)}; 
        } + \tikz[baseline=-.4ex]{\connectarrow{(0.,0.5)}{(-0.5,0.5)}; 
        }, \tikz[baseline=-.4ex]{\connectarrow{(0.,0)}{(0.5,0)}; 
        \node at (0,-0.2) {$r$};
        }]  &= \tikz[baseline=-.4ex]{\connectarrow{(-0.5,0.5)}{(0.,0.5)}; 
        {\connectarrow{(0.,0)}{(0.5,0)}}; 
        }
\end{aligned}
\end{equation}
Similarly,  $[H^x, \tikz[baseline=-.4ex]{\connectarrow{(0.,0)}{(0.5,0)}; 
        \node at (0,-0.2) {$i$};
        \node at (0.5,-0.2) {$j$};
        }]$ could be evaluated. The non-zero contributions are shown in Fig.~\ref{fig:sketch_current_II}(b).
Matching the equation 
\begin{equation}
    T^{xx}_{\rr+\ex} - T^{xx}_\rr = \partial_x T^{xx}_\rr = -i[\bar{H}^x,\bar{j}_\rr^x] - i[H^x_0,{j}_\rr^x]
\end{equation}
And exploiting the translational symmetry of the stress tensor, we obtain

\begin{equation}
    \begin{aligned}
        &i[\bar{H}^x,\bar{j}^x_{i,i+e_x}]  =  \tikz[baseline=-.4ex]{\connectarrow{(-0.5,0.5)}{(0,0.5)};
        \connectarrow{(0,0)}{(0.5,0)}; 
        \node at (0,-0.2) {$i$};
        } ~ + ~  \tikz[baseline=-.4ex]{\connectarrow{(-0.5,-0.5)}{(0,-0.5)};
        \connectarrow{(0,0)}{(0.5,0)}; 
        \node at (0,-0.2) {$i$};
        } ~+ ~\tikz[baseline=-.4ex]{\connectarrow{(-1,0)}{(-0.5,0)};
        \connectarrow{(0,0)}{(0.5,0)}; 
        \node at (0,-0.2) {$i$};
        } ~+ ~\tikz[baseline=-.4ex]{\connectarrow{(-0.5,0)}{(0,0)};
        \connectarrow{(0,0)}{(0.5,0)}; 
        \node at (0,-0.2) {$i$};
        } \\[1.5em] & ~-~ \tikz[baseline=-.4ex]{\connectarrow{(0.5,0.5)}{(1,0.5)};
        \connectarrow{(0,0)}{(0.5,0)}; 
        \node at (0,-0.2) {$i$};
        } ~-~ \tikz[baseline=-.4ex]{\connectarrow{(0.5,-0.5)}{(1,-0.5)};
        \connectarrow{(0,0)}{(0.5,0)}; 
        \node at (0,-0.2) {$i$};
        } ~-~ \tikz[baseline=-.4ex]{\connectarrow{(1,0)}{(1.5,0)};
        \connectarrow{(0,0)}{(0.5,0)}; 
        \node at (0,-0.2) {$i$};
        } ~-~ \tikz[baseline=-.4ex]{\connectarrow{(0.5,0)}{(1,0)};
        \connectarrow{(0,0)}{(0.5,0)}; 
        \node at (0,-0.2) {$i$};
        } + h.c.\\ &
        = B_{i+e_y-e_x,i+e_y}B_{i,i+e_x} +B_{i-e_y-e_x,i-e_y}B_{i,i+e_x}  \\ & + B_{i-2e_x,i-e_x}B_{i,i+e_x}  + B_{i-e_x,i}B_{i,i+e_x} - \\ &B_{i,i+e_x} B_{i+e_y,i+e_y+e_x} - B_{i,i+e_x} B_{i-e_y,i-e_y+e_x} -  \\ &B_{i,i+e_x} B_{i+2e_x,i+3e_x} - B_{i,i+e_x} B_{i+e_x,i+2e_x} + h.c.
        \end{aligned}
\end{equation}
Matching this expression with $T_{r+e_x}^{xx} - T_{r}^{xx}$, we derive the explicit expression,
\begin{equation}
    \begin{aligned}
        & T_r^{xx} =  \\ &\tikz[baseline=-.4ex]{\connectarrow{(-0.5,0.)}{(0.,0.)}; 
        {\connectarrow{(0,0.)}{(0.5,0.)};
        \node at (0,-0.2) {$r$};
        }} + \tikz[baseline=-.4ex]{\connectarrow{(-0.5,0.5)}{(0.,0.5)}; 
        {\connectarrow{(0,0.)}{(0.5,0.)};
        \node at (0,-0.2) {$r$};
        }} +
        \tikz[baseline=-.4ex]{\connectarrow{(-1,0.)}{(-0.5,0.)}; 
        {\connectarrow{(0,0.)}{(0.5,0.)};
        \node at (0,-0.2) {$r$};
        }} + \\[1.5em] & \tikz[baseline=-.4ex]{\connectarrow{(-0.5,-0.5)}{(0.,-0.5)}; 
        {\connectarrow{(0,0.)}{(0.5,0.)};
        \node at (0,-0.2) {$r$};
        }} + \tikz[baseline=-.4ex]{\connectarrow{(0,0.5)}{(0.5,0.5)}; 
        {\connectarrow{(-0.5,0.)}{(0,0.)};
        \node at (0,-0.2) {$r$};
        }} + \tikz[baseline=-.4ex]{\connectarrow{(0.5,0)}{(1,0.)}; 
        {\connectarrow{(-0.5,0.)}{(0,0.)};
        \node at (0,-0.2) {$r$};
        }} + \tikz[baseline=-.4ex]{\connectarrow{(0,-0.5)}{(0.5,-0.5)}; 
        {\connectarrow{(-0.5,0.)}{(0,0.)};
        \node at (0,-0.2) {$r$};
        }} + h.c.
    \end{aligned}
\end{equation}
Written explicitly,
\begin{equation}
    \begin{aligned}
    T^{xx}_{r}= A_r^{x,x}
    +\frac{B_{r-x}^x}{2}\left(\sum_{dr=\pm y,x} B_{r+dr}^x \right)+\mathrm{h.c}
    \end{aligned}
\end{equation}

By using rotational symmetry, $T_r^{yy}$ could be obtained by substituting all $x$ in the expression of $T_{xx}$ with $y$.

To derive $T_r^{xy}$, following similar calculations, we derive
\begin{equation}
    \begin{aligned}
        & i[\bar{H}^y,\bar{j}^x_{r,r+e_x}] =  \\ &\tikz[baseline=-.4ex]{
        {\connectarrow{(0,0.)}{(0.5,0.)};
        \connectarrow{(0,1)}{(0,0.5)};
        \node at (0,-0.2) {$r$};
        }}  + \tikz[baseline=-.4ex]{
        {\connectarrow{(0,0.)}{(0.5,0.)};
        \connectarrow{(-0.5,0.5)}{(-0.5,0)};
        \node at (0,-0.2) {$r$};
        }} - \tikz[baseline=-.4ex]{
        {\connectarrow{(0,0.)}{(0.5,0.)};
        \connectarrow{(0.5,0.5)}{(0.5,1)};
        \node at (0,-0.2) {$r$};
        }} \\[1.5em] &- \tikz[baseline=-.4ex]{
        {\connectarrow{(0,0.)}{(0.5,0.)};
        \connectarrow{(1,0)}{(1,0.5)};
        \node at (0,-0.2) {$r$};
        }} + \tikz[baseline=-.4ex]{
        {\connectarrow{(0,0.)}{(0.5,0.)};
        \connectarrow{(0,0.5)}{(0,0)};
        \node at (0,-0.2) {$r$};
        }} - \tikz[baseline=-.4ex]{
        {\connectarrow{(0,0.)}{(0.5,0.)};
        \connectarrow{(0.5,0)}{(0.5,0.5)};
        \node at (0,-0.2) {$r$};
        }}  \\[1.5em] & + \tikz[baseline=-.4ex]{
        {\connectarrow{(0,0.)}{(0.5,0.)};
        \connectarrow{(-0.5,-0.5)}{(-0.5,0)};
        \node at (0,-0.2) {$r$};
        }} + \tikz[baseline=-.4ex]{
        {\connectarrow{(0,0.)}{(0.5,0.)};
        \connectarrow{(0,-1)}{(0,-0.5)};
        \node at (0,-0.2) {$r$};
        }} - \tikz[baseline=-.4ex]{
        {\connectarrow{(0,0.)}{(0.5,0.)};
        \connectarrow{(1,0)}{(1,-0.5)};
        \node at (0,-0.2) {$r$};
        }} \\[1.5em] & - \tikz[baseline=-.4ex]{
        {\connectarrow{(0,0.)}{(0.5,0.)};
        \connectarrow{(0.5,-0.5)}{(0.5,-1)};
        \node at (0,-0.2) {$r$};
        }} + \tikz[baseline=-.4ex]{
        {\connectarrow{(0,0.)}{(0.5,0.)};
        \connectarrow{(0,-0.5)}{(0,0)};
        \node at (0,0.2) {$r$};
        }} - \tikz[baseline=-.4ex]{
        {\connectarrow{(0,0.)}{(0.5,0.)};
        \connectarrow{(0.5,0)}{(0.5,-0.5)};
        \node at (0,-0.2) {$r$};
        }}+ h.c.\\& = B_{r+2e_y,r+e_y}B_{r,r+e_x} +  B_{r+e_y -e_x,r-e_x}B_{r,r+e_x} - \\& B_{r,r+e_x}B_{r+e_y+e_x,r+2e_y+e_x}
        -B_{r,r+e_x}B_{r+2e_x,r+2e_x+e_y} +\\& B_{r+e_y,r}B_{r,r+e_x} - B_{r,r+e_x}B_{r+e_x,r+e_x+e_y} \\&+ B_{r-e_y,r}B_{r,r+e_x} + B_{r-2e_y,r-e_y}B_{r,r+e_x}\\& - B_{r,r+e_x}B_{r+2e_x,r+2e_x-e_y} - B_{r,r+e_x}B_{r+e_x - e_y,r+e_x -2e_y} \\&+ B_{r-e_y,r}B_{r,r+e_x} - B_{r,r+e_x}B_{r+e_x,r+e_x-e_y} + h.c.
    \end{aligned}
\end{equation}

We group terms in the first two lines as $T_{r+\frac{e_x}{2} + \frac{e_y}{2}}^{x,y}$ and the terms in the last two lines as $T_{r+\frac{e_x}{2} - \frac{e_y}{2}}^{x,y}$. Then the symmetrized $T_{r+\frac{e_x}{2} + \frac{e_y}{2}}^{x,y}$ using the translational symmetry of $T_{r+\frac{e_x}{2} - \frac{e_y}{2}}^{x,y}$, we derive,
\begin{equation}
    \begin{aligned}
        & T_{r+\frac{e_x}{2} + \frac{e_y}{2}}^{xy} =  \\ &\tikz[baseline=-.4ex]{
        {\connectarrow{(0,0.)}{(0.5,0.)};
        \connectarrow{(0,1)}{(0,0.5)};
        \node at (0,-0.2) {$r$};
        }}  + \tikz[baseline=-.4ex]{
        {\connectarrow{(0,0.)}{(0.5,0.)};
        \connectarrow{(-0.5,0.5)}{(-0.5,0)};
        \node at (0,-0.2) {$r$};
        }}  - \tikz[baseline=-.4ex]{
        {\connectarrow{(0,0.5)}{(0.5,0.5)};
        \connectarrow{(0,-0.5)}{(0,0)};
        \node at (0,0.7) {$r$};
        }} - \tikz[baseline=-.4ex]{
        {\connectarrow{(0,0.5)}{(0.5,0.5)};
        \connectarrow{(-0.5,0)}{(-0.5,0.5)};
        \node at (0,0.7) {$r$};
        }}   \\ & - \tikz[baseline=-.4ex]{
        {\connectarrow{(0,0.)}{(0.5,0.)};
        \connectarrow{(0.5,0.5)}{(0.5,1)};
        \node at (0,-0.2) {$r$};
        }} - \tikz[baseline=-.4ex]{
        {\connectarrow{(0,0.)}{(0.5,0.)};
        \connectarrow{(1,0)}{(1,0.5)};
        \node at (0,-0.2) {$r$};
        }} + \tikz[baseline=-.4ex]{
        {\connectarrow{(0,0.5)}{(0.5,0.5)};
        \connectarrow{(0.5,0)}{(0.5,-0.5)};
        \node at (0,-0.2) {$r$};
        }} + \tikz[baseline=-.4ex]{
        {\connectarrow{(0,0.5)}{(0.5,0.5)};
        \connectarrow{(1,0.5)}{(1,0.)};
        \node at (0,-0.2) {$r$};
        }} \\ & + \tikz[baseline=-.4ex]{
        {\connectarrow{(0,0.)}{(0.5,0.)};
        \connectarrow{(0,0.5)}{(0,0)};
        \node at (0,-0.2) {$r$};
        }} - \tikz[baseline=-.4ex]{
        {\connectarrow{(0,0.)}{(0.5,0.)};
        \connectarrow{(0.5,0)}{(0.5,0.5)};
        \node at (0,-0.2) {$r$};
        }}  - \tikz[baseline=-.4ex]{
        {\connectarrow{(0,0.5)}{(0.5,0.5)};
        \connectarrow{(0,0)}{(0,0.5)};
        \node at (0,-0.2) {$r$};
        }} + \tikz[baseline=-.4ex]{
        {\connectarrow{(0,0.5)}{(0.5,0.5)};
        \connectarrow{(0.5,0.5)}{(0.5,0.)};
        \node at (0,-0.2) {$r$};
        }}
    \end{aligned}
\end{equation}
In a similar way, we compute $i[H^x,J^y_{r,r+e_y}]$ and derive $T_{r+\frac{e_x}{2} + \frac{e_y}{2}}^{yx}$. The stress tensor is symmetrized by averaging $T^{xy}$ and $T^{yx}$. We derive the plaquette-centered stress tensor.
\begin{equation}
    \begin{aligned}
        & T_{r+\frac{1}{2}e_x + \frac{1}{2}e_y}^{xy} =  \frac{1}{2} (\tikz[baseline=-.4ex]{
        {\connectarrow{(0,0.)}{(0.5,0.)};
        \connectarrow{(0,0.5)}{(0,0)};
        \node at (0,-0.2) {$r$};
        }} - \tikz[baseline=-.4ex]{
        {\connectarrow{(0,0.)}{(0.5,0.)};
        \connectarrow{(0.5,0)}{(0.5,0.5)};
        \node at (0,-0.2) {$r$};
        }}  - \tikz[baseline=-.4ex]{
        {\connectarrow{(0,0.5)}{(0.5,0.5)};
        \connectarrow{(0,0)}{(0,0.5)};
        \node at (0,-0.2) {$r$};
        }} + \tikz[baseline=-.4ex]{
        {\connectarrow{(0,0.5)}{(0.5,0.5)};
        \connectarrow{(0.5,0.5)}{(0.5,0.)};
        \node at (0,-0.2) {$r$};
        }}) \\ & + \tikz[baseline=-.4ex]{
        {\connectarrow{(0,0.)}{(0.,0.5)};
        \connectarrow{(1,0.)}{(0.5,0.)};
        \node at (0,-0.2) {$r$};
        }} - \tikz[baseline=-.4ex]{
        {\connectarrow{(0,-0.5)}{(0.,0.)};
        \connectarrow{(0,0.5)}{(0.5,0.5)};
        \node at (0,0.2) {$r$};
        }} + \tikz[baseline=-.4ex]{
        {\connectarrow{(0,0.5)}{(0.5,0.5)};
        \connectarrow{(1,0.5)}{(1,0.)};
        \node at (0,-0.2) {$r$};
        }} - \tikz[baseline=-.4ex]{
        {\connectarrow{(0.5,0.)}{(0.,0.)};
        \connectarrow{(0.5,1)}{(0.5,0.5)};
        \node at (0,-0.2) {$r$};
        }} + h.c.
    \end{aligned}
\end{equation}
Finally, the site-centered stress tensor is written down using translational invariance. We note that upon summing over all the sites, platteque-centered convention is equivalent to the site-centered convention.
\begin{equation}
    \begin{aligned}
        & T_r^{xy} =  \frac{1}{2} (\tikz[baseline=-.4ex]{
        {\connectarrow{(0,0.)}{(0.5,0.)};
        \connectarrow{(0,0.5)}{(0,0)};
        \node at (0,-0.2) {$r$};
        }} - \tikz[baseline=-.4ex]{
        {\connectarrow{(-0.5,0.)}{(0.,0.)};
        \connectarrow{(0.,0)}{(0.,0.5)};
        \node at (0,-0.2) {$r$};
        }}  - \tikz[baseline=-.4ex]{
        {\connectarrow{(0,0.)}{(0.5,0.)};
        \connectarrow{(0,-0.5)}{(0,0.)};
        \node at (0,0.2) {$r$};
        }} + \tikz[baseline=-.4ex]{
        {\connectarrow{(-0.5,0.)}{(0.,0.)};
        \connectarrow{(0.,0.)}{(0.,-0.5)};
        \node at (0,0.2) {$r$};
        }}) \\ & + \tikz[baseline=-.4ex]{
        {\connectarrow{(0,0.)}{(0.,0.5)};
        \connectarrow{(1,0.)}{(0.5,0.)};
        \node at (0,-0.2) {$r$};
        }} - \tikz[baseline=-.4ex]{
        {\connectarrow{(0,-1)}{(0.,-0.5)};
        \connectarrow{(0,0.)}{(0.5,0.)};
        \node at (0,0.2) {$r$};
        }} + \tikz[baseline=-.4ex]{
        {\connectarrow{(-0.5,0.)}{(0.,0.)};
        \connectarrow{(0.5,0.)}{(0.5,-0.5)};
        \node at (0,-0.2) {$r$};
        }} - \tikz[baseline=-.4ex]{
        {\connectarrow{(0.,0.)}{(-0.5,0.)};
        \connectarrow{(0.,1)}{(0.,0.5)};
        \node at (0,-0.2) {$r$};
        }} + h.c.
    \end{aligned}
\end{equation}

\section{Details of building Periodic gauge and Hamiltonian}
\label{app:gauge_ham}
\begin{figure}[ht!]
    \centering
    \includegraphics[width=1\linewidth]{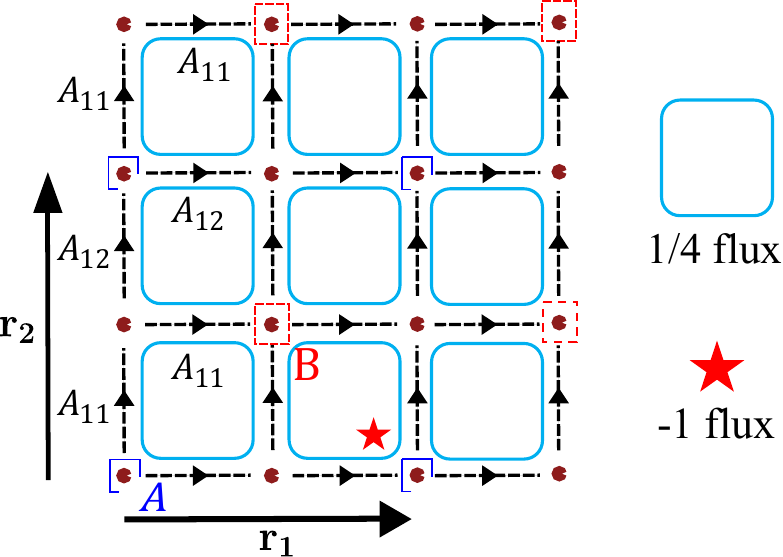}

    \caption{ Square lattice Hofstadter model with $1/4$ flux per plaquette. In the position of red star, we insert $-2\pi$ flux to make the unitcell zero net flux. $r_1$ and $r_2$ label the primitive lattice vectors of the 4-site unit cell. The A and B sublattices of the checkerboard sites are labeled with blue and red dashed rectangles, respectively. The black dashed arrow labels the direction of the vector potential, where we choose the gauge such that $A_{(i,j)\leftarrow(i,j-1)} = A_{(i+1,j)\leftarrow(i,j)}$, we also impose the periodic boundary condition $A_{(i+2,j+2)\leftarrow(i+2,j+1)} = A_{(i,j)\leftarrow(i,j-1)}$, same for the vector potential along $r_1$ directions.}
    \label{fig:single_particle}
\end{figure}
In this section, we detail how we build the periodic gauge and the 'site gauge' single particle Hamiltonian for a $1/4$ flux HH model. For $1/8$ flux HH model used in the main text, the gauge fixing could be done similarly. 

To fix a gauge choice of $\phi_{i,j}$, using a notation where sites are labeled as $i\equiv(x,y)$, we first impose a convention that the phase coming up to a site is the same as the phase going out of it toward the right. That is,
\begin{align}
     &\phi_{(x,y),(x,y-1)}= \phi_{(x+1,y),(x,y)}
\end{align}
Along with the periodic condition $\phi_{r_i,r_j} = \phi_{r_i + R,r_j+R}$ with $R$ the Bravais lattices of the HH model. We determine the vector potential by using Stokes' theorem
\begin{equation}
    \begin{aligned}
     &2\phi_{(x,y),(x,y-1)} +\phi_{(x+1,y-1),(x+1,y)}+\phi_{(x,y-1),(x+1,y-1)}\\ &=  \{2\pi n_\phi , 2\pi (n_\phi-1)\}
\end{aligned}
\end{equation}

The RHS is the magnetic flux of the plaquette, which equals $2\pi n_\phi$ if no $-2\pi$ flux is inserted and vice versa. Solving these linear equations within one unit cell, we fix the gauge $\phi_{i,j}$.

Specifically, for $1/4$ HH model. We have the following linear equations.
\begin{equation}
    \begin{aligned}
     &2A_{11} - A_{21} - A_{12}=  \frac{\pi}{2} \\
     &2A_{12} - A_{22} - A_{11}=  \frac{\pi}{2} \\
     &2A_{22} - A_{12} - A_{21}=  \frac{\pi}{2} \\
     &2A_{21} - A_{11} - A_{22}=  -\frac{3\pi}{2} \\
    \end{aligned}
\end{equation}
Fixing a global constant by setting $A_{11} = 0$. These equations could be solved by using,
\begin{equation}
    v = (M^TM)^{-1} M^T b
\end{equation}
Where $v = [A_{11} \ A_{12}\ A_{21}\ A_{22}]^T$ is the solution. $M$ is the coefficient matrix such that $M\cdot v = b$, with $b$ the RHS of the linear equations. The result is,
\begin{equation}
    A_{11} = 0 ,  A_{12} = -\frac{3}{4}\pi, A_{21} = -\frac{3}{4}\pi, A_{22} = -\pi
\end{equation}

With the gauge fixing, for a electron with momentum $k$, any translation operation with vector along $x(y)$ direction with $L = 2$ would accumulate extra $-\frac{7\pi}{4}(-\frac{3\pi}{4})$ phase. Therefore, we shift the zero (canonical) momentum point to $\Gamma = (-\frac{7\pi}{4},-\frac{3\pi}{4})$ to compensate for this phase accumulation. For $1/8$ flux HH model, this shift also occurs.

Having fixed the gauge, we now write down the single particle Hamiltonian momentum space in 'site gauge',  which respects the $C_4$ symmetry of the underlying lattice. In other words, the Fourier transformation of the hoppings within a unit cell $\sum_R{e^{\phi_{i,j}}a_{R,i}^\dagger a_{R,j}}$ becomes $\sum_k{e^{\phi_{i,j}}e^{i\mathbf{k}\cdot(\mathbf{r_i-r_j})}a_{k,i}^\dagger a_{k,j}}$. Here we use $R$ to label the unit cell and $i,j$ to denote the sublattices. Below, we write down the non-zero matrix elements of the upper triangular part of the single particle Hamiltonian matrix; the other terms could be obtained via Hermitian conjugate.
\begin{equation}
    \begin{aligned}
        H_{1,2} &= -te^{i\frac{k_y}{2}} -te^{i(-\frac{k_y}{2}+\frac{3\pi}{4})} \\ 
        H_{1,3} &=  -te^{i(\frac{k_x}{2} - \frac{3\pi}{4})} -te^{i(-\frac{k_x}{2} + \pi)}  \\ 
        H_{2,4} &= -te^{i\frac{k_x}{2}}  - t e^{i(-\frac{k_x}{2}+\frac{3}{4}\pi)} \\
        H_{3,4} & =  -te^{i(\frac{k_y}{2}-\frac{3}{4}\pi)} -te^{-i(\frac{k_y}{2}+\pi)} 
    \end{aligned}
\end{equation}

The construction of the vector potential for hoppings of the CB super lattice follows a similar idea of solving linear loop equations of the magnetic flux. For example, for the hopping of $A,B$ sublattices within the same unit cell, we have
\begin{equation}
    \phi_{B,A} - A_{21} - A_{12} = \frac{\pi}{4}
\end{equation}

We note that the position of the $-2\pi$ flux is $(1.75,0.25)$, which will be used in computing the magnetic flux enclosed by the loop.

\section{1/4 flux HH - CB interpolation}
\label{sec:appC}
\begin{figure}[ht!]
    \centering
    \begin{overpic}[width=1\linewidth]{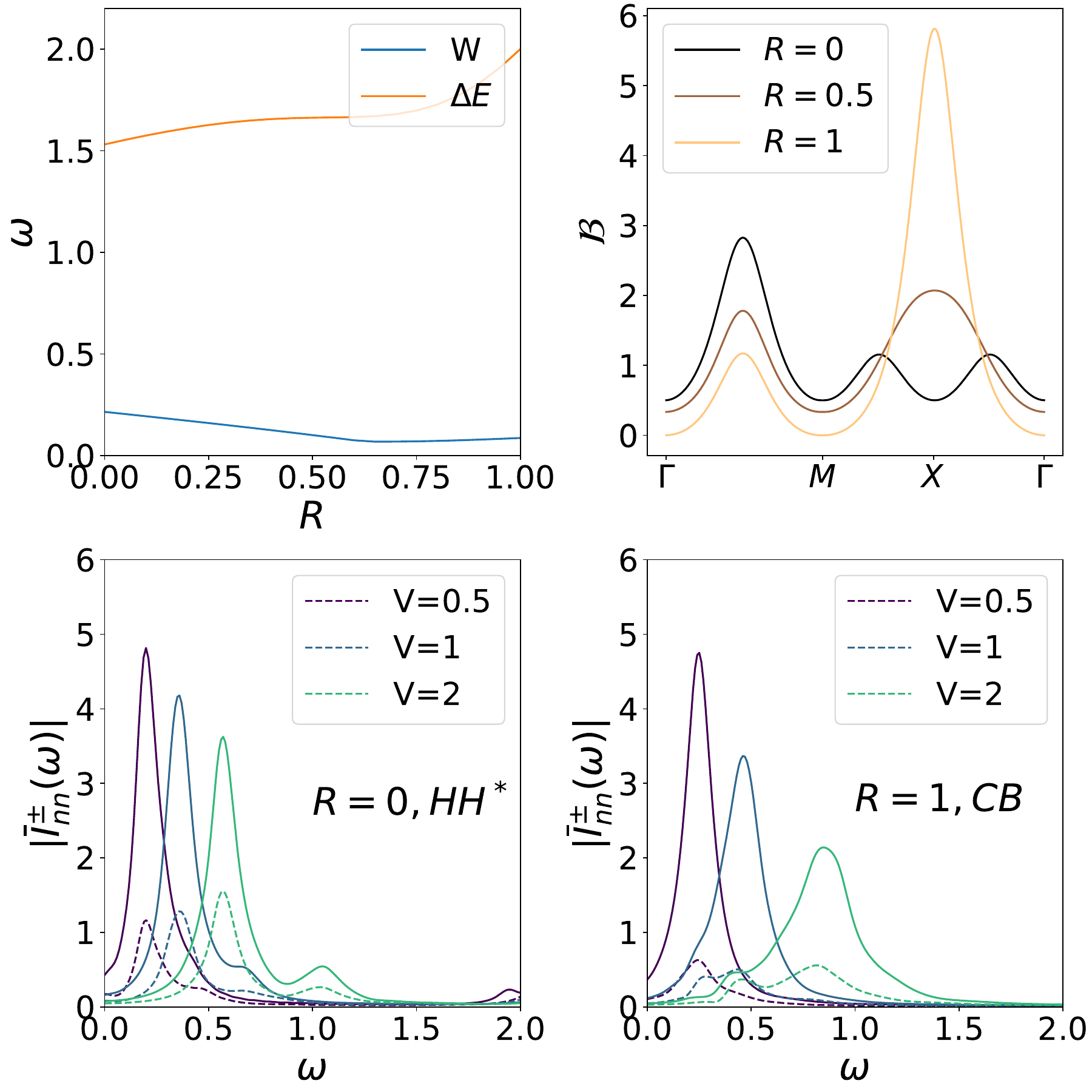}
    \put(10,96){(a)}
    \put(90,96){(b)}
    \put(10,45){(c)}
    \put(61,45){(d)}
    \end{overpic}

    \caption{Interpolation between $\frac{1}{4}$ flux Hofstadter model and Checkerboard Hamiltonian. (a) The band gap $\Delta E$ and band width $W$ of the lowest band as we tune from HH limit to the CB limit. Units of hopping $t=1$. (b) Berry curvature along the high symmetry point at different ratios $R$. (c) (HH limit) and (d)(CB limit) are the graviton spectrums on two limits from MPS simulations. With solid line labels $I_{nn}^-$ and dashed line labels $I_{nn}^+$.}
    \label{fig:single_particle1quater}
\end{figure}

The adiabatic transition could be built for $1/4$ flux HH model. In Fig.~\ref{fig:single_particle1quater}(a), we plot the band gap and the band width of the lowest band along the path. The gap remains intact and flat throughout the path. Signaling the adiabatic relation between two limits. Therefore, the connection between LL and FCI physics could also be connected through $1/4$ flux HH - CB path: We first increase the flux on the HH side, go from LL limit to $1/4$ HH, then we interpolate from $1/4$ HH to CB, this path is also adiabatic. 

The interpolation starting from $1/4$ HH to CB is different from that of $1/8$ interpolation. In $1/8$ case, we start with a band with uniform quantum geometry and good flatness, then tune it into a heterogeneous band. Therefore, the graviton in the former has a longer lifetime and a more prominent peak than the latter. While in $1/4$ interpolation, compared with the CB limit, the HH side has a relatively flatter band while the CB limit has relatively smoother quantum geometry(see Fig.~\ref{fig:single_particle1quater}(b)).

Therefore, a priori, we lack knowledge about which limits will host the longer-lived graviton mode. By MPS simulations, we obtained the graviton spectrum at two limits with different interaction strengths. As the interaction increases, the gravitons on both limits become broader.  The energy band in CB limit exhibits more flatness and a larger band gap compared with the HH side; however, with a broader peak. Therefore, we speculate that the intra-band scattering of the graviton mode is strong due to the non-uniform quantum geometry, while the energy dispersion is 'suppressed' by the interaction.   

\section{Details on MPS numerical results}
\label{sec:appD}
\begin{figure}[th]
    \centering
    \includegraphics[width=\linewidth]{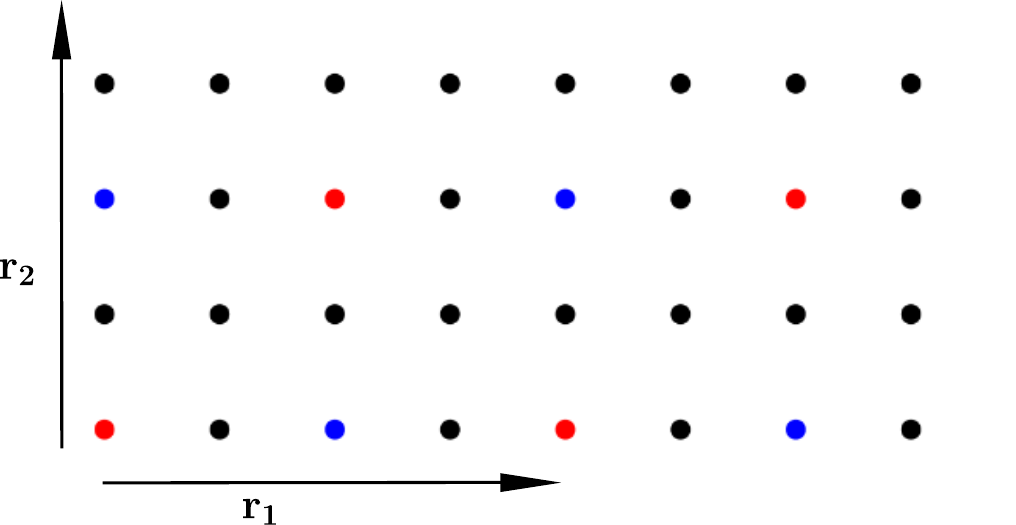}
    \caption{Schematic figure of the cylinder geometry we use in DMRG simulations for HH$_{\frac{1}{8}}^*$. The black and blue dots label the $A$ and $B$ sub-lattices. We impose periodic boundary conditions along $r_2$ direction and open boundary conditions along $r_1$ direction.}
    \label{fig:SI_cylindergeo}
\end{figure}

In this section, we describe the configuration of the finite cylinder geometry of the CB-HH Hamiltonian under investigation and the details of DMRG and TDVP simulations.

Fig.~\ref{fig:SI_cylindergeo} displays the schematic of $L_x\times L_y = 8 \times 4$ cylinders for demonstration. We use up to $L_x\times L_y = 48 \times 12$ in studying the 1/8 flux HH Hamiltonian. The black and blue dots label the $A$ and $B$ sub-lattices, respectively. We impose periodic boundary conditions along $\mathbf{r_2}$ direction and open boundary conditions along $\mathbf{r_1}$ direction. 

Charge $U(1)$ symmetry is implemented in both DMRG and TDVP calculations based on the TensorKit package~\cite{Tensorkit_web}. We keep the bond dimension up to $m=2700$ states in the DMRG simulation to ensure the maximum truncation error below $10^{-6}$. In TDVP simulation, we keep up to $m=600$ states ensuring maximum truncation error below $6\times10^{-4}$, and time evolve wavefunction upto $N_t = 10000$ steps with $\Delta t = 0.05$ in life time analysis to avoid broadening brought by finite time length, resulting energy resolution $\Delta \omega  = \frac{1}{N_t\Delta t}= 0.002$ upto $\max(\omega) = \frac{1}{\Delta t} = 20$, we set $\hbar = 1$ in time evolution.  

On the Checkerboard Limit, the result is obtained with bond dimension $D = 300$. We also enlarge $D$ to check the convergence of the result.

\subsubsection{Numerical stability scheme for TDVP}
In this section, we describe the numerical stabilization we added to the standard 2-site time evolution. 

This technique is introduced in Ref.~\cite{longSpectra2025}. The idea is to correctly fix the global phase of MPS, thereby enhancing the precision of the energy of excitations obtained from real-time dynamics.

After obtaining the ground state $|\psi_0\rangle$ from DMRG, we denote $f(\hat{n}_{\{i\}})$ any type of charge-neutral excitation that \textit{does not} change the $U(1)$ sector of MPS. $| \psi(t) \rangle = e^{iHt}f(\hat{n}_{\{i\}})|\psi_0\rangle$ is the time evolved MPS evaluated using TDVP. We have
\begin{equation}
    \langle \psi_0 | \psi(t) \rangle = e^{-iE_0t} \langle \psi_0 |f(\hat{n}_{\{i\}})| \psi_0 \rangle
\end{equation}
$E_0$ is the ground state energy. The right-hand side of the equation can be determined at $t = 0$, then after every TDVP sweep finishes, we rescale the canonical center of MPS by matching the above relation. We note that such a fixed scheme not only fix the norm of the wave function but also the global phase, which is crucial for the determination of the energy gap.

\subsubsection{Bond dimension}
\begin{figure}[th]
    \centering
    \includegraphics[width=\linewidth]{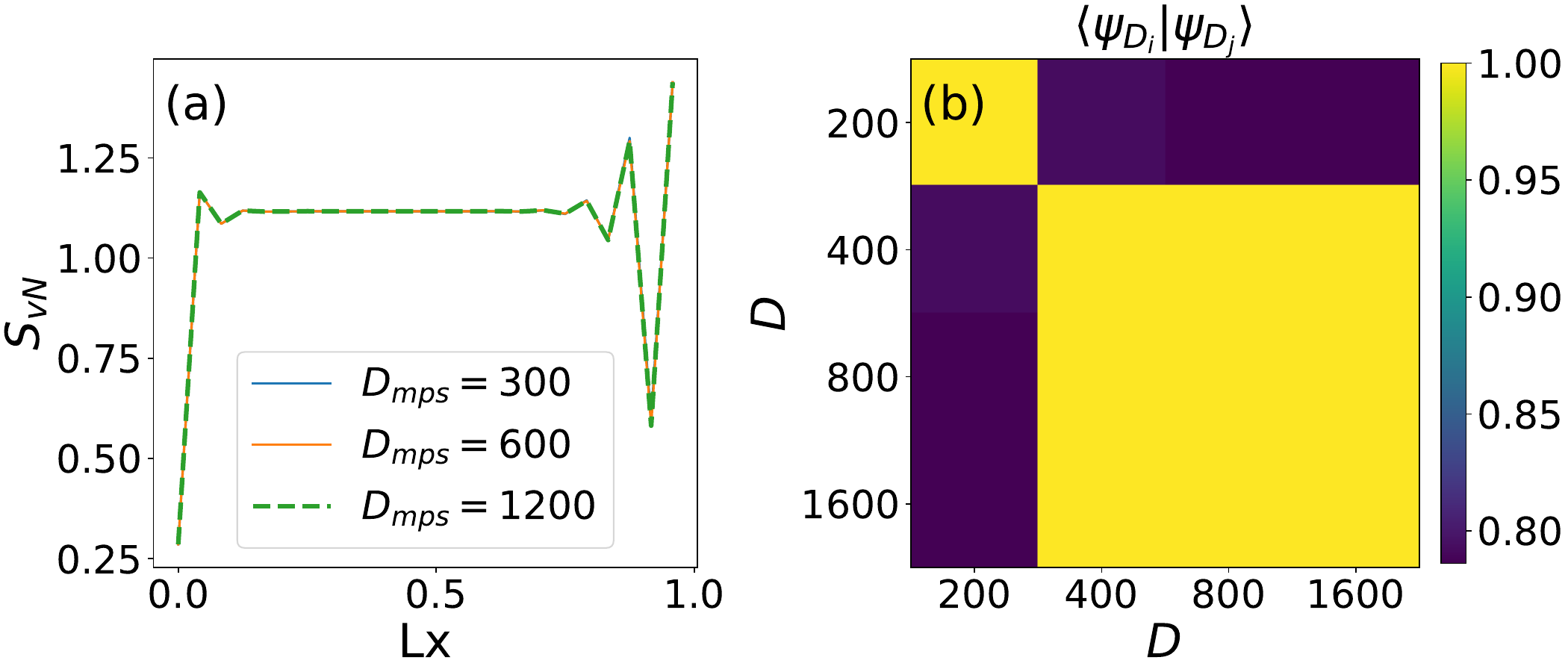}
    \caption{Ground state convergence of DMRG calculation. (a) Entanglement entropy between two partitions of the cylinder: a plateau develops at the bulk of the cylinder, remaining stable against increasing $D$, indicating a gapped phase and convergence of the DMRG simulations. (b) overlap of MPS obtained at different $D$, quantifying the fidelity of the states as we enlarge $D$. The nearly $1$ for the class of MPS with $D>300$ validates our TDVP simulation on MPS with moderate $D$.}
    \label{fig:SI_bond_DMRG}
\end{figure}
In this section, we discuss the finite bond dimension scaling of both DMRG and TDVP calculations. In Fig.~\ref{fig:SI_bond_DMRG}(a), we show the entanglement entropy as $D$ increases. A clear plateau develops at the bulk of the cylinder, i.e., a clear signature of a gapped phase, which remains stable at different bond dimensions. This indicates the convergence of our DMRG result. 

For TDVP calculation. Although the complexity scales as $D^3$, which remains the same as DMRG. The time cost could be much longer due to the large evolution time required desired energy resolution. Therefore, instead of using the MPS with the largest $D$ obtained in DMRG, we use an MPS with a moderate $D$ to leverage the speed and accuracy. In practice, we found that typically the bond dimension $D$ saturates after a threshold $D_0$ for these topological phases, indicating that MPS with moderate $D$ could not only approximate the true ground state in great precision, but also for excitation properties. We estimate $D_0$ basing on the fidelity of MPS as $D$ enlarges.

\begin{figure}[th]
    \centering
    \includegraphics[width=\linewidth]{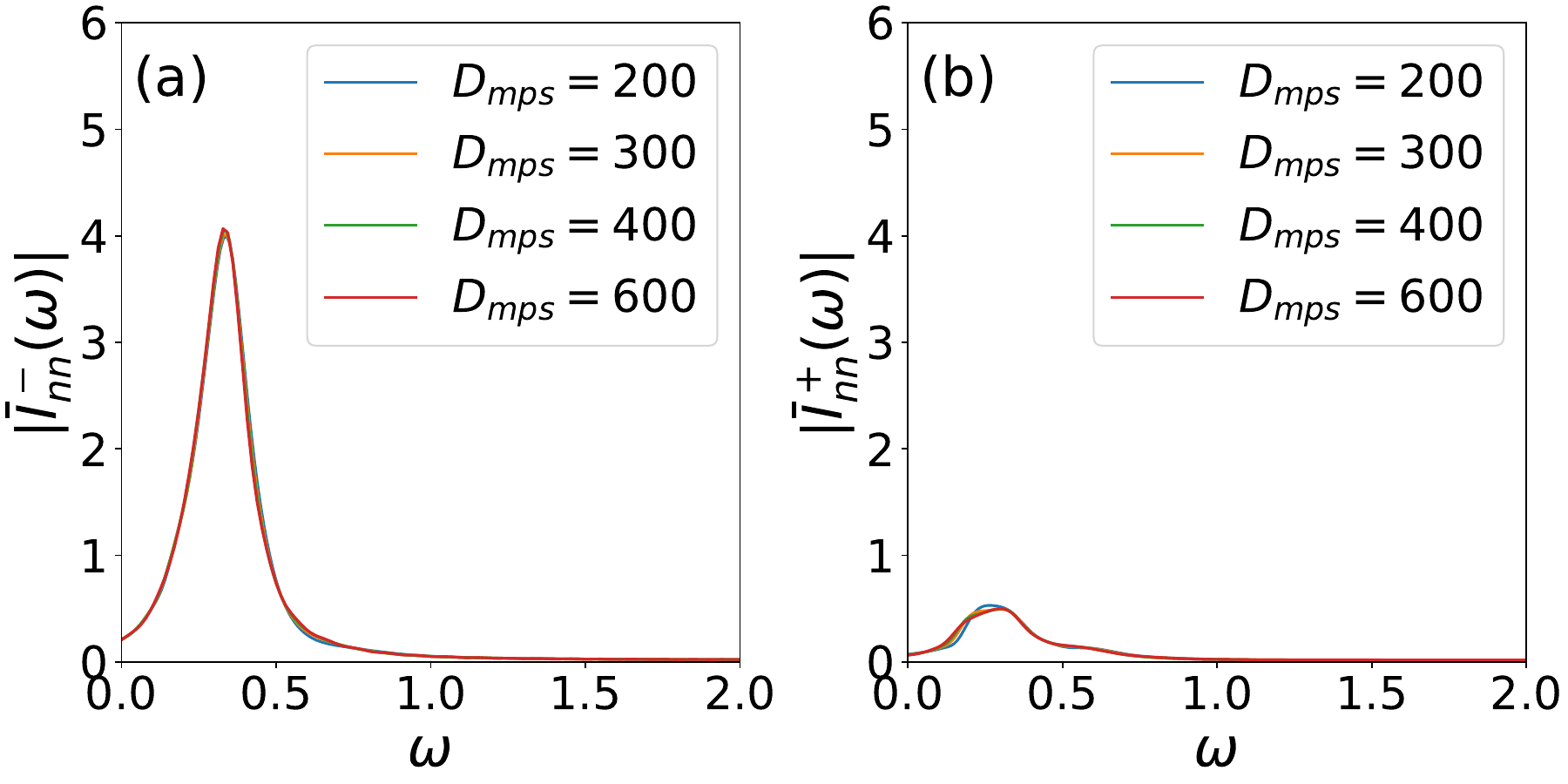}
    \caption{Graviton spectrum calculated at different bond dimensions for (a): $S = +2$ sector and (b): $S = -2$ sector. The result is computed at the CB limit. With NN interaction between two sublattices $V_{AB} = 0.71$. The system size is $N = 2N_x N_y$ with $N_x = 24, N_y = 3$. The chiral graviton mode is prominent and stable with increasing Bond dimension.Here we use $\eta = 0.05, T = 100$.}
    \label{fig:SI_bond_TDVP}
\end{figure}

As shown in Fig.~\ref{fig:SI_bond_DMRG}(b), the similarity between MPS obtained in different $D$, quickly converged after $D = 300$. For example, the difference $1-\langle \psi_{D = 300}|\psi_{D = 1200}\rangle$ is smaller than $10^{-4}$. We also examine the graviton spectrum obtained at different bond dimensions to check the convergence. In Fig.~\ref{fig:SI_bond_TDVP}, we display the graviton spectrum at both chiralities obtained at different bond dimensions. The spectrum converges for all the bond dimensions we show. This convergence trend matches our fidelity measurement.

\subsubsection{finite size scaling}
\begin{figure}[ht!]
  \centering
  \includegraphics[width=0.5\textwidth]{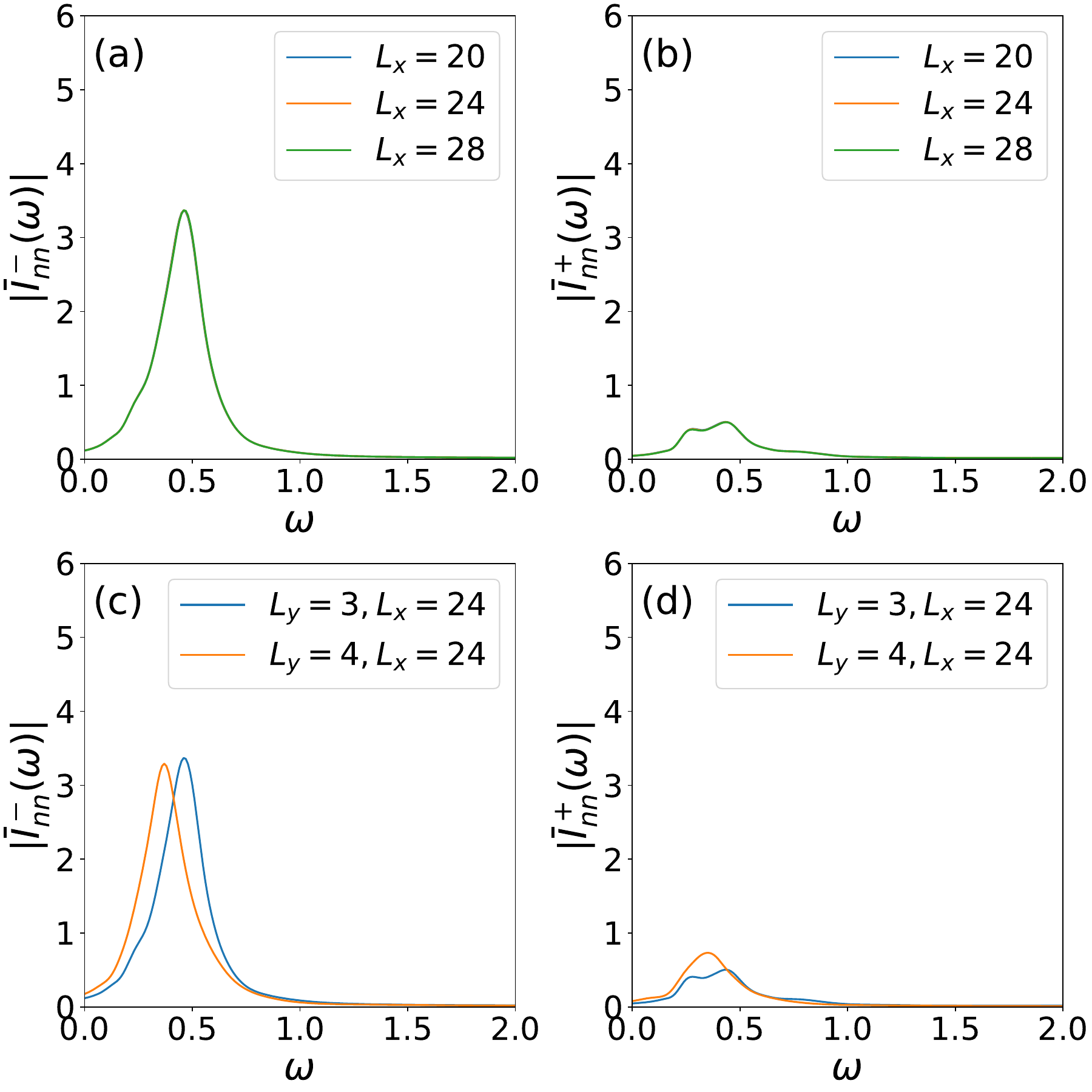}  
  \caption{Finite size scaling of the graviton spectrum with different chiralities on CB limit(R = 0). In (a)(b) we fix $L_y = 3$ and vary $L_x$ while in (c)(d) $L_x = 24$ and $L_y$ is tuned. We set $D_{mps} = 300$ in the finite-size scaling.}
  \label{fig:tdvp_finitesize}
\end{figure}
We also perform finite-size scaling for the graviton spectrum. 

As shown in Fig.~\ref{fig:tdvp_finitesize}(a) and (b), the result remains robust as we enlarge the cylinder length. 

We also compute the graviton spectrum at a larger cylinder circumference. The graviton exhibits a larger intrinsic scattering rate compared with $L_y = 3$, in the same trend with the result obtained using pED

\subsubsection{Evolution time}
\begin{figure}[th]
    \centering
    \includegraphics[width=\linewidth]{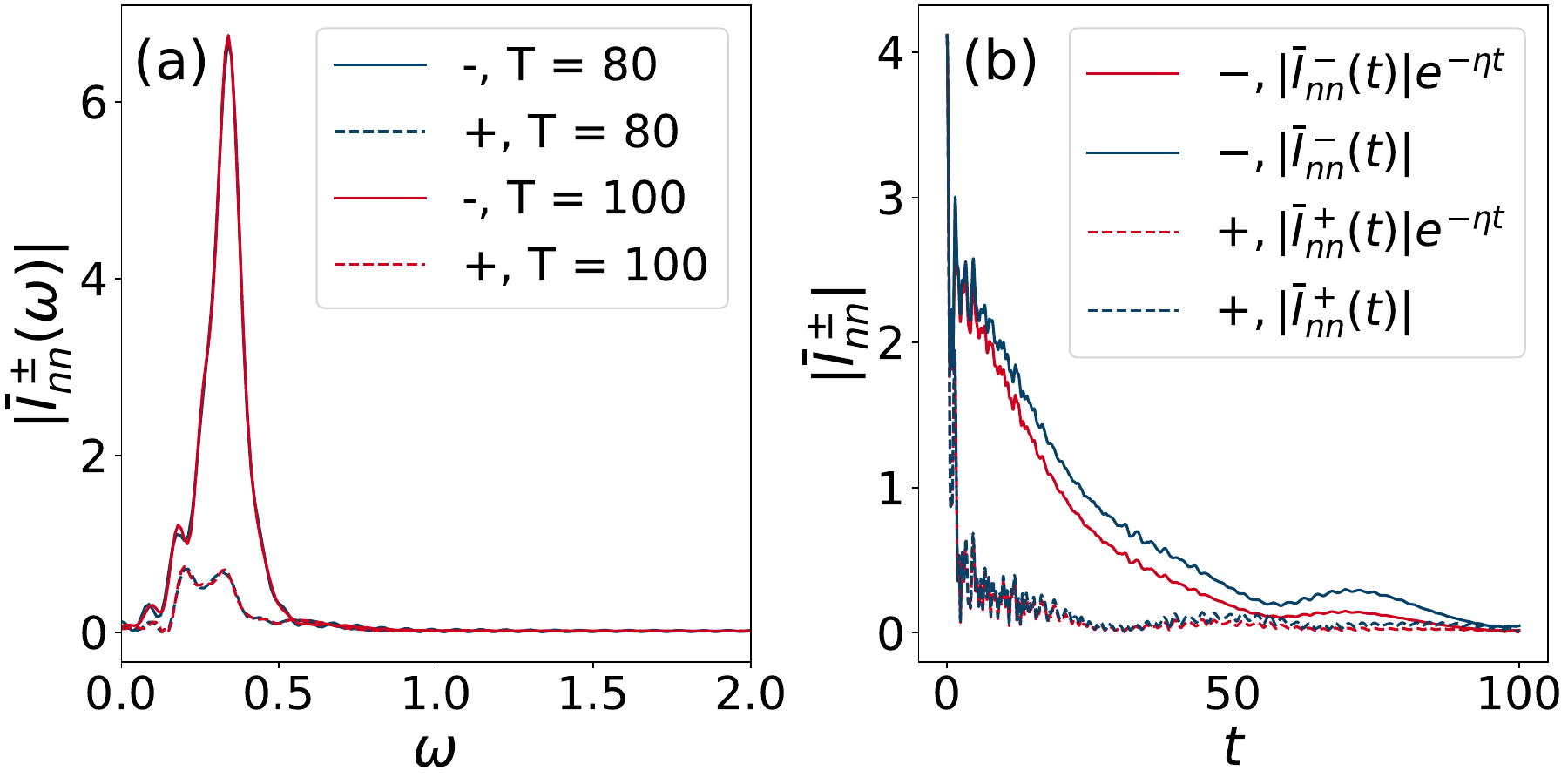}
    \caption{Graviton spectrum calculated at different time evolution lengths. The result is computed at the Checkerboard limit. With NN interaction between different sublattices $V_{AB} = 0.71$. The system size is $N = 2N_x N_y$ with $N_x = 24, N_y = 3$ and $D = 300$. To highlight the convergence of the evolution time, we use $\eta = 0.01$ for the data shown in this figure.  (a) displays the graviton mode computed at different time evolution lengths and is stable upon varying the evolution time. (b) shows the real-time profile, the Lorenzian broadening factor only slightly suppresses the long-time dynamics, whose decay, however, is due to the scattering of the graviton mode with edges on a finite cylinder simulation.}
    \label{fig:convergence_evolution_time}
\end{figure}

In this section, we check the convergence of the graviton spectrum against evolution time. The effect of a finite time window could be viewed as an extra step function $\theta(t-T)$ acting on the time-ordered correlator. Then in the frequency domain, $S(\omega) = F[\theta(t-T)G(t)] = F[\theta(t-T)]*F[G(t)]$, which effectively modifies the \textit{real} spectrum by its \textit{convolution} with a broadened delta function. 

In Fig. \ref{fig:convergence_evolution_time}(a), we compare the spectra obtained using different time-evolution lengths. Our results remain unchanged upon increasing the evolution time. In Fig.~\ref{fig:convergence_evolution_time}(b), we show the real-time profile; the long-time contribution of graviton dynamics decays even with a very small $\eta = 0.01$. Therefore, further evolution time only slightly changes the spectrum. We attribute the decay of the graviton to the scattering with edges, which contributes to a finite broadening of graviton peaks.

\subsubsection{Summing over sublattice}
\begin{figure}[h!]
    \centering
    \includegraphics[width=\linewidth]{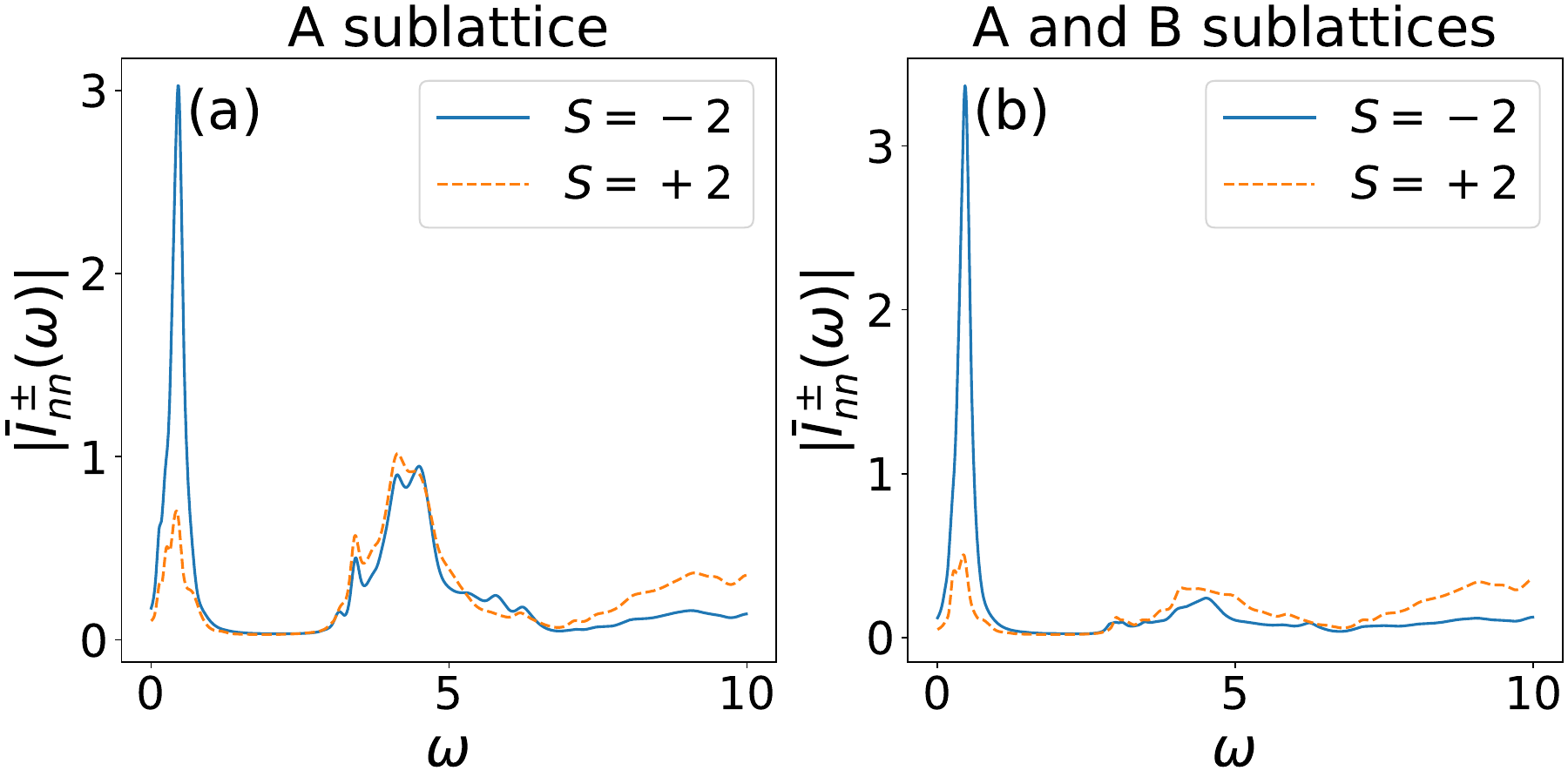}
    \caption{Graviton spectrum for $O^\pm_{nn}$ summing over (a): single sublattice or (b): summing both A and B sublattice. The result is computed at the Checkerboard limit. With NN interaction between different sublattices $V_{AB} = 1$. The system size is $N = 2N_x N_y$ with $N_x = 24, N_y = 3$ and $D = 300$. The low-energy dynamics are unaffected in either convention, while the interband modes with $w\sim 5$ are strongly suppressed by summing over all sites.}
    \label{fig:ab_sublattice_compare}
\end{figure}
In previous work,  a simplified version of the phenomenological density graviton operator was used in detecting the chiral graviton mode in a honeycomb lattice of bosonic $1/2$ FCI~\cite{longSpectra2025}.

In the simplified version, the summation is restricted to only include the next nearest neighbors, whose number is 6 and is sufficient to resolve the chirality, and the summation is restricted to a single sublattice. In this work, we propose the generalized density graviton operator, which is equivalent to the lattice stress tensor of the HH model in the continuum limit. In the generalized version, the summation contains all the neighbors with distances smaller than $r_c$ and \textit{all sublattices}. In this section, we compare the results of summing over different sublattices.

We compare the result on the CB limit. In Fig.~\ref{fig:ab_sublattice_compare}, we compare the spectrum obtained via summing on A sublattice (a) and all sites (b). The chiralities are both resolved. However, the contrast between $+2$ and $-2$ chiralities becomes more significant when summing over all sites. Moreover, there are contributions from both $\pm2$ sectors with an energy scale of the order of the band gap when summing over $A$ sublattices. This mode is greatly suppressed when we sum over all the sites. We note that the chiral graviton spectrums (modes with $\omega < \Delta$, $\Delta \approx 2$ is the band gap) obtained via these treatments are qualitatively in agreement. We attribute this mode to the two-roton-bound state that lives in the higher band.

\section{Details on ED numerical results}\label{sec:app_edtricks}
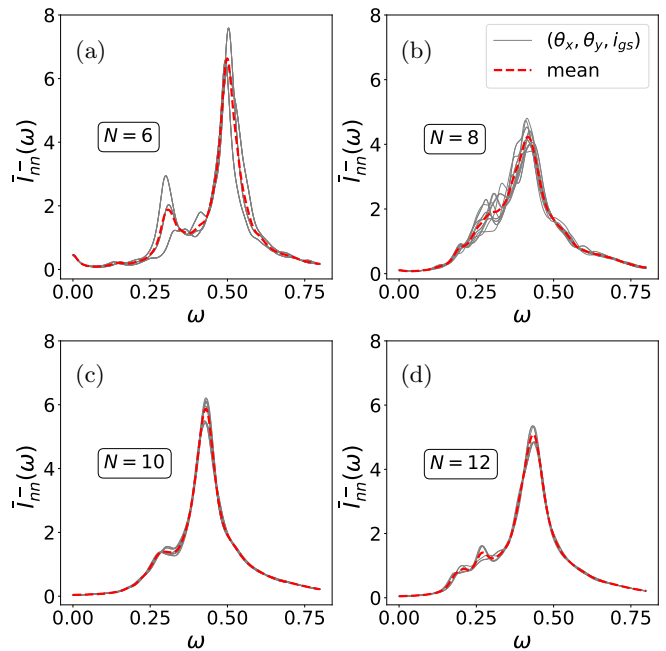
\begin{figure}[ht!]
    \centering
    \begin{overpic}
[width=\linewidth]{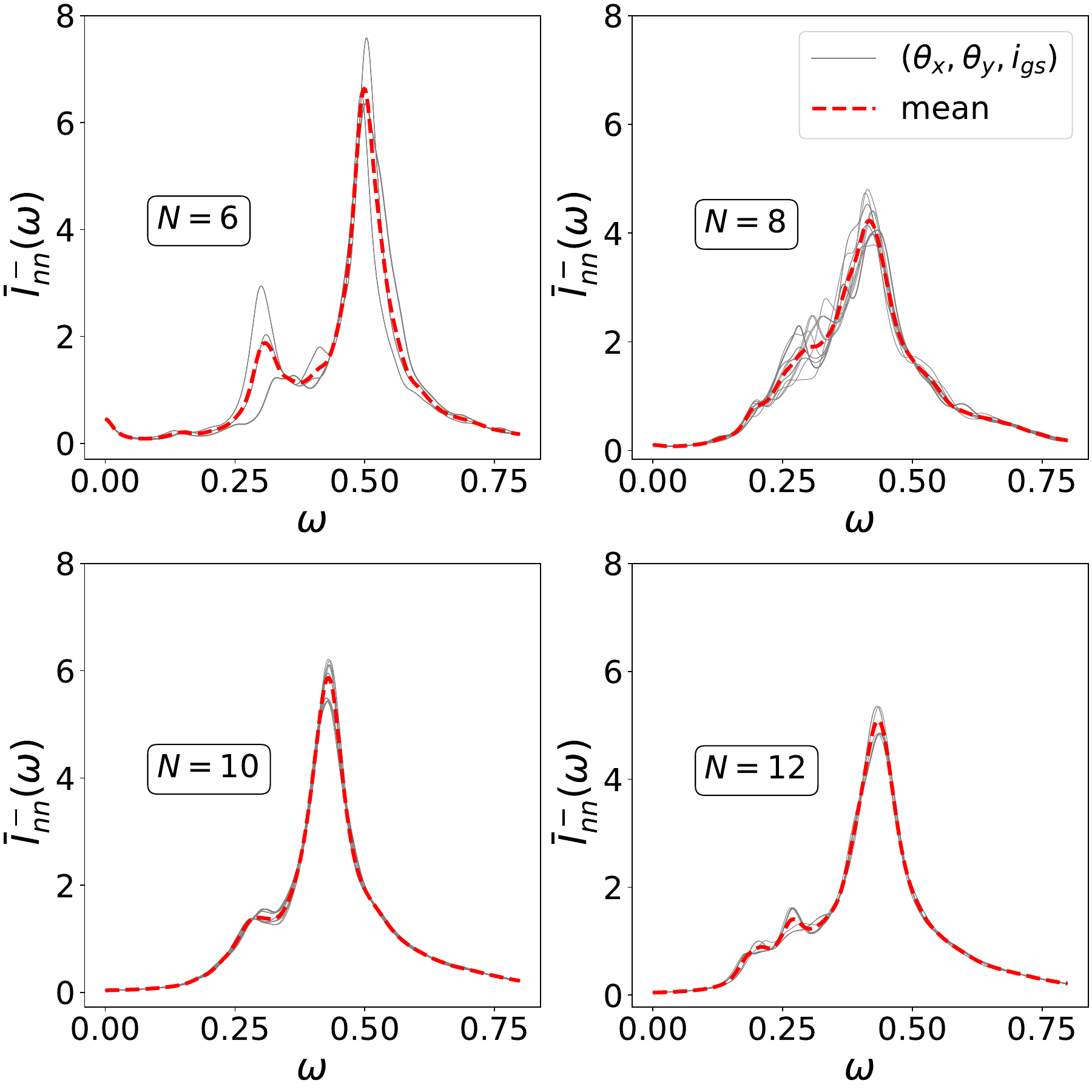}
\put(10,92){(a)}
\put(10,42){(c)}
\put(60,92){(b)}
\put(60,42){(d)}
    \end{overpic}
    \caption{pED graviton spectra at different system sizes $N=6,8,10,12$ for different twisted boundaries and ground states $(\theta_x,\theta_y,i_{gs})$ (gray lines). The average spectra  is shown with a thicker red dashed line. Parameters are $R=1$, $V=2$ and $\eta=0.02$.}
    \label{fig:tbc_gs_average}
\end{figure}
In this appendix, we expand on the role of averaging over different ground states and twisted boundary conditions $\theta=(\theta_x ,\theta_y)$ on the torus geometries studied in this work. In particular, the spectral functions we show are averaged as:
\begin{align}\label{eq:average_twisted}
    I_{O}(\omega)=\frac{1}{4\pi^2}&\int d\theta_xd\theta_y\frac{1}{3}\sum_{m=0}^2 \;\;\sum_n|\bra{n}_\theta\, O \ket{m}_\theta|^2\cdot\nonumber \\
    &\cdot \delta(\omega-(E_n+E_0(\theta))\;.
\end{align}
The average over different ground states is actually what one would get for a small finite temperature below the magnetoroton gap. We remark that the average over boundary conditions is actually quite important as it allows the finite system to artificially explore the full Brillouin zone. Indeed, twisted boundaries correspond to overall momentum shifts of the single-particle orbitals. While we find this to be irrelevant in the case of low flux HH models (which are close to the continuum), they become quite effective in the CB point.   

In figure \ref{fig:tbc_gs_average} we explicitly show an example of the averaging procedure for the $R=0$ case (CB) at different system sizes. We find that averaging over a small discrete set of twisted boundary conditions is actually enough to converge with respect to the integral in Eq. \eqref{eq:average_twisted}. As such we actually just average over $(\theta_x,\theta_y)=(n_x\pi,n_y\pi)$ with $n_{x/y}=0,1$. These, for all three different ground states, are shown in Fig. \ref{fig:tbc_gs_average} as gray lines.

While the averaging does not quantitatively change the peak position, it helps to make contributions from the two-magnetoroton continuum smoother, eventually allowing for the controlled decay rate extrapolation presented in the main text.
\section{Band truncation effects}\label{app:band_truncation}
\begin{figure}
    \centering
    \begin{overpic}[width=\linewidth]{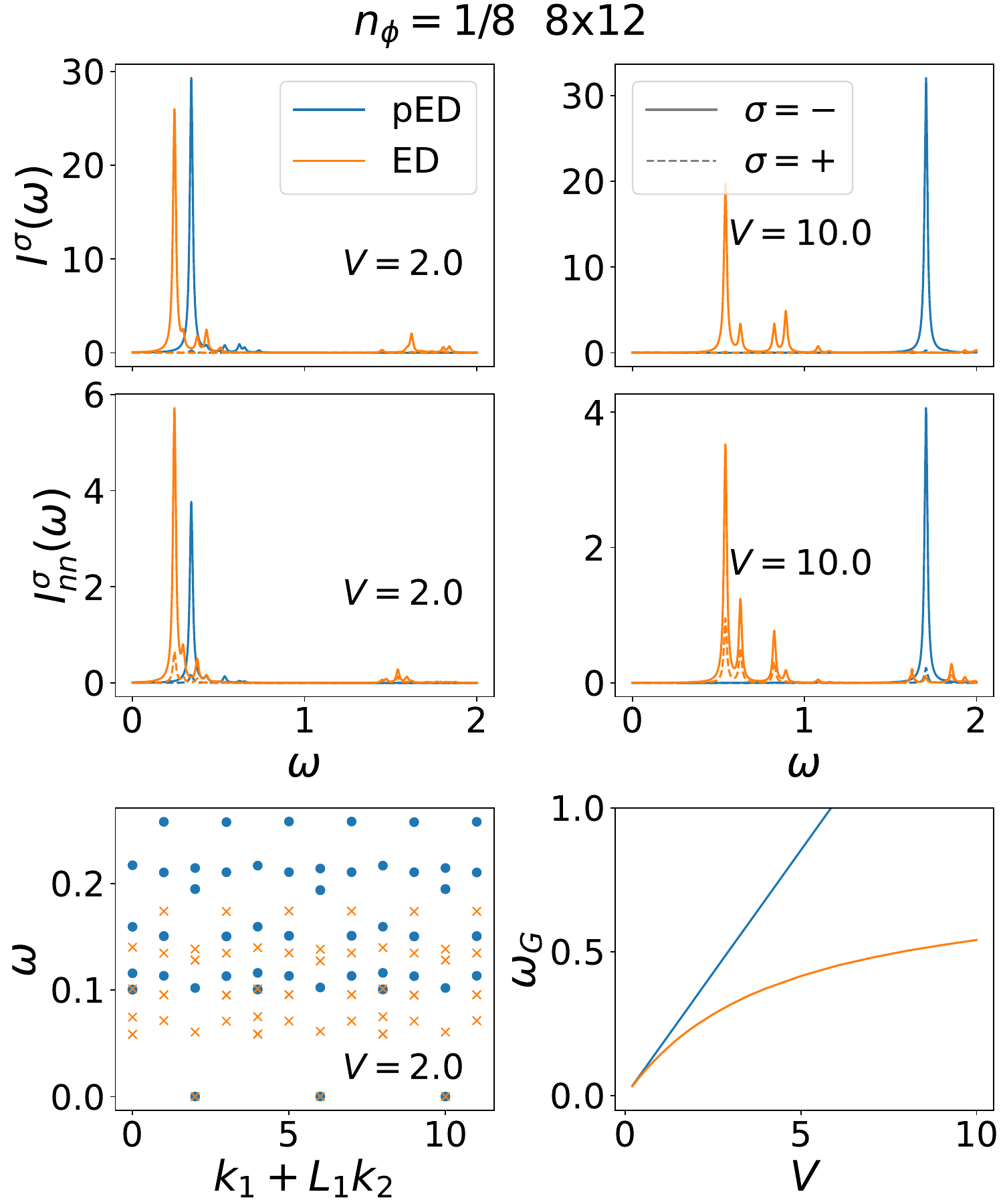}
\put(10,10){(e)}
\put(18,90){(a)}
\put(18,62){(c)}
\put(78,62){(d)}
\put(78,90){(b)}
\put(78,10){(f)}

    \end{overpic}
    \caption{Comparison of full ED (orange) and projected ED (blue) on the HH model at $n_\phi=1/8$. Graviton spectra obtained with $O_s$ (a,b) and $O_{nn}$ at different interaction strengths $V=2$ (a,c) and $V=10$ (b,d). (e) Low-lying neutral spectrum as a function of momentum showing the three degenerate ground states for both ED (blue) and pED (orange). (f) Energy of the graviton peak as a function of $V$. System size $8\times 12 $ ($N=4$)}
    \label{fig:EDvspED}
\end{figure}

In this section we explore the effects of band truncations, comparing directly pED results with full ED. We fix for convenience $r_c=2$ and $n_\phi=1/8$ on the HH model on a system of $N=4$ particles ($8\times12$). The results are shown in Fig. \ref{fig:EDvspED}. Top (a,b) and middle (c,d) panels show result for the graviton spectra with operators $O_s$ and $O_{nn}$ respectively. The high energy $\omega \sim 1.5$ features present in full ED (orange) correspond to interband transitions and are indeed absent in pED (blue). In the bottom right panel (f) we show how the graviton peak energy $\omega_G$ evolves as a function of $V$. In the bottom right panel (e) the full spectrum at an intermediate value $V=2$ shows that the system is in a FQH phase but that pED overestimates the energetics. 
\section{Lifetime extraction}
\label{sec:app_lifetime_extraction}
In this section we provide more details on the extraction of the decay rate $\Gamma_G$. First of all, as discussed in Sec. \ref{sec:HHnum_lifetime}, we use the following guess function to fit the spectra at different $\eta$:
\begin{align}
    I_0(\omega)= a_0+a_1\omega  + \frac{Z}{\pi} \frac{ \Gamma_{tot}}{(\omega-\omega_G)^2+\Gamma_{tot}^2}
\end{align}
The last term in the fitting formula is a broadened Lorentzian peak centered on $\omega_G$ with full width half maxima $2\Gamma_{tot}$. The first two terms instead account for the present of a smooth background. We empirically find that using higher order polynomials does not include the quality of the fit, while keeping at least the first order is important to capture small asymmetries of the peak. Once the total lifetime $\Gamma_{tot}$ is found for different values of $\eta$, we can extrapolate to $\eta \to 0$ with a linear ansatz:
\begin{align}
    \Gamma_{tot}=\Gamma_G+ c\,\eta
\end{align}
If the ansatz for the broadened spectral function is correct, then we expect $c=1$.

\begin{figure}
    \centering
    \includegraphics[width=\linewidth]{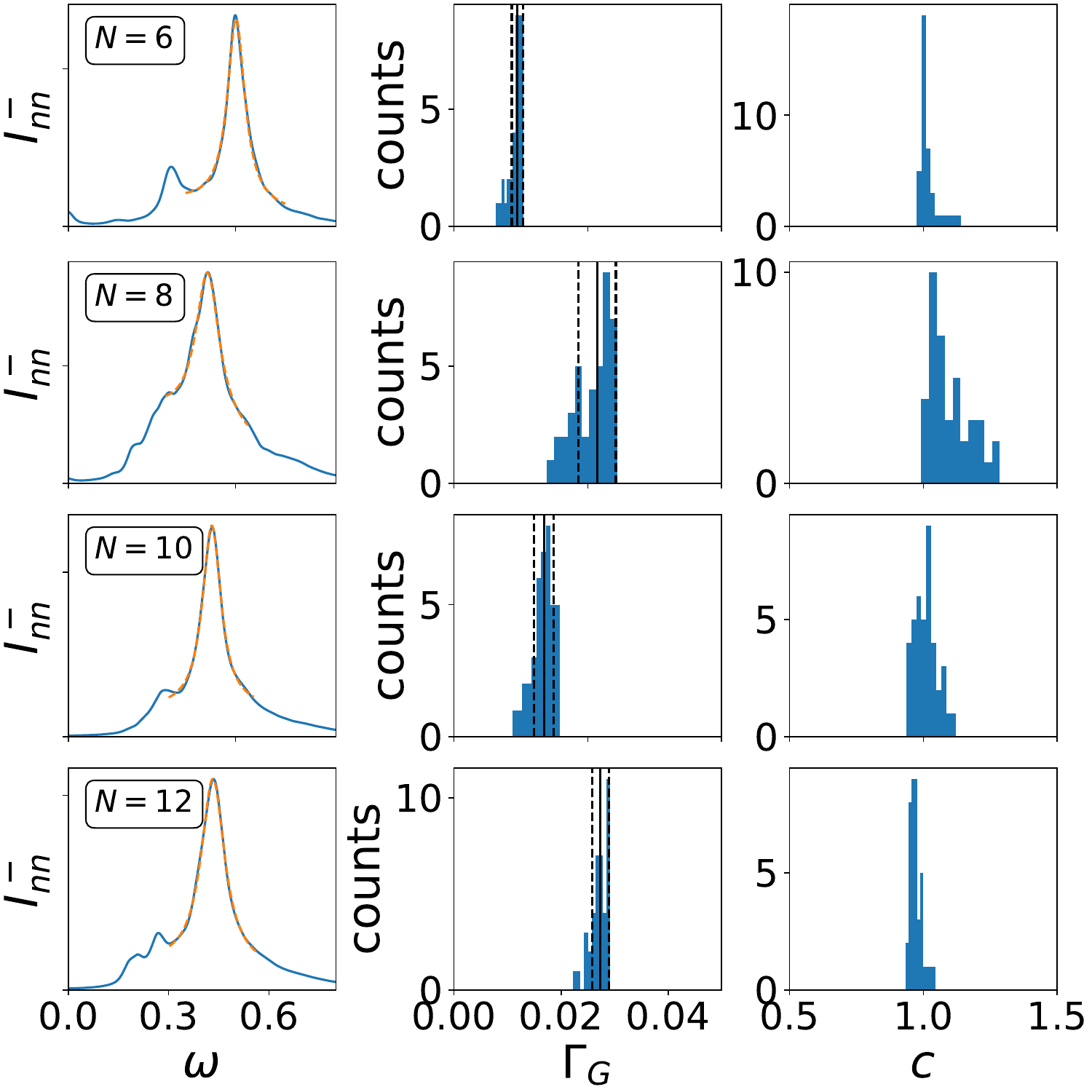}
    \caption{Fit of the spectral peaks for the CB case. The first column contains the graviton spectrum in negative chirality computed using pED for different numbers of electrons. The yellow lines are the fitting result with energy window $\delta = 0.2$. The second column is the fitted value of $\Gamma_{tot} - \eta$ from many random choices of $\eta$ and $\delta$. The third column shows the quasi-particle weight fitted from random collections of $\eta$ and $\delta$.}
    \label{fig:fit_detail}
\end{figure}

We have two important hyperparameters in the extraction of $\Gamma_G$, i) the fitting window for the spectral function $[\omega_{min},\omega_{max}]$ and ii) the window  $[\eta_{min},\eta_{max}]$ used for the extrapolation. Both of these quantities should be proportional to the graviton energy $\omega_G$ which sets the scale of the problem. We choose to take the peak position $\omega^*$ as a proxy for $\omega_G$ and vary the windows as follow:
\begin{align}
    &\omega \in[(1-\delta)\omega^* ,(1-\delta)\omega^*]\qquad \textrm{for}\qquad\delta=0.15,0.2,0.25,0.3\\
    &\eta \in [\eta_1,\eta_2]\qquad \textrm{for}\qquad\eta_{1/2}\in[0.02\omega^*,0.2\omega^*]
\end{align}
We sample different a total of 4 combinations for the frequency window and 10 combinations of $[\eta_1,\eta_2]$ and build a distribution that hence accounts for systematic errors in the procedure. The result of this is shown in Fig.\ref{fig:fit_detail}.

In the first column, we show the real spectrum and the fitting result for a specific sample. The center and right column shows instead the distribution of results for $\Gamma_G$ and $c$. The latter is close to one as expected. The value and error bars of $\Gamma_G$ used in the main text are shown as black vertical full and dashed lines.

\section{Tuning the graviton operator}\label{sec:app_fG}

\begin{figure}[h!]
    \centering    
           \begin{overpic}
    [width=\linewidth]{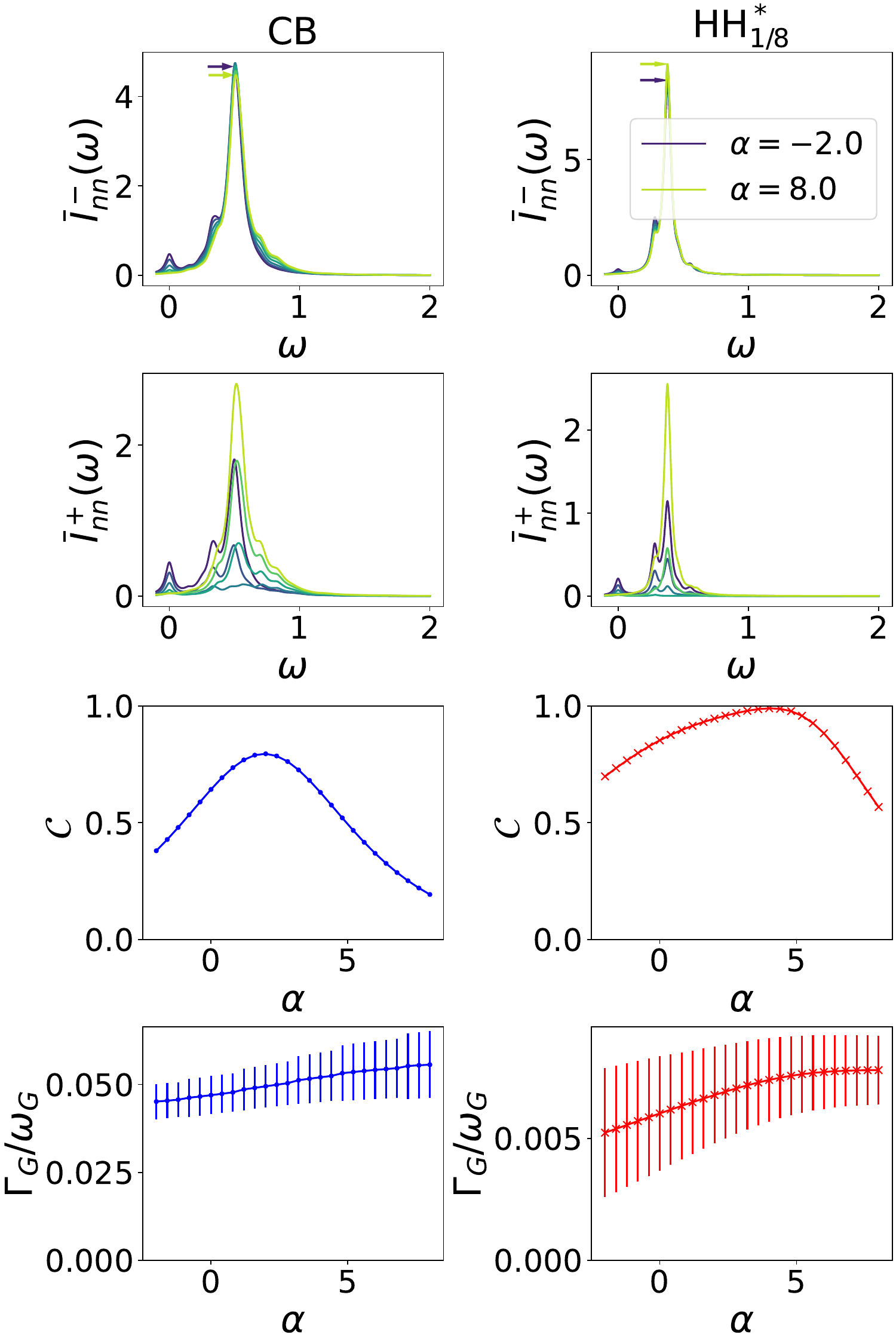}
    \put(11,93){(a)}
    \put(44,93){(b)}
    \put(11,69){(c)}
    \put(44,69){(d)}
    \put(11,44){(e)}
    \put(44,44){(f)}
     \put(11,21){(g)}
    \put(44,21){(h)}
    \end{overpic}
    \caption{(a) and (b), the normalized spectra in the $\sigma=-$ chiral channel from $\alpha=-2$ (dark color) to $\alpha=8$ (light color) for the Harper-Hofstadter point $R=1$ and the checkerboard point $R=0$. (c) and (d), same as (a) and (b) but for  $\sigma=+$ chiral channel. (e) and (f), the ratio between the total weight in the two chiralities $C = |\mathcal{N}^- - \mathcal{N}^+|/( \mathcal{N}^-+\mathcal{N}^+)$. And the scattering rate for $R=0$  (g) and $R=1$  (h) side}
    \label{fig:fG_tuning}
\end{figure}
In this appendix, we here explore different choices for the density-density graviton lattice operator encoded in the function $f_G(r)$ for both the Harper-Hofstadter point $R=1$ and the checkerboard point $R=0$. For the sake of simplicity, we stick with a finite range function parametrized by a single number $\alpha$:
\begin{align}
    f_G(r\leq2\sqrt{2})=\frac{1}{r^\alpha}  \qquad\mathrm{and}\qquad f_G(r>2\sqrt{2})=0
\end{align}
Note that for the $R=0$ case, the $r=2\sqrt{2}$ distance corresponds to the next nearest neighbor, while for the $R=1$ it is the $5^{th}$ nearest neighbor. 

In Fig. \ref{fig:fG_tuning} we show a sample of band-pED results obtained for a small system of $N=6$ particles. In particular, we explore a wide range of values of $\alpha$, from negative to positive. Note $\alpha=1$ has been used elsewhere. In the top panels (a,b) we show the normalized spectra in the $\sigma=-$ chiral channel from $\alpha=-2$ (dark color) to $\alpha=8$ (light color). Changes to the spectral distribution of the $\sigma=-$ channel are minor, particularly around the graviton peak. Instead, for the $\sigma=+$ channel, shown in panels (c,d), the spectra change substantially. As these spectra are normalized with respect to the $\sigma=-$ total weight in the lowest band, we also show in panels (e,f) the chirality of the spectra, i.e. the ratio between the total weight in the two chiralities $C = |\mathcal{N}^- - \mathcal{N}^+|/( \mathcal{N}^-+\mathcal{N}^+)$. Because of their definition, the two operators $O^-_{nn}$ and $O^+_{nn}$ become identical at $\alpha=+\infty$ and $\alpha=-\infty$ giving chirality 0 by definition. Away from these regimes, the chirality is quite strong for a broad regime of $\alpha$. This is overall much larger in the Harper Hofstadter case $R=1$ than the checkerboard case $R=0$, as lattice corrections are smaller. In panels (g,h), we also show the operator dependence of the scattering rate $\Gamma_G$ relative to the graviton frequency $\omega_G$. The lifetime in the Hofstadter case for all $\alpha$ is longer than the Checkerboard case and the operator dependence is minor. 

\begin{figure}[ht!]
    \centering
     \begin{overpic}
    [width=\linewidth]{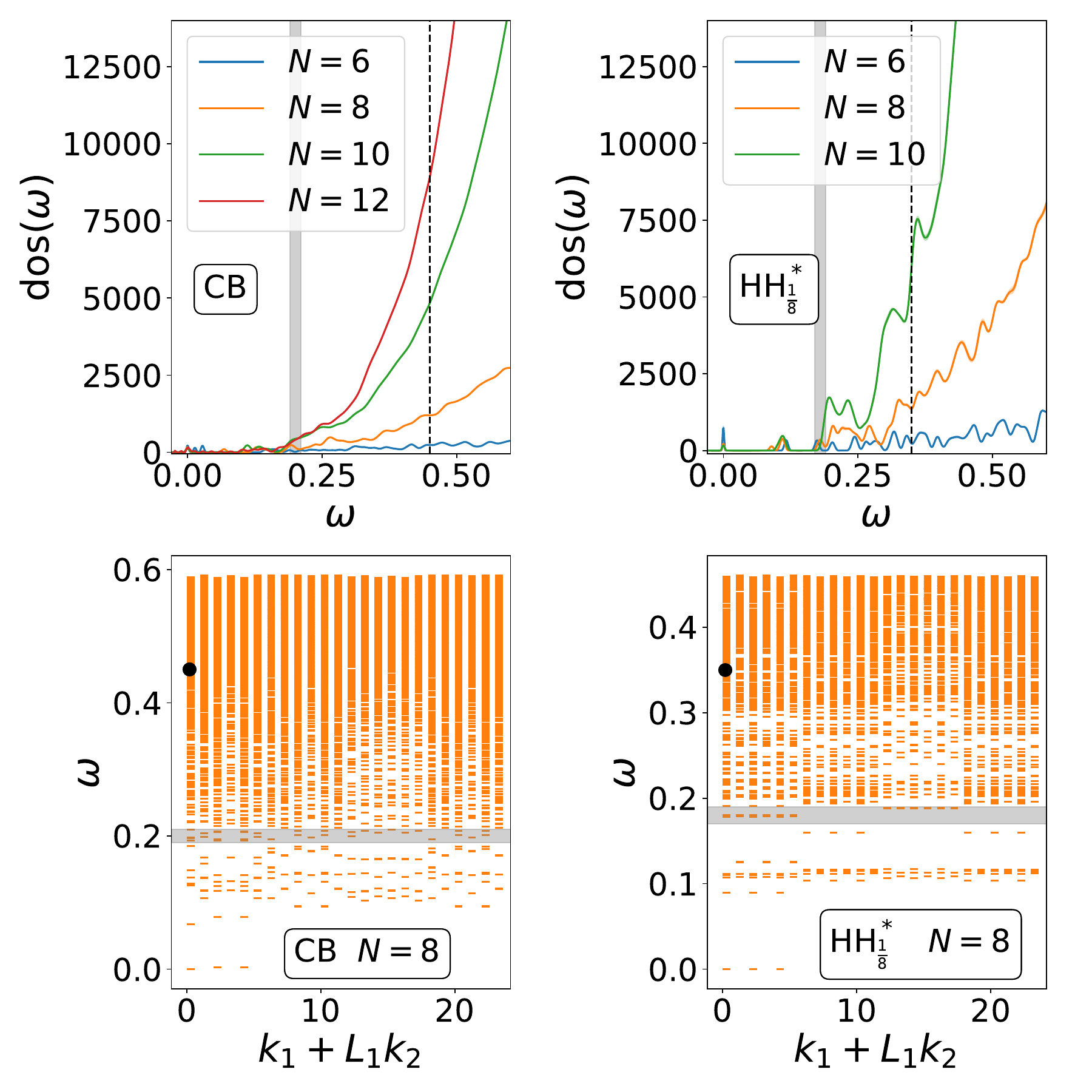}
    \put(41,95){(a)}
    \put(90,95){(b)}
    \put(41,45){(c)}
    \put(90,45){(d)}
    \end{overpic}
    \caption{(a) and (b). Density of states at the CB and HH$_{1/8}^*$ points for different system sizes obtained with pED in the ground state momentum sector. (c) and (d). Low-energy momentum resolved spectrum obtained with pED at the CB and HH$_{1/8}^*$ for $N=8$ ($6\times4$ unit cells). The graviton-mode energy is indicated by a dashed black line in the top panels and by a black dot in the bottom panels. The shaded gray areas represent the energy where the two magnetoroton continua starts.  }
    \label{fig:dos}
\end{figure}

\section{Density of states in the charge neutral sector}
\label{sec:app_dos}
In this section, we provide numerical results obtained with band pED (Fig. \ref{fig:dos}) on the density of states (DOS) at low energies in the charge neutral sector. Results in panels (a) and (b) are obtained with a Chebyshev expansion method~\cite{Weisse_rmp2008_kpm}, while those in panels (c) and (d) show 400 states obtained with Lanczos.

As shown in Fig.~\ref{fig:dos} (a) and (b), we find that as the system size increases, the density of states near the graviton energy is greatly enhanced, which is a natural trend when approaching the thermodynamic limit. Interestingly, the graviton energy (black dashed line in (a) and (b) and black dot in (c) and (d)) remains well inside the continuum excitations throughout the system sizes, whereas its lifetime remains long (see Fig. \ref{fig:lifetime_analysis}). This trend is different from the chiral gravitons pointed out in Ref.\cite{shenMagnetorotons2024}, where the energy of geometric excitations gets close to the lower bound of the continuum excitations region as the system size increases, and also different from Ref.\cite{wangDynamics2025}, where the graviton keeps lying deep inside the continuum excitations, which scatter with the graviton and rendering a short lifetime.
Although the enhanced density of states might give rise to possible decay channels for the chiral graviton, our observation that the chiral gravitons remain long-lived raised the intriguing question of what protects them from decay. We leave this question - as well as the quest for possible Hamiltonian perturbations that could lower the graviton energy below the continuum - for future study.

\section{Derivation for the continuum limit of density graviton operator using holomorphic properties of Landau Level}
\label{App: derivation_complex}

We start with the unprojected momentum space $O_{nn}$

\begin{align}
    O^{\pm}_{nn} & \to\sum_{\boldsymbol{q}}\rho(\boldsymbol{q})\rho(-\boldsymbol{q})\int d^2\delta \;\;e^{i\boldsymbol{q}\cdot\boldsymbol{\delta}} e^{\pm 2i \arg{[\delta]}}f_G(|\delta|). 
\end{align} 

We now express the integral over $\boldsymbol{\delta}$ and the momentum $\boldsymbol{q}$ in complex coordinates $\delta=\delta_x+i\delta_y$ and $\delta^*=\delta_x-i\delta_y$ such that we can express the phase factors as
\begin{align}
  e^{2i\mathrm{arg}[\delta]}e^{i\boldsymbol{q}\cdot\boldsymbol{\delta}}=-\frac{1}{4}\partial_{q^*}^2 \left[e^{i(q^* \delta+q\delta^*)/2}\frac{1}{|\delta|^2} \right],
\end{align}
where $\partial_{q*}=\frac{1}{2}(\partial_{q_x}+i\partial_{q_y})$ and for the opposite chirality the derivative $\partial_{q^*}$ is replaced by $\partial_q$. Using the fact that the remaining integrand in $\delta$ depends only on $|\delta|$, we conclude that its Fourier transform will also depend only on $|q|$
\begin{align}
    F_G(|q|)=-\int d\delta d\delta^* \frac{f_G(|\delta|)}{4|\delta|^2}e^{i(q^*\delta+q\delta^*)/2}.
\end{align}
Now, in order to recover the continuum stress tensor structure \cite{yang2016acoustic,liou2019chiral} of a local interaction potential, we define the following function:
\begin{align}
    V_G(|q|)=\frac{F''(|q|)}{4|q|^2} -\frac{F'(|q|)}{4|q|^3},
\end{align}
for which this identity holds
\begin{align}
\partial_{q^*}^2F_G(|q|)=q^2 V_G(|q|).
\end{align}

Finally, we arrive at an expression for the density-density operator in the continuum limit
\begin{align}
    O_{nn}^-\to \sum_q (q_x-iq_y)^2 V_G( |q| ) e^{-\frac{1}{2}(ql_B)^2} \bar{\rho}_q\bar{\rho}_{-q},
\end{align}
which coincide with the continuum stress tensor operators \cite{liou2019chiral,nguyen2022multiple} or an interaction potential whose Fourier transform is $V_G(|q|)$
up to irrelevant constant prefactors. Now we remark that if the original lattice function $f_G(|\delta|)$ is short-ranged, the above expression correctly captures the stress tensor operator of contact interactions in the continuum limit, e.g Eq.~\eqref{eq:op_s}.

\section{Ground state adiabatic connection}\label{app:adiabatic_gs}
\begin{figure}[h!]
    \centering
    \includegraphics[width=0.8\linewidth]{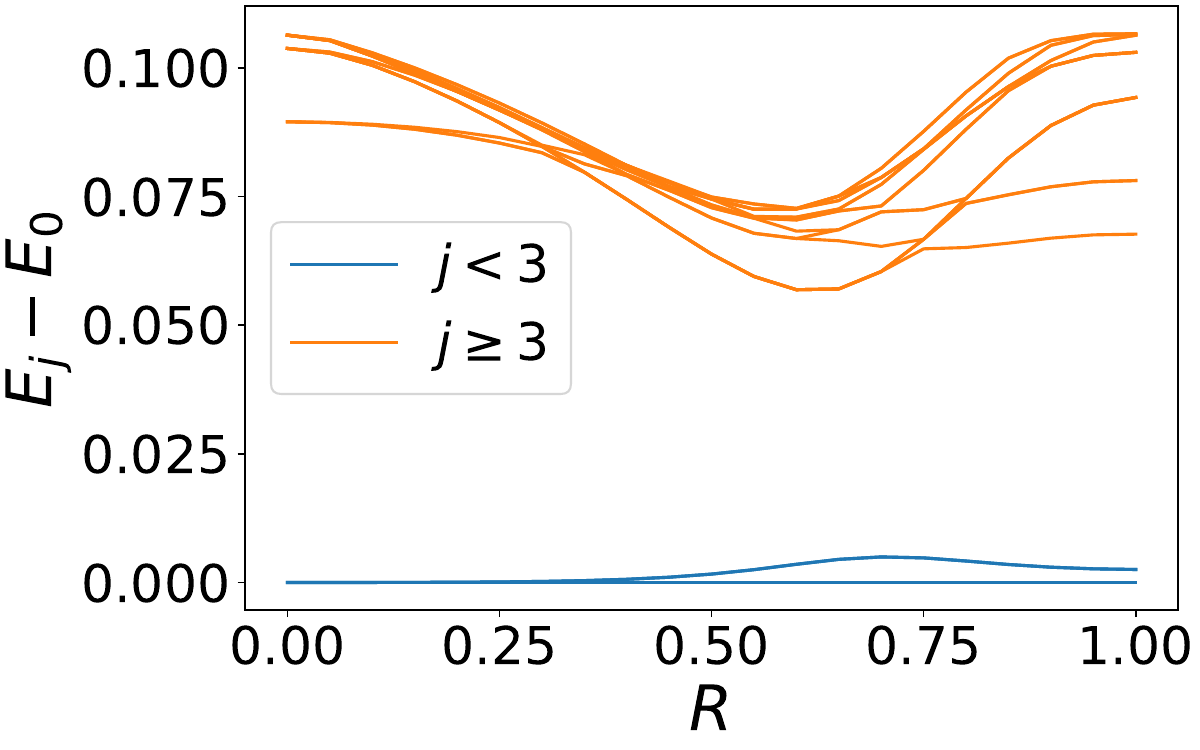}
    \caption{First few many-body states across the adiabatic path for $N=8$. The first three states are colored in blue, while the rest are in orange. }
    \label{fig:new_adiabaticgs}
\end{figure}
The adiabatic relation for the ground state of the $1/3$ Laughlin-like phase between the Harper-Hofstadter model and Checkboard lattice has been established in ~\cite{wuAdiabatic2012}. In order to provide self-contained material for the excitation adiabatic relation present in the main text. In this section, we show the three-fold ground state degeneracy across the adiabatic path. The $3-$fold degeneracy is clear and protected by a large energy gap throughout the interpolation, consistent with the adiabatic relation for graviton spectrum shown in the main text.

\bibliographystyle{longapsrev4-2}
\bibliography{graviton.bib}

\end{document}